\documentclass{article}
\usepackage[english]{babel}
\usepackage{amsmath,amssymb,graphicx,enumerate,theorem}
\usepackage{url}

\usepackage[breaklinks]{hyperref}
\usepackage{microtype}
\usepackage{hyphenat}
\newcommand\norm[1]{\left\lVert#1\right\rVert}
\usepackage[table]{xcolor}
\usepackage[margin=1in]{geometry}
\usepackage{csquotes} 
\usepackage{makecell}
\usepackage{comment}
\newcommand{\cdummy}{\cdot}
\usepackage{mathtools}
\newcommand{\mathD}{\mathrm{D}}
\newcommand{\mathd}{\mathrm{d}}
\newcommand{\mathe}{\mathrm{e}}
\newcommand{\nocomma}{}
\usepackage[export]{adjustbox}
\newcommand{\textdots}{...}

\newcommand{\tmmathbf}[1]{\ensuremath{\boldsymbol{#1}}}

\usepackage{float}

\newenvironment{enumerateroman}{\begin{enumerate}[i.] }{\end{enumerate}}
\newenvironment{itemizedot}{\begin{itemize} }{\end{itemize}}
{\theorembodyfont{\rmfamily}\newtheorem{remark}{Remark}}

\usepackage{cancel}

\usepackage[boxruled]{algorithm2e}

\begin{document}

\title{
An associated model of Coulomb friction based on the real area of contact
}

\author{
  Manon Thbaut, Laura De Lorenzis
}

\maketitle

\begin{abstract}
Coulomb friction is a non-associated law: it contains a threshold that depends on the contact pressure, i.e. a stress that is not conjugate to any internal variable, hence it admits no dissipation potential in the sense of generalized standard materials (GSM), and the incremental frictional contact problem is not a minimization problem. In this paper, we propose an alternative friction model that retains Coulomb-like behavior while being associated and fully GSM-compatible. Key to our formulation is the introduction of a new internal variable motivated by the microscopic contact state of rough surfaces, whose evolution is driven by the plastic deformation of the asperities. Under a non-decreasing contact pressure and below a threshold pressure, the model recovers Coulomb friction up to an asperity-scale penetration depth. This depth is physically meaningful and can be calibrated to reproduce known tribology results. Under decreasing contact pressures, the model exhibits a  memory effect due to irreversible plastic deformation of the asperities, whereas above the pressure threshold it yields a saturation of the frictional traction. We present both penalty and Lagrange multiplier formulations of the model. A 2D finite element implementation of the penalty approach confirms excellent agreement with the classical Coulomb predictions across two benchmark tests.
\end{abstract}


\section{Introduction} \label{Intro}
The framework of  generalized standard materials (GSM), also referred to as standard dissipative materials, is rooted in the fundamental results of Onsager~\cite{onsager1931-1,onsager1931} on the thermodynamics of irreversible processes. Onsager started from the classical description of the kinetics of irreversible processes by a system of linear equations relating the fluxes, i.e., the rates of these processes, such as heat fluxes and reaction rates, to their conjugate thermodynamic forces, such as temperature gradients and chemical affinities, and proved the symmetry of its coefficient matrix, expressed by what are now known as the Onsager reciprocal relations. This symmetry, in turn, allows the linear kinetic equations to be derived from a quadratic dissipation potential and to be cast in the form of a variational principle, known in the soft-matter community as Onsager's variational principle~\cite{doi2011}.

Onsager's framework was extended to material modeling by many authors. Meixner~\cite{meixner1954} and Biot~\cite{biot1954,biot1955} combined it with
internal variables to describe relaxation phenomena and viscoelasticity, and Biot also formulated the corresponding variational principles; Ziegler~\cite{ziegler1957,ziegler1958,ziegler1972} generalized it to nonlinear relations between thermodynamic forces and fluxes; Rice~\cite{rice1970,rice1971} identified the conditions under which the evolution laws of inelastic solids
derive from a flow potential; Edelen~\cite{edelen1972,edelen1973} showed that a
general nonlinear flux--force relation splits into a part derived from a dissipation potential and a non-dissipative part; and Moreau~\cite{moreau1970,moreau1971} extended the notion of dissipation potential to non-differentiable convex functions, thereby including rate-independent processes such as plasticity and Tresca friction. Building on these works, Halphen and Nguyen~\cite{halphen1975} introduced the notion of GSM; see \cite{germain1983,ziegler1987,maugin2015} for reviews. 

In the GSM framework, the behavior of a wide class of materials is fully characterized by two scalar functions or thermodynamic potentials: the free-energy density, which depends on the observable variables and on a set of internal variables, and the dissipation
potential, which depends on the rates of the internal variables. Unlike in Onsager's work, these potentials are convex but not necessarily quadratic, and may therefore lead to nonlinear relations between the thermodynamic forces and the rates of the internal variables, which play here the role of the fluxes. Convexity of the dissipation potential, together with its non-negativity and its vanishing at zero rates, guarantees thermodynamic consistency by construction. In the extended form of the framework \cite{hackl1997, mielke2003}, the dissipation potential may also depend on the internal variables themselves but, as in the original form, neither on the observables nor on the stresses. This is the sense in which GSM-compatibility is understood throughout the paper. Owing to its generality, the framework covers a wide range of possibly non-linear constitutive behaviors, including elasticity, viscosity, plasticity and damage (see \cite{steinmann2021} for a recent catalogue of one-dimensional models). Moreover, it is systematic, which makes the combination of several dissipative processes particularly simple \cite{arroyo2018}.
In addition, the evolution of the internal variables is governed by a minimization principle in rate form which, under suitable conditions, is equivalent to the
principle of maximum dissipation (see~{\cite{hackl2007}} for details). 
This rate formulation can be further discretized in time, which yields an incremental minimization principle: at each time step, the state at the end of the step minimizes a single convex functional~{\cite{ortiz1999,mielke2003}}. Such a structure is particularly well suited  to quasi-static bifurcation analysis~{\cite{nguyen1994}}, to rigorous proofs of the existence of solutions~{\cite{mielke2003}}, and
to numerical implementation and error estimation, including adaptive finite element methods~{\cite{radovitzky1999}}.

Although many material behaviors comply with this framework, some dissipation
laws, referred to as non-associated, remain incompatible with it. The Coulomb friction law is a case in point. The root cause is that the contact pressure
entering the friction threshold is not conjugate to any internal variable
in the free-energy density, since the displacement during slip is purely tangential
(see Section~\ref{sec:outline}). In the terms of Rice~\cite{rice1971}, a flow potential, and hence normality, is guaranteed when the rate of each internal variable depends on the stress only through its own conjugate force; Coulomb friction violates this condition, since the slip rate depends not only on its conjugate force, the tangential traction, but also on the contact pressure. This incompatibility makes the existence of solutions notably difficult to
establish~{\cite{ballard2024}}; it also yields a non-symmetric stiffness
matrix, and hence an increased computational cost of finite element
simulations~{\cite{laursen2003}}.
Several workarounds to this long-standing issue have been proposed, also in the context of non-associated plasticity. They fall into two broad categories.

The first category aims at rigorous existence results and gives up a minimization principle. The problem is  cast as a quasi-variational inequality, which amounts to a Tresca problem whose threshold depends on the unknown normal stress, and existence is obtained through a fixed-point argument on this threshold~{\cite{necas1980,panagiotopoulos1985,raous1988}}. This
approach does not provide the energy-based stability and bifurcation criteria available for GSM~\cite{nguyen1994} and, more importantly, the fixed-point argument yields existence only for small friction coefficients. This restriction can be relaxed by mollifying the normal stress through convolution with a smooth kernel, the point of view adopted in non-local friction~{\cite{duvaut1980,oden1983,cocu1984,francfort2018a}}, where the convolution product is interpreted as a way to mimic the stress
distribution at the scale of the asperities; a similar mollification underlies the existence results available for non-associated plasticity~\cite{babadjian2012, francfort2018a}. However, the regularization alters the
frictional behavior, and the convolution product makes the model significantly more complex and more expensive to simulate. More recently, existence for the time-incremental problem has been proved in
two dimensions, for arbitrarily large friction coefficients in the isotropic
case, by recasting it as a variational inequality with a nonlinear
operator~\cite{ballard2024}. This sharp result is an existence theory rather
than a computational framework, and it currently relies on a property of the
elastic Neumann-to-Dirichlet operator established only in two dimensions.

The second category aims primarily at restoring a variational structure that allows for a simple numerical implementation. Most of these works operate at the continuous level: they extend the framework of GSM to dissipation potentials that depend on the stress or on other state variables~{\cite{collins1997,houlsby2006,francfort2018b,ulloa2021}}, or generalize the principle of maximum dissipation~{\cite{polizzotto1998,petryk2022,hackl2020}}. The bipotential method, designed specifically for Coulomb friction and extended to non-associated plasticity, generalizes the GSM framework by
replacing the dissipation potential with a function of both the rates and the forces~\cite{desaxce1991,desaxce1998}. To our knowledge, none of these formulations has led to existence results without a regularization of the stress; see~\cite{stefanelli2025} for non-associated plasticity. A possible cause,
common to several of them~{\cite{desaxce1998,houlsby2006,francfort2018b,ulloa2021}}, is that a 
stress variable is used either as the main unknown or as a parameter of the
dissipation potential, which raises the same difficulties related
to the lack of smoothness of the stress as the classical variational
inequality~{\cite{necas1980}}. As an alternative to these continuous formulations, a dedicated numerical scheme has been proposed that restores
the variational structure directly in the time-discretized setting~{\cite{pandolfi2002}}. There, however, the incremental model coincides
with Coulomb friction only in the limit of infinitesimal time increments: for the finite time increments used in practical
numerical implementations, the scheme produces a spurious additional contact
force, which scales as $\mathcal{O} (\Delta t)$. A different route has recently been taken for fault gouge~\cite{agajanian2026}: the frictional response is not postulated as a contact law but derived, with associated flow, from a critical-state (Cam-clay) model of the sheared granular layer, in which the elastic domain is scaled by a consolidation pressure that depends on the volumetric plastic strain; a shear stress proportional to the normal stress then emerges at the critical state, i.e.\ only after sufficient slip and in the rate-independent limit.

In summary, none of the above approaches has reconciled Coulomb friction, understood as a contact law between two surfaces, with the classical framework of GSM. The present paper addresses this open issue by proposing a new friction model that is associated, in the sense that the evolution of its internal variables obeys normality with respect to a convex domain in the space of their conjugate forces, parametrized by the internal variables alone, and hence GSM-compatible. Our core idea is
close to the one underlying non-local
friction~{\cite{duvaut1980,oden1983,cocu1984,francfort2018a}}, namely to keep track
of the state of contact  at the microscopic level, but the strategy we follow is different: motivated by  classical models of rough-surface contact {\cite{bowden1954,greenwood1966,greenwood1967}}, we introduce a new internal variable, which represents the local mean ratio of the real area of contact, accounting for
the asperities present on the contact surface, to the nominal one. We then combine two weakly coupled dissipative processes, each of them associated: we reformulate the Coulomb law as a Tresca friction whose threshold is parametrized by the contact area ratio, while the evolution of the contact area ratio itself is assumed to be dictated by the perfectly plastic deformation of the asperities. The pressure dependence of the friction threshold is thus carried by an internal variable rather than by a stress, as in the critical-state model of~\cite{agajanian2026}; unlike there, slip remains purely tangential, since normality is not imposed in the space of the contact tractions, and the Coulomb law is recovered from the onset of slip rather than at a steady state only.

With our approach, we dispense with the cumbersome convolution product of non-local friction and obtain a model that inherits all the convenient properties of the GSM framework: it is thermodynamically consistent and variational, in both the rate and the incremental sense, and can therefore be implemented easily in a finite element framework. The variational structure of the problem also naturally restores the symmetry of the stiffness matrix. As long as the contact pressure does not decrease and remains below a saturation threshold, our model coincides with Coulomb friction, up to an additional asperity-scale penetration. As a direct consequence of the irreversible nature of the asperity deformation, the model exhibits a memory effect when the pressure decreases and a saturation of the frictional traction at sufficiently large pressure, which are both absent from the classical Coulomb friction model.

In its time-discretized form, the model departs from Coulomb friction by one further quantity, of order $\Delta t$: the friction threshold is evaluated with the internal state of the previous time step and thus lags the current pressure by one increment. Both this lag and the discrepancy of the scheme of~\cite{pandolfi2002} are $\mathcal{O} ( \Delta t )$, but they differ in nature. There, the variational structure exists only at the incremental level; the limit $\Delta t \rightarrow 0$ corresponding to Coulomb friction admits no such structure, and the price paid at finite $\Delta t$ is a spurious force. In our model, the variational structure exists already at the continuous level, as a rate principle; the incremental problem is the time discretization of that principle, so that its limit $\Delta t \rightarrow 0$ is itself variational, and the price paid at finite $\Delta t$ is a lag of the friction threshold, which vanishes with the time step.

We develop our approach for a three-dimensional elastic body in frictional contact with a flat rigid substrate. Section~\ref{sec:outline} states the problem, the incompatibility between Coulomb friction and GSM, and the results of~\cite{bowden1954,greenwood1966,greenwood1967} on which the new model relies. In Section~\ref{sec:formulation},
we derive penalty formulations for the new model in two variants. For each variant, we present the corresponding contact problem in strong form and its continuous variational formulation, and then discretize the latter in time to obtain a global incremental energy minimization, following~{\cite{ortiz1999,mielke2003}}. We also develop a Lagrange multiplier formulation, which enforces the contact constraints exactly. In Section~\ref{sec:numerical-results}, we implement the penalty formulations in a 2D finite element framework and compare their predictions with those of classical Coulomb friction. 

We consider mechanical fields that may depend
both on the time $t$ and on the space coordinate $\tmmathbf{x} \in
\mathbb{R}^3$ in the cartesian frame $(\tmmathbf{e}_1, \tmmathbf{e}_2,
\tmmathbf{e}_3)$. We use Einstein summation rule on repeated indices. Simple
and double contractions are denoted by $\cdot$ and $:$, respectively, and 
the Euclidean norm by $\| \cdummy \|$. 
Time derivatives are indicated by a superposed dot, while $\nabla$ is the gradient operator and $\otimes$ the dyadic product. We also denote by $\langle \cdummy \rangle_+$ the Macaulay bracket, which returns the
positive part of its operand,
\begin{equation}
  \langle x \rangle_+ = \left\{\begin{array}{ll}
    x \enspace & \text{if } x \geqslant 0,\\
    0 & \text{if } x < 0.
  \end{array}\right. \label{eq:macaulay-bracket}
\end{equation}
For smooth functionals, differentiability is understood in the sense of the
G{\^a}teaux derivative and we denote as $\mathD_{\tmmathbf{u}} \Phi
(\tmmathbf{u}, \tmmathbf{v}) [\tmmathbf{h}]$ the directional derivative with
respect to $\tmmathbf{u}$, evaluated in direction $\tmmathbf{h}$.
For non-smooth functionals, differentiability is understood in the sense of the 
subdifferential~\cite{rockafellar1970}.

  \section{The Coulomb frictional contact problem and its lack of variational structure}\label{sec:outline}
  
This section states the problem addressed in the paper. Section~\ref{sec:contact-problem} gives the frictional contact problem in strong form. Section~\ref{sec:objective} formulates the question: can the contact conditions of Coulomb friction be derived from a free-energy density and a dissipation potential, as in the framework of GSM? We show that this question reduces to identifying energy contributions localized on the contact boundary. Section~\ref{sec:lack-of-normality} explains why Coulomb friction resists such a derivation and lacks normality, and examines the variant obtained by imposing normality in the space of the contact tractions. Finally, Section~\ref{sec:tribology-result} recalls the classical results on the real area of contact of rough surfaces~\cite{bowden1954,greenwood1966, greenwood1967} on which the new model of Section~\ref{sec:formulation} relies.

\subsection{Coulomb frictional contact problem}
\label{sec:contact-problem}
\begin{figure}[H]
\begin{center}
  \raisebox{-0.48093916914528\height}{\includegraphics[width=7.32714318509773cm,height=4.18442542306179cm]{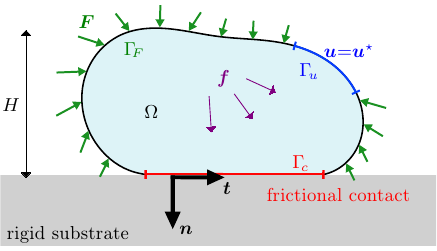}}
  \caption{Scheme of the contact problem: a three-dimensional elastic body $\Omega$ in
  frictional contact with a flat rigid surface. Contact occurs on the
  part~$\Gamma_c$ of the boundary and the slip direction is denoted by $\tmmathbf{t}$. \label{fig:scheme}}\end{center}
\end{figure}

 We consider a 3D linear elastic
body occupying the set of points $\tmmathbf{x} \in \Omega$, see
Figure~\ref{fig:scheme}, of characteristic height $H$ and with unit outward normal $\tmmathbf{n}$ on its boundary $\partial \Omega$. 
The boundary is partitioned into three disjoint parts: $\partial
\Omega = \Gamma_c \bigcup \Gamma_u \bigcup \Gamma_F$. On $\Gamma_u$ and $\Gamma_F$, Dirichlet and Neumann boundary conditions are imposed, with prescribed displacement $\tmmathbf{u}^{\star} (\tmmathbf{x}, t)$ on $\Gamma_u$ and applied  traction $\tmmathbf{F} (\tmmathbf{x}, t)$ on $\Gamma_F$. In addition, a body force $\tmmathbf{f}(\tmmathbf{x}, t)$  
is applied in the bulk of the elastic body.
The body may come into contact with a flat rigid substrate along the part $\Gamma_c$ of its boundary. Consistently with the linearized kinematics adopted for the bulk, the normal $\tmmathbf{n}$ on $\Gamma_c$ is identified with the constant unit normal of the substrate, pointing into it\footnote{Where $\Gamma_c$ is curved, as in the benchmarks of Section~\ref{sec:numerical-results}, the outward normal of the body differs from the substrate normal in the contact zone by terms of the order of the ratio of the contact width to the radius of curvature of $\Gamma_c$, which are neglected.}. The initial gap $g_0$ is measured along $\tmmathbf{n}$, and so are the normal components of displacements and tractions introduced below.

The displacement of the body, $\tmmathbf{u}
(\tmmathbf{x}, t)$, depends on both the spatial coordinate $\tmmathbf{x}$ and
the time $t$, and the strain tensor $\tmmathbf{\varepsilon}
(\tmmathbf{x}, t)$ is defined as its symmetrized gradient,
$\tmmathbf{\varepsilon}= 1/2(\nabla \tmmathbf{u}+ \nabla
   \tmmathbf{u}^T).$
The stress tensor $\tmmathbf{\sigma} (\tmmathbf{x}, t)$ is related to the
strain through
$\tmmathbf{\sigma}=\tmmathbf{c}: \tmmathbf{\varepsilon}$. The elasticity tensor $\tmmathbf{c}$
of the body possesses the minor and major
symmetries and is strictly positive in the sense of quadratic forms,
\begin{equation}
  \forall \,\text{symm.}\, \delta \tmmathbf{\varepsilon} \neq \tmmathbf{0}, \hspace{4em}
  \delta \tmmathbf{\varepsilon}: \tmmathbf{c}: \delta \tmmathbf{\varepsilon}>
  0. \label{eq:bulk-NRJ_CVX}
\end{equation}

The behavior on the
contact boundary $\Gamma_c$ is described by the following set of
inequalities,
\begin{equation}
  \begin{array}{lll}
    \forall t, \hspace{1.5em} \forall \tmmathbf{x} \in \Gamma_c, \qquad & g_0
    - u_n \geqslant 0, & \sigma_{n \nocomma n} \leqslant 0, \hspace{3em} (g_0
    - u_n) \sigma_{n \nocomma n} = 0,\\
    &  & \\
    & \dot{\tmmathbf{u}}_t = - \dot{\gamma}  \frac{\tmmathbf{\sigma}_{n
    \nocomma t}}{\| \tmmathbf{\sigma}_{n \nocomma t} \|} & \enspace
    \text{with} \quad \left\{\begin{array}{l}
      \dot{\gamma} = 0 \hspace{0.7em} \text{if } \hspace{0.7em} \|
      \tmmathbf{\sigma}_{n \nocomma t} \| < \mu | \sigma_{n \nocomma n} |,\\
      \dot{\gamma} \geqslant 0 \hspace{0.7em} \text{if } \hspace{0.7em} \|
      \tmmathbf{\sigma}_{n \nocomma t} \| = \mu | \sigma_{n \nocomma n} |.
    \end{array}\right.
  \end{array} \label{coulomb-law}
\end{equation}
In the above expressions, we have denoted the (uniquely defined) normal and tangential components of the
displacement $\tmmathbf{u}$ and of the traction vector $\tmmathbf{\sigma} \cdot \tmmathbf{n}$ on the contact boundary as $u_n=\tmmathbf{u} \cdot \tmmathbf{n}$ and $\tmmathbf{u}_t = \tmmathbf{u}- (\tmmathbf{u} \cdot \tmmathbf{n})
      \tmmathbf{n}$, and $\sigma_{n \nocomma n}=\tmmathbf{n} \cdot \tmmathbf{\sigma} \cdot
      \tmmathbf{n}$ and $\tmmathbf{\sigma}_{n \nocomma t}=\tmmathbf{\sigma} \cdot \tmmathbf{n}- (\tmmathbf{n} \cdot
      \tmmathbf{\sigma} \cdot \tmmathbf{n}) \tmmathbf{n}$, respectively.
By construction, $\tmmathbf{u}_t \cdot \tmmathbf{n}=\tmmathbf{\sigma}_{n \nocomma t} \cdot \tmmathbf{n}= 0$. 
The first
line of~\eqref{coulomb-law} corresponds to classical Signorini contact
conditions: $g_0$ is the initial gap, $g_0 - u_n$ is the \textit{current }normal gap between the two bodies, and the inequality $g_0 - u_n \geqslant 0$ rules out interpenetration, while $\sigma_{n \nocomma n} \leqslant 0$ enforces a
compressive normal stress whenever contact is active. The second line is the
Coulomb friction law. The amount of sliding $\dot{\gamma}$ is governed by the ratio of the norm of the tangential stress vector $\| \tmmathbf{\sigma}_{n \nocomma t}
\|$ to the normal pressure $| \sigma_{n \nocomma n} |$ at the contact boundary:
sliding $\left( \dot{\gamma} > 0 \right)$ can occur only once this ratio 
reaches the friction coefficient~$\mu$, in which case the material point slides
in the direction $\tmmathbf{t}$, opposite to the tangential stress vector, see Figure~\ref{fig:scheme}.

The displacement of the elastic body is obtained by solving the following boundary-value problem in strong form, for all $t \geqslant 0$:
\begin{equation}
  \left\{\begin{array}{lll}
    \text{div} \,\tmmathbf{\sigma}+\tmmathbf{f}=\tmmathbf{0}, &  \text{in}
    \hspace{0.3em} \Omega & \text{(bulk equilibrium)}\\[2pt]
    
    \tmmathbf{u}=\tmmathbf{u}^{\star}, &  \text{on} \hspace{0.3em}
    \Gamma_u  \quad & \text{(Dirichlet bound.)}\\[2pt]
    
    \tmmathbf{\sigma} \cdot \tmmathbf{n}=\tmmathbf{F}, & \text{on}
    \hspace{0.3em} \Gamma_F & \text{(Neumann bound.)}\\[2pt]
    
    g_0 - u_n \geqslant 0, \quad \sigma_{n \nocomma n} \leqslant 0, \quad
    (g_0 - u_n) \sigma_{n \nocomma n} = 0, &\text{on} \hspace{0.3em}
    \Gamma_c  & \text{(normal contact)}\\[2pt]
    
\| \tmmathbf{\sigma}_{n t} \| \leqslant \mu | \sigma_{n n} |, \qquad
\dot{\tmmathbf{u}}_t \cdot \tmmathbf{\sigma}_{n t} = - \mu | \sigma_{n n} | \, \| \dot{\tmmathbf{u}}_t \| & \text{on}
    \hspace{0.3em} \Gamma_c & \text{(Coulomb
    friction)}
  \end{array}\right. \label{strong-form}
\end{equation}
supplemented by the initial condition $\tmmathbf{u}(\cdot, 0) = \tmmathbf{u}_0$. The friction condition~\eqref{strong-form}$_5$ is equivalent to the second line of~\eqref{coulomb-law}: by the Cauchy--Schwarz inequality, equality in~\eqref{strong-form}$_5$ holds if and only if $\dot{\tmmathbf{u}}_t$ is positively collinear with $- \tmmathbf{\sigma}_{n t}$ and vanishes unless $\| \tmmathbf{\sigma}_{n t} \| = \mu | \sigma_{n n} |$.

\subsection{Objective and reduction to an interface problem} \label{sec:objective}
The strong form~\eqref{strong-form} is not suited to numerical implementation nor to the derivation of proofs of the existence of solutions. In practice, the contact problem is recast in the following weak form, for all $t\geqslant 0$:
\begin{equation}
  \forall \delta \tmmathbf{u} \in \mathcal{C}_0, \hspace{3em} \int_{\Omega}
  (\tmmathbf{\varepsilon} (\tmmathbf{u}) : \tmmathbf{c}:
  \tmmathbf{\varepsilon} (\delta \tmmathbf{u}) -\tmmathbf{f} \cdot \delta
  \tmmathbf{u}) \mathd \tmmathbf{x}- \int_{\Gamma_F} \tmmathbf{F} \cdot \delta
  \tmmathbf{u}\, \mathd s + \int_{\Gamma_c} (P_n \delta u_n +\tmmathbf{T}_t
  \cdot \delta \tmmathbf{u}_t) \,\mathd s = 0, \label{eq:usual-weak-form}
\end{equation}
where $\mathcal{C}_0$ is the set of kinematically admissible perturbations,
\begin{equation}
  \mathcal{C}_0 = \left\{ \delta \tmmathbf{u} (\tmmathbf{x}, t) \in H^1(\Omega)^3 \enspace
  \text{such that} \enspace \delta \tmmathbf{u}=\tmmathbf{0} \enspace
  \text{on } \Gamma_u \right\} . \label{eq:kinematic-admissible-set}
\end{equation}
Integrating the first term by parts and invoking the fundamental lemma of the calculus of
variations shows that~\eqref{eq:usual-weak-form} implies the first three
equations of the strong form~\eqref{strong-form}, together with the following
relations between the contact tractions $P_n, \tmmathbf{T}_t$ and the Cauchy
stress $\tmmathbf{\sigma}$ at the contact boundary,
\begin{equation}
  \begin{array}{lll}
    P_n & = & - \sigma_{n \nocomma n},\\
    \tmmathbf{T}_t & = & -\tmmathbf{\sigma}_{n \nocomma t} .
  \end{array} \label{link-Cauchy-contact}
\end{equation}
The contact tractions $P_n, \tmmathbf{T}_t$ are then constructed so as to ensure
 that~\eqref{strong-form}$_{4,
5}$ are satisfied as well. Their explicit expression depends on the method adopted for the enforcement of the contact constraints (Lagrange
multipliers, penalty, or others) as well as on the tangential contact state, stick
 ($\dot{\gamma} = 0$) or slip ($\dot{\gamma} > 0$). In all cases, once the contact tractions are updated according to \eqref{strong-form}$_{4,5}$,
the weak form \eqref{eq:usual-weak-form} is not the stationarity condition of a potential in $\tmmathbf{u}$, as shown in the discrete setting by the non-symmetric tangent stiffness matrix of the classical finite element implementation given in Section~\ref{C2}.

The goal of the present paper is to overcome this
limitation by identifying a GSM-compatible model of the contact interface, i.e. a free-energy density and a dissipation potential of internal variables, that underlies the strong form~\eqref{strong-form} or an approximation of it: the weak form~\eqref{eq:usual-weak-form} and the contact tractions $P_n, \tmmathbf{T}_t$ then follow from the free-energy density, the evolution of the internal variables from a rate principle, and the time discretization of the latter yields an incremental energy minimization, as in~{\cite{ortiz1999,mielke2003}}.  

The first two terms of the weak form~\eqref{eq:usual-weak-form}, which yield
the first three equations in~\eqref{strong-form}, have a straightforward
variational origin: they follow from the differentiation (in the G{\^a}teaux sense) of the elastic potential
\begin{equation}
  \Phi^{\text{el}} (\tmmathbf{u}) = \int_{\Omega} \left( \frac{1}{2}
  \tmmathbf{\varepsilon} (\tmmathbf{u}) : \tmmathbf{c}: \tmmathbf{\varepsilon}
  (\tmmathbf{u}) -\tmmathbf{f} \cdot \tmmathbf{u} \right) \mathd \tmmathbf{x}-
  \int_{\Gamma_F} \tmmathbf{F} \cdot \tmmathbf{u} \,\mathd s.
  \label{eq:bulk-energy}
\end{equation}
We assume that the Dirichlet boundary $\Gamma_u$ has positive measure, which excludes rigid-body motions: by Korn's inequality~\cite{duvaut1972} and~\eqref{eq:bulk-NRJ_CVX}, the elastic potential~\eqref{eq:bulk-energy} is then strictly convex and coercive on the set of admissible displacements, $\tmmathbf{u} = \tmmathbf{u}^{\star}$ on $\Gamma_u$. All uniqueness statements of the paper rely on this property; the corresponding proofs are collected in Appendix~\ref{appendix:incremental}.

The challenge is to identify a contact energy, additional to $\Phi^{\text{el}}$ and localized
on the contact surface $\Gamma_c$, together with the internal variables on which it depends, such that
\begin{enumerateroman}
  \item \textit{the stationarity of this energy with respect to $\tmmathbf{u}$
  yields the third term of the weak form~\eqref{eq:usual-weak-form}},
  
  \item \textit{the evolution of the internal variables derives from a dissipation potential depending on these variables and on their rates, i.e. the interface model is GSM-compatible, and the resulting evolution laws imply the remaining contact conditions in~\eqref{strong-form}$_{4, 5}$, or an approximation of these conditions.}
\end{enumerateroman}
\begin{figure}[H]
\begin{center}
  \begin{tabular}{ll}
    \raisebox{-0.477454656698016\height}{\includegraphics[width=6.25553423848878cm,height=3.53769841269841cm]{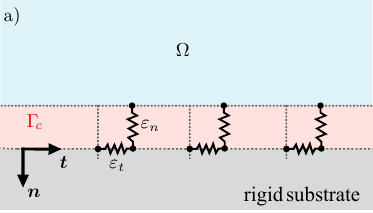}}
    &
    \qquad\raisebox{-0.477976899288686\height}{\includegraphics[width=6.70175783812147cm,height=3.62158926931654cm]{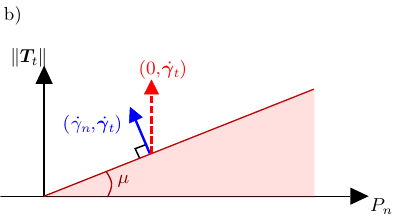}}
  \end{tabular}
  \caption{a) Scheme of the penalty approach: a free-energy density,
  corresponding to distributed springs, is assigned to the contact boundary
  $\Gamma_c$. b) Coulomb cone $\| \tmmathbf{T}_t \| \leqslant \mu P_n$ in the space of contact tractions (light red). The internal variables can evolve only when the tractions reach the boundary of the cone (solid dark
  red line). The
  arrows represent the rates of the internal variables for the two laws discussed in Section~\ref{sec:lack-of-normality}: for Coulomb friction (red), the slip rate is purely tangential and is not normal to the cone; for its dilatant variant (blue), normality is enforced, at the price of a normal opening rate proportional to the slip rate. \ \label{fig:penalty}}    
\end{center}
\end{figure}

In order to identify such a contact energy, we
 first develop a penalty approach (Sections \ref{sec:formulation} and \ref{sec:numerical-results}), in which the contact boundary is assumed to behave as an assembly of springs in both the normal and the tangential directions,  with respective spring constants $\varepsilon_n,\varepsilon_t>0$, see Figure~\ref{fig:penalty}a. This allows us to treat the
contact interface independently of the bulk elasticity, which is convenient for exploring different contact formulations, and yields a simple numerical implementation.
The major drawback of penalty formulations is that they reproduce the 
inequality constraints, such as those in~\eqref{strong-form}$_{4, 5}$, only in the
limit
\[ \varepsilon_n, \varepsilon_t \rightarrow + \infty . \]
For large but finite values of the penalty parameters, the constraints are
enforced only approximately, \textit{i.e.} small violations of the
constraints are permitted.
Section~\ref{LM} removes this limitation by developing a Lagrange multiplier formulation, which enforces the inequality constraints exactly.

\subsection{A long-standing issue: ``lack of
normality''}\label{sec:lack-of-normality}

\paragraph{Pressure-dependent law incompatible with GSM} The contact free-energy density used in classical penalty formulations takes the form
\begin{equation}
  \psi (u_n, \tmmathbf{u}_t, \tmmathbf{\gamma}_t) = \frac{\varepsilon_n}{2} 
  \langle u_n - g_0 \rangle^2_+ + \frac{\varepsilon_t}{2}  \| \tmmathbf{u}_t
  -\tmmathbf{\gamma}_t \|^2, \label{classical-free-energy}
\end{equation}
and depends on the normal
displacement $u_n$, on the tangential displacement $\tmmathbf{u}_t$, and on a new
variable~$\tmmathbf{\gamma}_t$ representing the amount of tangential slip, see Remark~\ref{abusive-notation} for details. In GSM terminology, $(u_n, \tmmathbf{u}_t)$ are \textit{observables}, which satisfy balance laws, while  $\tmmathbf{\gamma}_t$ is an \textit{internal variable}, hidden and governed by an evolution law.

The observables are the primary variables of the contact problem: at any time $t$ and $\textit{\text{for a
given }} \tmmathbf{\gamma}_t$, they solve the equilibrium problem
\begin{equation}
  \underset{\tmmathbf{u}=\tmmathbf{u}^{\star} \text{on}\,
  \Gamma_u}{\underset{\tmmathbf{u}}{\text{Min}}} \Phi^{\text{el}}
  (\tmmathbf{u}) + \int_{\Gamma_c} \psi (u_n, \tmmathbf{u}_t, \ldots) \mathd s
  \label{v}
\end{equation}
where the ellipsis $\ldots$ stands for the internal variables on which $\psi$ depends (here $\tmmathbf{\gamma}_t$ only) and $\Phi^{\text{el}} (\tmmathbf{u})$ is the elastic potential~\eqref{eq:bulk-energy}. 
Since $\Phi^{\mathrm{el}}$ is strictly convex and coercive (Section~\ref{sec:objective}) and $\int_{\Gamma_c} \psi \, \mathrm{d} s$ is continuous and convex in $\tmmathbf{u}$, this problem has a unique solution, and its stationarity condition is the weak form~\eqref{eq:usual-weak-form} with
\begin{equation}
  \begin{array}{lllll}
    P_n & = & \frac{\partial \psi}{\partial u_n} & = & \varepsilon_n  \langle
    u_n - g_0 \rangle_+,\\
    \tmmathbf{T}_t & = & \frac{\partial \psi}{\partial \tmmathbf{u}_t} & = &
    \varepsilon_t  (\tmmathbf{u}_t -\tmmathbf{\gamma}_t) .
  \end{array} \label{eq:conventional-contact-stresses}
\end{equation}
In GSM terminology, the derivatives of the free-energy density with respect to its arguments are the \textit{energetic forces}: those with respect to the observables are the contact tractions~\eqref{eq:conventional-contact-stresses}, and those with respect to the internal variables define, with the opposite sign, the \textit{dissipative forces}. For the free-energy density~\eqref{classical-free-energy}, the dissipative force conjugate to $\tmmathbf{\gamma}_t$ is the tangential traction itself,
\begin{equation}
  - \frac{\partial \psi}{\partial \tmmathbf{\gamma}_t} = \varepsilon_t  (\tmmathbf{u}_t -\tmmathbf{\gamma}_t) = \tmmathbf{T}_t .
  \label{eq:dissipative-force-classical}
\end{equation}
The evolution law of an internal variable is a relation between its rate and its dissipative force, and the products of the dissipative forces with the rates add up to the dissipation $\mathcal{D}$. In this case, $\mathcal{D} = \tmmathbf{T}_t \cdot \dot{\tmmathbf{\gamma}}_t$.

The free-energy density~\eqref{classical-free-energy} thus
meets requirement i. It is also consistent with the normal contact conditions~\eqref{strong-form}$_4$ in the limit $\varepsilon_n \rightarrow + \infty$:
by~\eqref{link-Cauchy-contact}$_1$, $- \sigma_{n \nocomma n} = P_n
\geqslant 0$, and the term $\varepsilon_n  \langle u_n - g_0 \rangle^2_+ / 2$ penalizes the violations of the constraint~$g_0 - u_n
\geqslant 0$.

Requirement ii., by contrast, cannot be fulfilled with the variables $u_n,  
\tmmathbf{u}_t, \tmmathbf{\gamma}_t$ only. To see this, 
we write the Coulomb law~\eqref{strong-form}$_5$ in
terms of the slip $\tmmathbf{\gamma}_t$ and of the contact tractions~\eqref{eq:conventional-contact-stresses}, using~\eqref{link-Cauchy-contact} and replacing the tangential velocity $\dot{\tmmathbf{u}}_t$ by the slip rate $\dot{\tmmathbf{\gamma}}_t$, with which it coincides in the original problem (Remark~\ref{abusive-notation}):
\begin{equation}
\forall t, \ \forall \tmmathbf{x} \in \Gamma_c, \qquad
\| \tmmathbf{T}_t \| \leqslant \mu P_n, \qquad
\tmmathbf{T}_t \cdot \dot{\tmmathbf{\gamma}}_t = \mu P_n \, \| \dot{\tmmathbf{\gamma}}_t \| .
  \label{eq:coulomb-friction-for-penalty}
\end{equation}
The slip can evolve only for traction states on the boundary of the cone $\| \tmmathbf{T}_t \| \leqslant \mu P_n$ (Figure~\ref{fig:penalty}b, red line); by~\eqref{eq:coulomb-friction-for-penalty}$_2$ and the Cauchy--Schwarz inequality, $\dot{\tmmathbf{\gamma}}_t$
is positively collinear with $\tmmathbf{T}_t$, and the dissipation reads
\begin{equation}
  \mathcal{D}= \mu P_n  \| \dot{\tmmathbf{\gamma}}_t \| \geqslant 0.
  \label{eq:coulomb-dissipatin}
\end{equation}
GSM-compatibility requires the evolution of~$\tmmathbf{\gamma}_t$ to derive from a minimum
principle of the form: \textit{for given}~$\tmmathbf{u},
\tmmathbf{\gamma}_t, \dot{\tmmathbf{u}}$, \textit{and given}~$(t, \tmmathbf{x}) \in\mathbb{R}^+ \times \Gamma_c$, \textit{the rate }$\dot{\tmmathbf{\gamma}_t}$ \textit{solves}
\begin{equation}
  \underset{\dot{\tmmathbf{\gamma}}_t}{\text{Min}} \enspace \dot{\psi}
  (u_n, \tmmathbf{u}_t, \tmmathbf{\gamma}_t, \dot{u}_n, \dot{\tmmathbf{u}}_t,
  \dot{\tmmathbf{\gamma}}_t) + \varphi (\dot{\tmmathbf{\gamma}}_t ;
  \tmmathbf{\gamma}_t), \label{eq:max-dissipation-principle}
\end{equation}
where $\varphi (\dot{\tmmathbf{\gamma}}_t ; \tmmathbf{\gamma}_t)$ is a convex
dissipation potential depending on the rate~$\dot{\tmmathbf{\gamma}}_t$ and possibly on~$\tmmathbf{\gamma}_t$ itself~\cite{hackl1997,mielke2003}.
Since $\partial \dot{\psi} / \partial \dot{\tmmathbf{\gamma}}_t = \partial \psi / \partial \tmmathbf{\gamma}_t$, 
the stationarity condition of~\eqref{eq:max-dissipation-principle} yields 
\begin{equation}
  \tmmathbf{T}_t = - \frac{\partial \dot{\psi}}{\partial
  \dot{\tmmathbf{\gamma}}_t} \in \partial \varphi (\dot{\tmmathbf{\gamma}}_t ;
  \tmmathbf{\gamma}_t), \label{eq:conventional-evolution-law}
\end{equation}
where $\partial \varphi$ is
the subdifferential of $\varphi$: the dissipative force~\eqref{eq:dissipative-force-classical} belongs to the subdifferential of the dissipation potential, which is the normality rule. Since~$P_n$ is absent
from~\eqref{eq:conventional-evolution-law}, no choice of $\varphi$ can reproduce the pressure-dependent threshold
in~\eqref{eq:coulomb-friction-for-penalty}$_1$ or the dissipation~\eqref{eq:coulomb-dissipatin}. These would require the pressure-dependent potential
\begin{equation}
  \varphi ( \dot{\tmmathbf{\gamma}}_t ; P_n ) = \mu P_n \, \| \dot{\tmmathbf{\gamma}}_t \| ,
  \label{eq:coulombpot}
\end{equation}
which the GSM framework precludes (Section~\ref{Intro}): a dissipation potential may be parametrized by the internal variables, but not by a stress, and a stress can drive the evolution of the internal variables only if it is a dissipative force, i.e. if it is conjugate to an internal variable, which $P_n$ is not. This is the long-standing incompatibility between pressure-dependent laws and GSM recalled in the introduction~{\cite{maier1979,collins1997,desaxce1998}}.

Note that, at fixed pressure, \eqref{eq:coulombpot} is a perfectly acceptable potential, convex and positively homogeneous in~$\dot{\tmmathbf{\gamma}}_t$, and the normality rule $\tmmathbf{T}_t \in \partial \varphi ( \dot{\tmmathbf{\gamma}}_t ; P_n )$ with $P_n$ frozen is exactly the Coulomb law~\eqref{eq:coulomb-friction-for-penalty}~\cite{michalowski1978,curnier1984}; likewise, the equilibrium problem~\eqref{v} is well posed at frozen internal variables. What is lost is the coupling between the two, since $P_n$ depends on the unknown observable $u_n$.

\begin{remark}
  \label{rmk:evacuate-u} The remainder of this section and Section~\ref{sec:formulation}  explore alternative expressions for the contact free-energy density $\psi$
  in~\eqref{classical-free-energy}. They all share the  tangential
  elastic term $\varepsilon_t  \| \tmmathbf{u}_t -\tmmathbf{\gamma}_t \|^2 /
  2$ and are quadratic in~$u_n$. For all these models, the observable $\tmmathbf{u}$ is obtained by solving the same minimization problem~\eqref{v}, with all internal variables frozen, and the weak form~\eqref{eq:usual-weak-form} remains valid, with a model-dependent expression of the contact pressure~$P_n$, so that requirement i. is automatically satisfied. In the remainder, we focus on deriving the expression of the contact pressure~$P_n$  for each model, checking its consistency with the normal contact conditions in~\eqref{strong-form}$_4$,
  and identifying a dissipation potential~$\varphi$ from which the evolution laws for the internal variables follow through a minimization principle of the form~\eqref{eq:max-dissipation-principle}.
\end{remark}

\begin{remark}
  \label{abusive-notation}The distinction between the tangential
  displacement~$\tmmathbf{u}_t$, an observable, and the amount of
  slip~$\tmmathbf{\gamma}_t$, an internal variable, is meaningful only within a
  penalty approach; the two quantities coincide  in the   original contact problem~\eqref{strong-form}. The penalty formulation
  presented here, like any other penalty approach,  is a regularized version of
  the problem~\eqref{strong-form} that accounts for an additional elastic
  displacement $\tmmathbf{u}_t -\tmmathbf{\gamma}_t$, and the original
  problem~\eqref{strong-form} is recovered only formally in the limit
  $\varepsilon_t \rightarrow \infty$, which yields $\tmmathbf{u}_t \rightarrow
  \tmmathbf{\gamma}_t$. The distinction is not needed when an exact constraint enforcement approach, such
 as Lagrange multipliers, is used, see Section~\ref{LM}.
\end{remark}


\paragraph{A dilatant version with normality in traction space} Pressure-dependent laws such as Coulomb are termed ``non-associated'' or ``lacking normality'' because the slip rate is not normal to the friction cone in the space of the contact tractions $( P_n, \tmmathbf{T}_t )$ (Figure~\ref{fig:penalty}b, red arrow). It is instructive to examine the variant obtained by imposing normality in that space, which restores GSM-compatibility by construction.

We consider the alternative free-energy density
\begin{equation}
  \psi (u_n, \tmmathbf{u}_t, \gamma_n, \tmmathbf{\gamma}_t) =
  \frac{\varepsilon_n}{2}  \langle u_n - g_0 - \gamma_n \rangle^2_+ +
  \frac{\varepsilon_t}{2}  \| \tmmathbf{u}_t -\tmmathbf{\gamma}_t \|^2,
  \label{eq:free-nrj-dilatant}
\end{equation}
which involves an additional dissipative internal variable $\gamma_n$ accounting for the evolution of the displacement in the normal direction. The contact pressure becomes
\[ \begin{array}{lllll}
     P_n & = & \frac{\partial \psi}{\partial u_n} & = & \varepsilon_n  \langle
     u_n - g_0 - \gamma_n \rangle_+,
   \end{array} \]
while the tangential traction $\tmmathbf{T}_t$ is unchanged, given by~\eqref{eq:conventional-contact-stresses}$_2$, and the dissipative forces conjugate to the two internal variables are the contact tractions themselves,
\begin{equation}
  - \frac{\partial \psi}{\partial \gamma_n} = P_n, \hspace{4em}
  - \frac{\partial \psi}{\partial \tmmathbf{\gamma}_t} = \tmmathbf{T}_t.
  \label{eq:dissipative-forces-dilatant}
\end{equation}
The pressure dependence of the friction law can now be carried by an internal variable. The minimum principle~\eqref{eq:max-dissipation-principle} extends to dissipation
potentials $\varphi (\dot{\gamma}_n, \dot{\tmmathbf{\gamma}}_t ; \gamma_n,
\tmmathbf{\gamma}_t)$ that depend on the new internal variable and on its rate: \textit{for given} $\tmmathbf{u}, \gamma_n, 
\tmmathbf{\gamma}_t, \dot{\tmmathbf{u}}$, \textit{and given} $(t, \tmmathbf{x}) \in\mathbb{R}^+ \times \Gamma_c$, \textit{the rates }$(\dot{\gamma}_n, \dot{\tmmathbf{\gamma}_t})$ \textit{solve the joint minimization}
\[    \underset{(\dot{\gamma}_n, \dot{\tmmathbf{\gamma}}_t)}{\text{Min}}
   \enspace \dot{\psi} (u_n, \tmmathbf{u}_t, \gamma_n, \tmmathbf{\gamma}_t,
   \dot{u}_n, \dot{\tmmathbf{u}}_t, \dot{\gamma}_n, \dot{\tmmathbf{\gamma}}_t)
   + \varphi (\dot{\gamma}_n, \dot{\tmmathbf{\gamma}}_t ; \gamma_n,
   \tmmathbf{\gamma}_t) . \]
The corresponding stationarity condition reads
\begin{equation}
  \left(\begin{array}{c}
    P_n\\
    \tmmathbf{T}_t
  \end{array}\right) = - \left(\begin{array}{c}
    \partial \dot{\psi} / \partial \dot{\gamma}_n\\
    \partial \dot{\psi} / \partial \dot{\tmmathbf{\gamma}}_t
  \end{array}\right) \in \partial \varphi (\dot{\gamma}_n,
  \dot{\tmmathbf{\gamma}}_t ; \gamma_n, \tmmathbf{\gamma}_t),
  \label{eq:associated-flow-rule}
\end{equation}
 which implies that the rates $(\dot{\gamma}_n, \dot{\tmmathbf{\gamma}}_t)$ are normal to the Coulomb cone (Figure~\ref{fig:penalty}b, blue arrow).

The resulting model is, however, \emph{not} Coulomb friction. We show in Appendix~\ref{appendix:dilatant-coulomb} that the only rate-independent dissipation potential compatible with the internal variables $(\gamma_n, \tmmathbf{\gamma}_t)$ and whose
elastic domain is the Coulomb cone $\| \tmmathbf{T}_t \| \leqslant \mu P_n$ of~\eqref{eq:coulomb-friction-for-penalty}$_1$ is the indicator function of its polar cone $\left\{ (\dot{\gamma}_n
, \dot{\tmmathbf{\gamma}}_t), \enspace \dot{\gamma}_n + \mu \| \dot{\tmmathbf{\gamma}}_t \| \leqslant 0 \right\}$, so that active
contact ($P_n > 0$) imposes
\begin{equation}
  \forall (t, \tmmathbf{x}) \in \mathbb{R}^+ \times \Gamma_c, \hspace{3em}
  \dot{\gamma}_n = - \mu \| \dot{\tmmathbf{\gamma}}_t \| \leqslant 0.
  \label{eq:dilatancy}
\end{equation}
Slip is accompanied by a normal opening, or dilatancy, of the same order of magnitude as the slip itself, which contradicts the normal contact condition in~\eqref{strong-form}$_4$. We refer to this law as the ``dilatant Coulomb law''. A major issue is that, since the Coulomb cone has its apex at the origin, the rates, normal to it, are orthogonal to the tractions, and the work of the friction force is exactly canceled by the work of the pressure against the opening,
\begin{equation}
  \mathcal{D} = P_n \dot{\gamma}_n + \tmmathbf{T}_t \cdot \dot{\tmmathbf{\gamma}}_t
  = - \mu P_n \| \dot{\tmmathbf{\gamma}}_t \| + \mu P_n \| \dot{\tmmathbf{\gamma}}_t \| = 0,
  \label{eq:zero-dissipation}
\end{equation}
the well-known property of associated Mohr--Coulomb plasticity without cohesion, which dissipates no energy.

The main features of the classical and of the dilatant Coulomb laws are summarized in Table~\ref{table1}. 
\begin{table}[H]
  \centering
  \resizebox{\textwidth}{!}{%
  \begin{tabular}{|c|c|c|c|c|}
    \hline
\textbf{} & \textbf{\makecell{classical\\ Coulomb}} & \textbf{\makecell{dilatant \\ Coulomb}} & \textbf{\makecell{new model \\ (linear)}} &\textbf{\makecell{new model \\ (logarithmic)}} \\ \hline
    internal variables & $\tmmathbf{\gamma}_t$ & $\gamma_n, \tmmathbf{\gamma}_t$ & $\alpha, \tmmathbf{\gamma}_t$ & $\alpha, \tmmathbf{\gamma}_t$
\\ \hline
   \rule{0pt}{20pt} free-energy density $\psi$ & $ \begin{array}{c}
    
        \frac{\varepsilon_n}{2}  \langle u_n - g_0 \rangle^2_+ \\
       +
  \frac{\varepsilon_t}{2}  \| \tmmathbf{u}_t -\tmmathbf{\gamma}_t \|^2
    \end{array}$ & $ \begin{array}{c}
    
        \frac{\varepsilon_n}{2}  \langle u_n - g_0 - \gamma_n \rangle^2_+ \\
         +
  \frac{\varepsilon_t}{2}  \| \tmmathbf{u}_t -\tmmathbf{\gamma}_t \|^2
    \end{array}$ & $ \begin{array}{c}
    
        \frac{\varepsilon_n}{2}  \langle u_n - g_0 -\eta \alpha \rangle^2_+ \\
       +
  \frac{\varepsilon_t}{2}  \| \tmmathbf{u}_t -\tmmathbf{\gamma}_t \|^2
    \end{array}$ & $ \begin{array}{c}
    
        \frac{\varepsilon_n}{2}  \langle u_n - g_0 -\eta \ln (\alpha/\alpha^{\star})\rangle^2_+ \\
       +
  \frac{\varepsilon_t}{2}  \| \tmmathbf{u}_t -\tmmathbf{\gamma}_t \|^2
    \end{array}$ \\[12pt] \hline
    
   \rule{0pt}{25pt} energetic forces & $\begin{array}{cl}
       P_n=&\varepsilon_n  \langle
    u_n - g_0 \rangle_+
     \\
        T_t \hspace{0.25em} =& \varepsilon_t  (\tmmathbf{u}_t -\tmmathbf{\gamma}_t)
    \end{array}$&$ \begin{array}{cl}
       P_n=&\varepsilon_n  \langle
    u_n - g_0 -\gamma_n\rangle_+
     \\
        T_t \hspace{0.25em} =& \varepsilon_t  (\tmmathbf{u}_t -\tmmathbf{\gamma}_t)
    \end{array}$ & $\begin{array}{cl}
       P_n=&\varepsilon_n  \langle
    u_n - g_0 - \eta \alpha \rangle_+
     \\
        T_t \hspace{0.25em} =&\varepsilon_t  (\tmmathbf{u}_t -\tmmathbf{\gamma}_t)
    \end{array}$ & $\begin{array}{cl}
       P_n=&\varepsilon_n  \langle
    u_n - g_0 - \eta \ln(\alpha/\alpha^{\star})\rangle_+
     \\
        T_t \hspace{0.25em} =& \varepsilon_t  (\tmmathbf{u}_t -\tmmathbf{\gamma}_t)
    \end{array}$ \\[16pt] \hline
    dissipative forces & $T_t$& $(P_n, T_t)$ & $(\eta P_n, T_t)$ & $(\eta P_n/\alpha, T_t)$ \\ \hline
    dissipation potential $\varphi$ & $ \mu P_n  \|
  \dot{\tmmathbf{\gamma}}_t \|$ & $I_{\left\{
  \hspace{0.em} \dot{\gamma}_n + \| \dot{\tmmathbf{\gamma}}_t \| \mu \leqslant
  0 \right\}}$ &$ 
         \eta \sigma_0 \alpha \dot{\alpha} +
  I_{\mathbb{R}^+} (\dot{\alpha})  + \sigma_0 \mu \alpha \|
  \dot{\tmmathbf{\gamma}}_t \| 
    $ & $ 
          \eta \sigma_0 \dot{\alpha} +
  I_{\mathbb{R}^+} (\dot{\alpha}) 
         + \sigma_0 \mu \alpha \|
  \dot{\tmmathbf{\gamma}}_t \| 
    $ \\ \hline
    intrinsic dissipation $\mathcal{D}$  & $\mu P_n  \| \dot{\tmmathbf{\gamma}}_t \|$ & 0 &  \makecell{ \\ $ P_n (\eta  \dot{\alpha} + \mu  \|
\dot{\tmmathbf{\gamma}}_t \| )$ \\[0.3em] \small{(active plastic flow)}} & \makecell{ \\ $ \eta \sigma_0 \dot{\alpha}  + \mu P_n  \|
\dot{\tmmathbf{\gamma}}_t \| $ \\[0.3em] \small{(active plastic flow)}} \\[23pt] \hline
    penetration ($\varepsilon_n \rightarrow +\infty$) & 0 &  $\gamma_n \leqslant 0$  & $\eta \alpha \geqslant 0$ & $\eta \ln(\alpha/\alpha^{\star}) \geqslant 0$ \\ \hline
    \makecell{normal behavior \\ consistent with Coulomb} & yes & no & yes & yes \\ \hline
    
    GSM-compatible & no & yes & yes & yes \\ \hline
  \end{tabular}%
  }
  \caption{Summary of model properties.}
  \label{table1}
\end{table}

\subsection{Classical results on the real area of contact}\label{sec:tribology-result}

The appearance of a normal displacement during slip is not in itself unphysical. On the contrary, tribologists have long shown that macroscopic
contact laws originate from a combination of contact and elasto-plastic
deformation at the level of the asperities~{\cite{bowden1954,greenwood1966}}, and that the evolution of the deformation of the asperities, along with asperity ride-overs, leads to small changes in the normal displacement. The issue with the dilatant Coulomb law is therefore not so much the appearance of dilatancy during slip, but the fact that the dilatancy amplitude is \textit{comparable} to the amount of slip.

The key idea of our formulation is to take advantage of the plastic deformation of the asperities to restore GSM-compatibility. To do so, we introduce an additional internal variable $\alpha$, with conjugate force proportional to the pressure $P_n$, that leads only to small evolutions of the normal displacement during contact. By small evolutions we mean that the dissipation associated with this new variable $\alpha$ is small compared with that
produced by friction during a macroscopic slip.

The nature of this new internal variable is motivated by two classical results of tribology. The first was obtained by Bowden and Tabor~\cite{bowden1954}: when the asperities deform plastically, the real pressure on each of them equals the indentation hardness $h$ of the material, about three times its yield strength~\cite{tabor1951}, so that the nominal pressure is proportional to the \textit{local ratio between the real area of contact and the nominal one}, also denoted in the following as real-area fraction $\alpha$,
\begin{equation}
  P_n = \sigma_0 \alpha, \qquad \sigma_0 = h, \label{eq:greenwood-and-williamson-result}
\end{equation}
whatever the shape and the height distribution of the asperities. The second result is Greenwood and Williamson's statistical description of a rough surface~\cite{greenwood1966,greenwood1967}, which relates $\alpha$ to the normal gap. We summarize it as presented in~\cite{greenwood1967} (\textit{local} formulation), which assumes the existence of a small representative volume element whose intersection with the contact boundary~$\Gamma_c$ contains a large number of asperities; the nominal pressure $P_n (\tmmathbf{x}, t)$, the normal gap $(g_0 - u_n) (\tmmathbf{x}, t)$ and the real-area fraction $\alpha (\tmmathbf{x}, t)$ are quantities averaged over the representative volume element at position $\tmmathbf{x}$. The asperities are assumed not to interact and to have spherical summits of the same radius of curvature $\beta$, with density $\delta$ per unit nominal area and heights distributed exponentially, with density $\eta^{-1} \mathe^{- z / \eta}$ for $z \geqslant 0$ measured from the lowest summit; $\eta$ is thus both the mean and the standard deviation of the heights. For a flat at distance $g_0 - u_n \geqslant 0$ from the lowest summit, the asperities of height $z > g_0 - u_n$ are in contact, with interference $w = z - ( g_0 - u_n )$, and a plastically flattened asperity has the contact area of its geometric intersection with the flat, $2 \pi \beta w$. Summing over the asperities and using~\eqref{eq:greenwood-and-williamson-result} gives
\begin{equation}
  \begin{array}{lll}
    \alpha & = & \alpha^{\star} \mathe^{(u_n - g_0) / \eta},\\
    P_n & = & \sigma_0 \alpha^{\star} \mathe^{(u_n - g_0) / \eta},
  \end{array} \label{eq:exp-distrib-Pn-alpha}
\end{equation}
with the reference area fraction
\begin{equation}
  \alpha^{\star} = 2 \pi \delta \beta \eta , \label{eq:ref-area-for-exp}
\end{equation}
the value reached at $u_n = g_0$, where every asperity touches the flat with a mean interference $\eta$. Relations~\eqref{eq:exp-distrib-Pn-alpha} hold for $g_0 - u_n \geqslant 0$, i.e.\ for $\alpha \leqslant \alpha^{\star}$, and, since the asperities are assumed not to interact, for small real-area fractions; for a Gaussian height distribution, which fits most measured surfaces, they describe the upper tail of the distribution.

\begin{remark}\label{rmk:elastic-asperities}
The main point of~\cite{greenwood1966} is that the proportionality~\eqref{eq:greenwood-and-williamson-result} does not require plastic deformation. For Hertzian asperities with the same exponential height distribution, both the real area and the load are proportional to $\mathe^{(u_n - g_0)/\eta}$, so that \eqref{eq:exp-distrib-Pn-alpha} holds with
\begin{equation}
  \sigma_0 = \frac{1}{\sqrt{\pi}} \frac{E \eta^{1/2}}{(1 - \nu^2) \beta^{1/2}}, \qquad \alpha^{\star} = \pi \delta \beta \eta ,
  \label{eq:elastic-constants}
\end{equation}
where $\sigma_0$, which the authors call the elastic contact hardness, plays for elastic asperities the role that $h$ plays for plastic ones; for a Gaussian distribution the proportionality holds only approximately, $P_n/\alpha$ varying slowly with the separation. This is how these authors explain that Coulomb friction is observed for elastic
as well as for plastic contacts. We restrict ourselves to plastic asperities, which the same authors observe to be the frequent case, and use~\eqref{eq:elastic-constants} only as an estimate of the orders of magnitude of $\sigma_0$ and $\alpha^{\star}$ in Section~\ref{sec:numerical-results}.
\end{remark}

Where~\eqref{eq:greenwood-and-williamson-result}$_1$ holds, i.e. during active plastic flow of the asperities, the classical Coulomb law~\eqref{eq:coulomb-friction-for-penalty} can be rewritten as
\begin{equation}
\forall (t, \tmmathbf{x}) \in \mathbb{R}^+ \times \Gamma_c, \qquad
\| \tmmathbf{T}_t \| \leqslant \mu \sigma_0 \alpha, \qquad
\tmmathbf{T}_t \cdot \dot{\tmmathbf{\gamma}}_t = \mu \sigma_0 \alpha \, \| \dot{\tmmathbf{\gamma}}_t \| .
  \label{eq:Tresca-reformulation}
\end{equation}
Coulomb friction thus appears as a Tresca friction whose threshold $\mu \sigma_0 \alpha$ is parametrized by the real-area fraction~$\alpha$. This has important consequences for the variational structure of the problem, since Tresca friction is known to be compatible with GSM~\cite{moreau1971,duvaut1972}. In the next section, we rely on the Tresca formulation~\eqref{eq:Tresca-reformulation} to derive two associated friction models that we will show, under certain conditions, to be akin to Coulomb friction.

\section{A new associated model of Coulomb friction}\label{sec:formulation}

This section presents the main contribution of the paper: a new variational friction model that approximates the contact problem~\eqref{strong-form}. The formulation is based on three
ingredients: a new internal variable~$\alpha$ representing the real-area fraction, the assumption that its evolution is dictated by the plastic deformation of the asperities, and the Tresca formulation \eqref{eq:Tresca-reformulation} for the
frictional behavior. Sections~\ref{sec:goal}
to~\ref{sec:incremental-energy} derive the simplest penalty formulation based on these
ingredients and clarify in what sense it approximates the original
problem~\eqref{strong-form}. Section~\ref{sec:log-model} proposes a more
sophisticated variant, which reproduces the exponential scaling law~\eqref{eq:exp-distrib-Pn-alpha}. Section~\ref{LM} presents a version of the model based on the Lagrange multiplier method.

\subsection{Objective: approximation of the contact problem}\label{sec:goal}

Our key idea is to introduce a new internal variable~$\alpha$, representing the local mean ratio of the real area of contact to the nominal one and motivated by the classical results discussed in
Section~\ref{sec:tribology-result}. Clearly, it satisfies $\alpha \in [0, 1]$. We also introduce two new microscopic material parameters: the mean height of the asperities $\eta$ and their hardness
 $\sigma_0>0$. 
The former is assumed very small compared to the characteristic height~$H$ of the elastic
body (Figure~\ref{fig:scheme}); using the bounds on $\alpha$ we have
\begin{equation}
  0 \leqslant \eta \alpha \leqslant \eta \ll H. \label{eq:small-penetration}
\end{equation}
In the limit of infinitely large penalty parameters ($\varepsilon_n,
\varepsilon_t \rightarrow \infty$), the new model yields the strong
form, for all~$ t \geqslant 0$,
\begin{equation}
  \left\{\begin{array}{lll}
    \text{div} \tmmathbf{\sigma}+\tmmathbf{f}=\tmmathbf{0}, &\text{in}
    \hspace{0.3em} \Omega & \text{(bulk equilibrium)}\\[2pt]
  
    \tmmathbf{u}=\tmmathbf{u}^{\star}, & \text{on} \hspace{0.3em}
    \Gamma_u   & \text{(Dirichlet bound.)}\\[2pt]
 
    \tmmathbf{\sigma} \cdot \tmmathbf{n}=\tmmathbf{F}, & \text{on}
    \hspace{0.3em} \Gamma_F & \text{(Neumann bound.)}\\
    
    P_n = - \sigma_{n \nocomma n}, \hspace{3em} \tmmathbf{T}_t =
    -\tmmathbf{\sigma}_{n \nocomma t}, &\text{on} \hspace{0.3em}
    \Gamma_c & \text{(contact tractions)}\\[2pt]
   
    g_0 - u_n + \eta \alpha \geqslant 0, \quad P_n \geqslant 0, \quad (g_0 -
    u_n + \eta \alpha) P_n = 0, & \text{on} \hspace{0.3em} \Gamma_c
     & \text{(normal contact)}\\[8pt]
  
    \left\{\begin{array}{ll}
      \text{if } 0 \leqslant \alpha < 1, & \quad \eta P_n \leqslant \eta \sigma_0 \alpha,
      \quad \dot{\alpha} \geqslant 0, \quad \eta (P_n -
      \sigma_0 \alpha)  \dot{\alpha} = 0,\\[3pt]
       \alpha = 1 & \quad \text{otherwise} .
    \end{array}\right. &  \text{on} \hspace{0.3em} \Gamma_c &
    \text{(plastic asperities)}\\[15pt]
    
      \| \tmmathbf{T}_t \| \leqslant \mu \sigma_0 \alpha, \qquad
\tmmathbf{T}_t \cdot \dot{\tmmathbf{u}}_t = \mu \sigma_0 \alpha \, \| \dot{\tmmathbf{u}}_t \|
     & \text{on} \hspace{0.3em} \Gamma_c  &
    \text{(Tresca friction)}\\
  \end{array}\right. \label{new-strong-form}
\end{equation}
supplemented by the initial conditions $\tmmathbf{u}(\cdot,0)=\tmmathbf{u}_0$ and $\alpha(\cdot,0)=0$ on $\Gamma_c$.
This problem \textit{approximates} the original contact
problem~\eqref{strong-form} in the following sense.
\begin{itemizedot}
  \item The first three equations in~\eqref{new-strong-form} coincide
  with those of~\eqref{strong-form}.
  
  \item Combining~\eqref{new-strong-form}$_4$
  and~\eqref{new-strong-form}$_5$ yields normal contact conditions that
  differ from~\eqref{strong-form}$_4$ only by a microscopic penetration $\eta
  \alpha \ll H$, see~\eqref{eq:small-penetration}. 
  
  \item In the new model, the frictional behavior is given by the Tresca law~\eqref{new-strong-form}$_7$, whose threshold is \textit{parametrized}
  by the new internal variable~$\alpha$, as
  in~\eqref{eq:Tresca-reformulation}. The evolution of $\alpha$
  is dictated by the plastic deformation of the asperities and obeys the flow rule~\eqref{new-strong-form}$_6$. As expected from plasticity, this evolution
  is irreversible~$(\dot{\alpha} \geqslant 0)$ and the hardness $\sigma_0$ is the threshold of the \textit{real} contact
  pressure~$P_n / \alpha$. Whenever the asperities are at the plastic threshold, $P_n = \sigma_0
  \alpha$ and the Tresca law~\eqref{new-strong-form}$_7$ coincides exactly
  with Coulomb friction~\eqref{strong-form}$_5$; the two friction laws can
  differ, however, when $P_n < \sigma_0 \alpha$ or once the cap $\alpha = 1$ is reached.
\end{itemizedot}
Unlike the original problem~\eqref{strong-form}, the contact problem~\eqref{new-strong-form} is GSM-compatible: it derives from a free-energy density and a dissipation potential of the internal variables, as we show in the following. 

\subsection{Free-energy density}\label{sec:free-energy}

We consider the free-energy density
\begin{equation}
  \psi (u_n, \tmmathbf{u}_t, \alpha, \tmmathbf{\gamma}_t) =
  \frac{\varepsilon_n}{2}  \langle u_n - g_0 - \eta \alpha \rangle^2_+ +
  \frac{\varepsilon_t}{2}  \| \tmmathbf{u}_t -\tmmathbf{\gamma}_t \|^2 ,
  \label{new-free-energy}
\end{equation}
that has the same structure as the classical free-energy density~\eqref{classical-free-energy}, up to the additional internal
variable~$\alpha$ representing the real-area fraction.
Eq.~\eqref{eq:small-penetration} shows that the new term $\eta \alpha$ is
bounded and very small compared with the macroscopic height $H$ of the elastic
body. This new variable plays a role similar to that of $\gamma_n$ in
dilatant Coulomb friction~\eqref{eq:free-nrj-dilatant}, in that it accounts for the evolution of
the normal displacement during contact; however, it appears with the small prefactor
$\eta$. In contrast to $\gamma_n$
in~\eqref{eq:dilatancy}, the evolution of $\alpha$
produces only microscopic variations of the normal displacement and, since $\alpha
\geqslant 0$, it
can lead only to a small penetration, never to a dilatancy. This is consistent with the plastic asperity-crushing mechanism introduced in the next subsection.

As already highlighted in Remark~\ref{rmk:evacuate-u}, the observables 
($u_n$, $\tmmathbf{u}_t$)  are determined by solving the same minimization
problem~\eqref{v} as in the classical approach,  with $\tmmathbf{\gamma}_t$ and $\alpha$ frozen. This automatically ensures consistency with
the bulk equilibrium~\eqref{new-strong-form}$_1$, the boundary
conditions~\eqref{new-strong-form}$_{2, 3}$ and the
relations~\eqref{new-strong-form}$_4$ between the Cauchy stress
tensor~$\tmmathbf{\sigma}$ and the contact tractions~$(P_n, \tmmathbf{T}_t)$,
which are the energetic forces conjugate to $(u_n, \tmmathbf{u}_t)$.

For the free-energy density in~\eqref{new-free-energy}, these contact tractions take the form
\begin{equation}
  \begin{array}{lllll}
    P_n & = & \frac{\partial \psi}{\partial u_n} & = & \varepsilon_n  \langle
    u_n - g_0 - \eta \alpha \rangle_+,\\
    \tmmathbf{T}_t & = & \frac{\partial \psi}{\partial \tmmathbf{u}_t} & = &
    \varepsilon_t  (\tmmathbf{u}_t -\tmmathbf{\gamma}_t) .
  \end{array} \label{eq:new-contact-stresses}
\end{equation}
The tangential traction $\tmmathbf{T}_t$ is identical to that of the classical
approach~\eqref{classical-free-energy}, whereas the pressure $P_n$ contains the
additional contribution~$- \eta \alpha$. As in any penalty formulation, the
presence of the Macaulay bracket~\eqref{eq:macaulay-bracket} guarantees $-
\sigma_{n \nocomma n} = P_n \geqslant 0$. In addition, the elastic term
$\varepsilon_n  \langle u_n - g_0 - \eta \alpha \rangle^2_+ / 2$ penalizes
violations of the constraint $g_0-u_n + \eta \alpha \geqslant 0$. In the limit
$\varepsilon_n \rightarrow + \infty$, we recover the normal contact conditions
announced in~\eqref{new-strong-form}$_5$.

We have just shown that the first five lines of the contact
problem~\eqref{new-strong-form} can be obtained from the free-energy
density~\eqref{new-free-energy} and the minimization problem~\eqref{v}.
The aim of the next subsection is to identify a dissipation potential from which~\eqref{new-strong-form}$_{6, 7}$ derive by normality.

\subsection{Dissipation potential and maximum dissipation principle}\label{sec:dissipation}

To close the model within the GSM framework, we must define a dissipation potential that encapsulates the evolution laws for the internal variables $\tmmathbf{\gamma}_t$ and~$\alpha$. In dilatant Coulomb friction, the evolution of both internal variables is dictated by a single dissipative mechanism, the same convex domain in the space of contact tractions,
see Figure~\ref{fig:penalty}b; here, by contrast, we assign a distinct dissipative process to each of $\tmmathbf{\gamma}_t$ and $\alpha$.

This is made possible by the decoupling of the internal variables $(\tmmathbf{\gamma}_t, \alpha)$ in the free-energy density~\eqref{new-free-energy}.
The dissipative forces conjugate to $\alpha$ and $\tmmathbf{\gamma}_t$ are, respectively, the rescaled contact pressure $\eta P_n$ and the tangential
stress $\tmmathbf{T}_t$ of~\eqref{eq:new-contact-stresses},
\begin{equation}
  - \frac{\partial \psi}{\partial \alpha} = \eta P_n, \hspace{4em} -
  \frac{\partial \psi}{\partial \tmmathbf{\gamma}_t} =\tmmathbf{T}_t .
  \label{eq:irreversible-forces}
\end{equation}
On the one hand, we assume that the evolution of $\alpha$ is governed by perfect
plastic deformation of the asperities as long as $\alpha \leqslant 1$, and that it satisfies the flow rule~\eqref{new-strong-form}$_6$. The first
inequality, $\eta P_n \leqslant \eta \sigma_0 \alpha$, defines the plastic
domain, parametrized by the\textit{ current} value of the real-area fraction
$\alpha$; the second, $\dot{\alpha} \geqslant 0$, encodes the irreversibility of the plastic deformation; and the
complementarity condition $\eta (P_n - \sigma_0 \alpha)  \dot{\alpha} = 0$
expresses that plastic deformation ($\dot{\alpha} > 0$) can occur only when the real pressure $P_n / \alpha$ reaches the yield threshold~$\sigma_0$.
Once the real-area fraction $\alpha$ reaches 1, no further crushing of the 
asperities is possible and $\alpha$ is capped at that value.

On the other hand, the amount of slip $\tmmathbf{\gamma}_t$
satisfies the Tresca friction law in~\eqref{eq:Tresca-reformulation}, whose
threshold $\mu \sigma_0 \alpha$ is \textit{parametrized} by the real-area fraction $\alpha$. By construction, this frictional behavior matches
Coulomb friction~\eqref{eq:coulomb-friction-for-penalty} whenever the asperities
 undergo active plastic deformation ($P_n = \sigma_0 \alpha$). In the
limit $\varepsilon_t \rightarrow + \infty$, which imposes
$\dot{\tmmathbf{\gamma}}_t \rightarrow \dot{\tmmathbf{u}}_t$, the Tresca
friction \eqref{eq:Tresca-reformulation} takes the
form~\eqref{new-strong-form}$_7$.

The evolution of the system is thus assumed to result from the combination of
two dissipative mechanisms with distinct characteristic length scales, a
microscopic plastic deformation and a macroscopic Tresca friction, coupled exclusively through the presence of~$\alpha$ in the
friction threshold. The dissipation potential $\varphi$ is obtained by
adding the support function~$\varphi_p$ of the plastic domain~\eqref{new-strong-form}$_6$ to the support function $\varphi_f$ of the friction
domain~\eqref{new-strong-form}$_7$,
\begin{equation}
  \varphi (\dot{\alpha}, \dot{\tmmathbf{\gamma}_t} ; \alpha) = \varphi_p
  (\dot{\alpha} ; \alpha) + \varphi_f (\dot{\tmmathbf{\gamma}_t} ; \alpha),
  \label{eq:new-dissipation-potential}
\end{equation}
where
\begin{equation}
  \varphi_p (\dot{\alpha} ; \alpha) = \eta \sigma_0 \alpha \dot{\alpha} +
  I_{\mathbb{R}^+} (\dot{\alpha}), \hspace{6em} \varphi_f
  (\dot{\tmmathbf{\gamma}_t} ; \alpha) = \sigma_0 \mu \alpha \|
  \dot{\tmmathbf{\gamma}}_t \|, \label{eq:expr-dissipation}
\end{equation}
with $I_{\mathbb{R}^+} (\dot{\alpha})$ as the indicator function of $\mathbb{R}^+$, 
\begin{equation}
  I_{\mathbb{R}^+} (\dot{\alpha}) = \left\{\begin{array}{ll}
    0 & \text{if } \dot{\alpha} \geqslant 0,\\
    + \infty & \text{if } \dot{\alpha} < 0.
  \end{array}\right. \label{eq:indicator-function}
\end{equation}
As is standard in rate-independent processes, the dissipation potential~$\varphi$
is convex and positively one-homogeneous with respect to the pair~$(\dot{\alpha},\dot{\boldsymbol{\gamma}}_t)$; it also satisfies $\varphi(\dot{\alpha},\dot{\tmmathbf{\gamma}}_t;\alpha)\geqslant \varphi(0,\tmmathbf{0};\alpha)=0$. The intrinsic dissipation reads
$\mathcal{D}= \eta \sigma_0 \alpha \dot{\alpha} + \sigma_0 \mu \alpha \|
\dot{\tmmathbf{\gamma}}_t \| \geqslant 0$, so that, unlike the dilatant Coulomb law, the proposed model does not suffer from the limitation~\eqref{eq:zero-dissipation}: any non-zero slip $\| \dot{\tmmathbf{\gamma}}_t \| > 0$ at $\alpha>0$ produces a strictly positive dissipation~$\mathcal{D}$. In addition, as anticipated in
Section~\ref{sec:tribology-result}, the dissipation produced by the plastic deformation of the asperities is small compared with that produced by a macroscopic slip: since $\alpha \leqslant 1$, the former is bounded by $\eta \sigma_0 / 2$ over the whole loading history, whereas a slip $\| \tmmathbf{\gamma}_t \| \sim H$ at $\alpha \sim 1$ dissipates $\mu \sigma_0 H$, larger by a factor $2 \mu H / \eta \gg 1$ by~\eqref{eq:small-penetration}.

The new potential~\eqref{eq:new-dissipation-potential} is parametrized by the
\emph{current value of the internal variable}~$\alpha$. Such a parametrization is perfectly \textit{compatible with the GSM framework}~{\cite{hackl1997,mielke2003}}, by contrast to pressure-dependent
potentials; this is a key advantage over the
classical formulation \eqref{eq:coulombpot} presented
in~Section~\ref{sec:lack-of-normality}.

Since the proposed model is GSM-compatible by construction, the evolution
laws in~\eqref{eq:Tresca-reformulation} and in the first line of~\eqref{new-strong-form}$_{6}$ can
be derived from a minimization principle
(see~Appendix~\ref{appendix:linear-model} for a rigorous check): 
\textit{for given}~$\tmmathbf{u}, \alpha, \tmmathbf{\gamma}_t, \dot{\tmmathbf{u}}$\textit{, and given }$(t, \tmmathbf{x}) \in \mathbb{R}^+ \times \Gamma_c$, \textit{the rates }$(\dot{\alpha}, \dot{\tmmathbf{\gamma}_t})$ \textit{are solutions to}
\begin{equation}
  \underset{\dot{\alpha}, \dot{\tmmathbf{\gamma}}_t}{\text{Min}}
  \enspace \dot{\psi} (u_n, \tmmathbf{u}_t, \alpha, \tmmathbf{\gamma}_t,
  \dot{u}_n, \dot{\tmmathbf{u}}_t, \dot{\alpha}, \dot{\tmmathbf{\gamma}}_t) +
  \varphi (\dot{\alpha}, \dot{\tmmathbf{\gamma}_t} ; \alpha) .
  \label{eq:new-max-dissipation-principle}
\end{equation}
Eqs.~\eqref{v} and~\eqref{eq:new-max-dissipation-principle} are the continuous
variational principles underlying the contact problem~\eqref{new-strong-form}. The model is thus associated: the rates $(\dot{\alpha}, \dot{\tmmathbf{\gamma}}_t)$ are normal to the elastic domain $\{ \eta P_n \leqslant \eta \sigma_0 \alpha \} \times \{ \| \tmmathbf{T}_t \| \leqslant \mu \sigma_0 \alpha \}$ in the space of the dissipative forces $(\eta P_n, \tmmathbf{T}_t)$, a domain that grows with $\alpha$ as the yield surface of a hardening material grows with its hardening variable. In the space of the contact
tractions $(P_n, \tmmathbf{T}_t)$, the Coulomb cone presented in Section~\ref{sec:lack-of-normality} (Figure~\ref{fig:penalty}b) is replaced by a cylinder capped at the plastic threshold $\sigma_0 \alpha$. The friction domain $\| \tmmathbf{T}_t \| \leqslant \mu \sigma_0 \alpha$ corresponds to the cylindrical part, which is why the slip remains purely tangential, in contrast to the dilatant law~\eqref{eq:dilatancy}.
In the next subsection we combine~\eqref{v} and~\eqref{eq:new-max-dissipation-principle} into a single global incremental
energy minimization principle, which is more convenient both for the numerical 
implementation and for deriving rigorous proofs of the existence of solutions. 

Note that~\eqref{eq:new-max-dissipation-principle} yields only the plastic flow rule in the first line of \eqref{new-strong-form}$_{6}$, and is valid solely while $\alpha$ may evolve, that is, as long as $\alpha <1$. Once the cap $\alpha=1$ is reached, $\dot{\alpha}=0$ and only the minimization over $\dot{\boldsymbol{\gamma}}_t$ is to be performed in~\eqref{eq:new-max-dissipation-principle}. The incremental formulation established in the next subsection requires no such distinction between the plastic and the cap regimes: both regimes are governed by the same incremental variational principle.

The main features of the new model are summarized in Table~\ref{table1}.

\begin{remark}
Since the plastic potential \eqref{eq:expr-dissipation}$_1$ is an exact differential,
$\varphi_p ( \dot{\alpha} ; \alpha ) = \mathrm{d} w / \mathrm{d} t + I_{\mathbb{R}^+} ( \dot{\alpha} )$ with
$w ( \alpha ) = \eta \sigma_0 \alpha^2 / 2$, the crushing work can equivalently be included in the free-energy density rather than in the dissipation: with $\tilde{\psi} = \psi + w ( \alpha ) + I_{[0, 1]} ( \alpha )$ and $\tilde{\varphi}_p ( \dot{\alpha} ) = I_{\mathbb{R}^+} ( \dot{\alpha} )$, the plastic part of the model is a GSM in the original sense of~\cite{moreau1971,halphen1975}, its flow rule $0 \in \partial_{\alpha} \tilde{\psi} + \partial \tilde{\varphi}_p ( \dot{\alpha} )$ reproduces \eqref{new-strong-form}$_6$ including the cap regime (at $\alpha = 1$, the subdifferential of $I_{[0, 1]}$, i.e. the reaction of the constraint $\alpha \leqslant 1$, absorbs the excess force $\eta ( P_n - \sigma_0 )$), and the incremental problems of Section~\ref{sec:incremental-energy} are unchanged. The two options are mechanically indistinguishable, because $\dot{\alpha} \geqslant 0$ implies that $w$ is never recovered; they differ only in whether the crushing work is counted as dissipated or as locked-in
energy, exactly as for the dissipated energy $w ( \alpha )$ of gradient damage models \cite{pham2011}, and we retain the former, more physically natural, choice.
\end{remark}

\subsection{Incremental energy minimization}\label{sec:incremental-energy}

We follow the classical approach to identify an incremental energy minimization principle based on the continuous GSM formulation~\eqref{v},~\eqref{eq:new-max-dissipation-principle}, see~{\cite{ortiz1999,mielke2003}} among others. We first introduce a time
discretization $t_k = k \Delta t$ with time step $\Delta t > 0$ and index $k
= 0, 1, \ldots$. Discretizing~\eqref{eq:new-max-dissipation-principle} in time over one increment, i.e. integrating the energy rate and the dissipation potential over
$[t_{k - 1}, t_k]$ $(k > 0)$ then leads to the alternative\textit{ local}
incremental minimization principle: \textit{for given} $\tmmathbf{u}^{(k)},
 \alpha^{(k - 1)}, \tmmathbf{\gamma}_t^{(k - 1)}$, \textit{and given $\tmmathbf{x} \in \Gamma_c,$} \textit{the internal variables $(\alpha,\tmmathbf{\gamma}_t)$ at the current time $t_k$ are solutions to }
\begin{equation}
  \underset{\substack{\alpha, \tmmathbf{\gamma}_t \\ \alpha^{(k - 1)} \leqslant \alpha \leqslant 1}}{\mathrm{Min}}
  \ \psi ( u_n^{(k)}, \tmmathbf{u}_t^{(k)}, \alpha, \tmmathbf{\gamma}_t )
  + \frac{\eta \sigma_0}{2} \bigl[ \alpha^2 - ( \alpha^{(k - 1)} )^2 \bigr]
  + \sigma_0 \mu \alpha^{(k - 1)} \, \bigl\| \tmmathbf{\gamma}_t - \tmmathbf{\gamma}_t^{(k - 1)} \bigr\| .
  \label{eq:local-incremental}
\end{equation}
The superscripts $(k)$ and $(k - 1)$ denote evaluation at the current time step $t_k$ and at the previous one~$t_{k - 1}$, respectively. The constraint $\alpha^{(k - 1)} \leqslant \alpha$ is the discrete counterpart of the
irreversibility constraint $\dot{\alpha} \geqslant 0$, previously accounted for by the indicator function $I_{\mathbb{R}^+} ( \dot{\alpha} )$ in~\eqref{eq:expr-dissipation}$_1$.
The last two terms in \eqref{eq:local-incremental} are the plastic and frictional contributions $\varphi_p$, $\varphi_f$ of the dissipation potential \eqref{eq:new-dissipation-potential}, integrated over the time increment; they are obtained differently.

For the plastic part the integration is exact. The potential
$\varphi_p ( \dot{\alpha} ; \alpha )$ in \eqref{eq:expr-dissipation}$_1$ is the time derivative of $w ( \alpha ) = \eta \sigma_0 \alpha^2 / 2$ along every admissible path, $\eta \sigma_0 \alpha \dot{\alpha} = \mathrm{d} w / \mathrm{d} t$ for $\dot{\alpha} \geqslant 0$, so that
\begin{equation}
  \int_{t_{k - 1}}^{t_k} \varphi_p ( \dot{\alpha} ; \alpha ) \, \mathrm{d} t
  = w ( \alpha^{(k)} ) - w ( \alpha^{(k - 1)} )
  = \frac{\eta \sigma_0}{2} \bigl[ ( \alpha^{(k)} )^2 - ( \alpha^{(k - 1)} )^2 \bigr]
  \label{eq:exact-integration}
\end{equation}
for every path with $\dot{\alpha} \geqslant 0$, irrespective of the time step; the indicator function $I_{\mathbb{R}^+} ( \dot{\alpha} )$ in \eqref{eq:expr-dissipation}$_1$ becomes the lower bound $\alpha^{(k - 1)} \leqslant \alpha$ of the constraint set in \eqref{eq:local-incremental}, whose upper bound is the cap $\alpha \leqslant 1$.

The frictional potential \eqref{eq:expr-dissipation}$_2$, by contrast, is parametrized by $\alpha$, and the integral $\int_{t_{k - 1}}^{t_k} \sigma_0 \mu \alpha \| \dot{\tmmathbf{\gamma}}_t \| \, \mathrm{d} t$
depends on the evolution of $\alpha$ within the increment. It is approximated in \eqref{eq:local-incremental} by freezing $\alpha$ at its known value $\alpha^{(k - 1)}$ at the beginning of the step, which is a first-order (explicit) approximation; for frozen $\alpha$, $\sigma_0 \mu \alpha^{(k - 1)} \| \tmmathbf{\gamma}_t - \tmmathbf{\gamma}_t^{(k - 1)} \|$ is the minimum of the integrated dissipation over all paths joining $\tmmathbf{\gamma}_t^{(k - 1)}$ to $\tmmathbf{\gamma}_t$, the dissipation distance of \cite{mielke2003}. Evaluating the friction threshold at the current value $\alpha^{(k)}$ instead would introduce the term $\sigma_0 \mu \| \tmmathbf{\gamma}_t - \tmmathbf{\gamma}_t^{(k - 1)} \|$ into the stationarity condition of \eqref{eq:local-incremental} with respect to $\alpha$, a coupling of
order $\Delta t$ between slip and asperity crushing with no counterpart in \eqref{new-strong-form}$_6$, and would destroy the convexity of \eqref{eq:local-incremental} in $(\alpha, \tmmathbf{\gamma}_t)$, since
$(\alpha, \tmmathbf{\gamma}_t) \mapsto \alpha \| \tmmathbf{\gamma}_t - \tmmathbf{\gamma}_t^{(k - 1)} \|$
is not jointly convex. The lagged choice preserves both the consistency with \eqref{new-strong-form}$_6$ and the convexity on which the results of Appendix~\ref{appendix:incremental} rely, at the price of a one-step lag of the friction threshold.

We show in Appendix~\ref{appendix:incremental-linear-model} that the first-order stationarity conditions of~\eqref{eq:local-incremental} are fully consistent with the continuous evolution laws~\eqref{eq:Tresca-reformulation},~\eqref{new-strong-form}$_6$ up to the one-step lag of the friction threshold. In particular, owing to the exact integration \eqref{eq:exact-integration}, the plastic flow rule \eqref{new-strong-form}$_6$ is satisfied exactly by the current variable $\alpha^{(k)}$, and the plastic and frictional dissipations remain decoupled.

Finally, we derive a \textit{global} incremental energy minimization principle for the whole contact problem, by combining the primal minimization problem~\eqref{eq:bulk-energy},~\eqref{v} with the local
incremental one~\eqref{new-free-energy},~\eqref{eq:expr-dissipation},~\eqref{eq:local-incremental}
integrated over the contact boundary~$\Gamma_c$. The observable $\tmmathbf{u}^{(k)} : \Omega \rightarrow \mathbb{R}^3$ and the internal
variables~$\alpha^{(k)} : \Gamma_c \rightarrow [\alpha^{(k - 1)}, 1]$ and $
\tmmathbf{\gamma}_t^{(k)} : \Gamma_c \rightarrow \mathbb{R}^2$, at
time step~$t_k$ with $k > 0$, are obtained as the solution of the convex minimization problem
\begin{equation}
  \begin{array}[t]{ll}
    \underset{\alpha^{(k - 1)} \leqslant \alpha \leqslant
    1}{\underset{\tmmathbf{u}=\tmmathbf{u}^{\star} \text{on}
    \Gamma_u}{\underset{\tmmathbf{u}, \alpha,
    \tmmathbf{\gamma}_t}{\text{Min}}}} &
    \begin{array}[t]{l}
    \int_{\Omega} \left( \frac{1}{2}
    \tmmathbf{\varepsilon} (\tmmathbf{u}) : \tmmathbf{c}:
    \tmmathbf{\varepsilon} (\tmmathbf{u}) -\tmmathbf{f} \cdot \tmmathbf{u}
    \right) \mathd \tmmathbf{x}- \int_{\Gamma_F} \tmmathbf{F} \cdot
    \tmmathbf{u} \,\mathd s\\[7pt]
     \quad + \int_{\Gamma_c} \left( \frac{\varepsilon_n}{2}  \langle u_n -
    g_0 - \eta \alpha \rangle^2_+ + \frac{\varepsilon_t}{2}  \| \tmmathbf{u}_t
    -\tmmathbf{\gamma}_t \|^2 + \frac{\eta \sigma_0 \alpha^2}{2} + \sigma_0
    \mu \alpha^{(k - 1)}  \| \tmmathbf{\gamma}_t -\tmmathbf{\gamma}_t^{(k -
    1)} \| \right) \mathd s,
    \end{array}
  \end{array} \label{eq:global-incremental}
\end{equation}
where $\alpha^{(k - 1)}$ and $\tmmathbf{\gamma}_t^{(k - 1)}$ are assumed to be known. The constant term $- \eta \sigma_0 ( \alpha^{(k - 1)} )^2 / 2$ of \eqref{eq:exact-integration} has been dropped from~\eqref{eq:global-incremental}, which does not affect the minimizer. The internal variables are initialized as $\alpha^{(0)} = 0$ and $\tmmathbf{\gamma}_t^{(0)} = \tmmathbf{0}$; the resulting treatment of the first time step in the boundary-value
problems is discussed in Section~\ref{sec:numerical-results}.

\begin{remark}\label{rmk:open-gap}
In~\eqref{eq:global-incremental} the tangential spring and the friction term act on the whole of $\Gamma_c$. Where the gap is open ($P_n = 0$), the classical penalty formulation~\eqref{classical-free-energy}, \eqref{eq:coulombpot} automatically yields $\tmmathbf{\gamma}_t = \tmmathbf{u}_t$ and $\tmmathbf{T}_t = \tmmathbf{0}$, because the Coulomb threshold $\mu P_n$ vanishes; the threshold $\mu \sigma_0 \alpha^{(k - 1)}$ of~\eqref{eq:global-incremental} does not, since $\alpha$ is irreversible and keeps the memory of the crushed asperities after the surfaces separate. Taken literally, \eqref{eq:global-incremental} would therefore transmit friction across an open gap. As in any penalty formulation, its tangential terms are to be understood as restricted to the contact zone $\{ P_n > 0 \}$: on that zone the incremental problem is the one analysed in Appendix~\ref{appendix:incremental}, while outside it the slip simply follows the displacement, $\tmmathbf{\gamma}_t = \tmmathbf{u}_t$, and $\tmmathbf{T}_t = \tmmathbf{0}$. The corresponding treatment of the nodes entering and leaving contact, specified in Section~\ref{sec:return-mapping}, introduces for the contact status the same one-step lag as for $\alpha^{(k - 1)}$. The same remark applies to~\eqref{eq:global-incremental-LOG}, \eqref{eq:Lagrange} and~\eqref{eq:Lagrange-LOG}.
\end{remark}

The lagged threshold $\mu \sigma_0 \alpha^{(k - 1)}$ is the only approximation introduced beyond the incremental framework itself, since the plastic dissipation is integrated exactly, see \eqref{eq:exact-integration}, and so is the frictional one for frozen $\alpha$; its effect can be quantified. Under active plastic flow at time $t_{k - 1}$, the plastic law \eqref{new-strong-form}$_6$ gives $\sigma_0 \alpha^{(k - 1)} = P_n^{(k - 1)}$, so that the friction limit of \eqref{eq:global-incremental} at time $t_k$ is $\mu P_n^{(k - 1)}$ instead of the Coulomb limit $\mu P_n^{(k)}$. The incremental model thus reproduces Coulomb friction up to a lag of one time step in the threshold, that is, up to an error $\mu ( P_n^{(k)} - P_n^{(k - 1)} ) = \mathcal{O} ( \Delta t )$, which vanishes as $\Delta t \rightarrow 0$ and is negligible whenever the contact pressure varies little over a time step. This lag is the price of the incremental variational structure; it dictates the refinement of the time steps during the establishment of contact in Section~\ref{sec:rela-BVP}.

The functional in~\eqref{eq:global-incremental} is the sum of the elastic potential $\Phi^{\text{el}}$ (first line), the contact 
energy $\int_{\Gamma_c} \psi \mathd s$ (first two terms of the second line) and
the integrated total dissipation $\int_{\Gamma_c} \int_{t_{k - 1}}^{t_k}
\varphi \mathd t \mathd s$ (last two terms of the second line). This functional is strictly convex, and problem~\eqref{eq:global-incremental} has a unique solution (Appendix~\ref{appendix:incremental-linear-model}).

Since the dissipation (last two terms) does not depend on the displacement, minimization with respect to $(u_n, \tmmathbf{u}_t)$ is fully equivalent to
the equilibrium problem~\eqref{v}, which was shown to produce the first four equations in~\eqref{new-strong-form}, or equivalently the weak form~\eqref{eq:usual-weak-form}. Similarly, minimization with respect
to~$(\alpha, \tmmathbf{\gamma}_t)$ is unaffected by the elastic potential (first line), and its solution coincides with that of the local incremental principle~\eqref{eq:local-incremental}, which satisfies the plastic law~\eqref{new-strong-form}$_6$
exactly and the Tresca law~\eqref{new-strong-form}$_7$ with the lagged threshold $\mu \sigma_0
\alpha^{(k-1)}$, in the limit $\varepsilon_t \rightarrow + \infty$, see Appendix~\ref{appendix:incremental-linear-model}. We have already shown, in the argument below
Eq.~\eqref{eq:new-contact-stresses}, that the free-energy density~\eqref{new-free-energy} encapsulates the normal contact condition~\eqref{new-strong-form}$_5$. The entire contact
problem~\eqref{new-strong-form} has thus been recast, in the limit
$\varepsilon_n, \varepsilon_t \rightarrow + \infty$ and up to the lag of the friction threshold, as  a single incremental energy minimization~\eqref{eq:global-incremental}.

\subsection{Alternative formulation with enhanced microscopic
behavior}\label{sec:log-model}

In this subsection we propose a more sophisticated variant of the model, which allows us to reproduce the exponential scaling relations~\eqref{eq:exp-distrib-Pn-alpha} of Section~\ref{sec:tribology-result}.
 The logarithmic model below extends
this exponential law to real-area fractions in the range $\alpha^{\star}\leqslant\alpha\leqslant 1$ as a constitutive assumption. The main features of the model are summarized in Table~\ref{table1}.

\paragraph{Free-energy density}To reproduce this exponential scaling, we propose to replace the linear term $- \eta \alpha$ in the contact free-energy density~\eqref{new-free-energy} by a logarithmic term $- \eta \ln (\alpha /
\alpha^{\star})$,
\begin{equation}
  \psi (u_n, \tmmathbf{u}_t, \tmmathbf{\gamma}_t, \alpha) =
  \frac{\varepsilon_n}{2}  \langle u_n - g_0 - \eta \ln (\alpha /
  \alpha^{\star}) \rangle^2_+ + \frac{\varepsilon_t}{2}  \| \tmmathbf{u}_t
  -\tmmathbf{\gamma}_t \|^2 . \label{new-free-energy-LOG}
\end{equation}
This free-energy density depends more closely on the surface topography, that is, on the asperity
density and geometry, through the reference area fraction~$\alpha^{\star}$
defined in~\eqref{eq:ref-area-for-exp}. To avoid stability issues in the numerical
simulations, typically caused by the strong variations of the logarithm, we restrict
the domain of admissible~$\alpha$ to
\begin{equation}
  \alpha \in [\alpha^{\star}, 1] \subset (0,1], \label{eq:alpha-domain-LOG}
\end{equation}
so that the penetration $\eta \ln (\alpha / \alpha^{\star})$ is non-negative and the logarithm remains bounded: $g_0$ is then the position of the substrate at which the real-area fraction equals $\alpha^{\star}$, and $\alpha$ takes this value as soon as contact is established.
The tangential traction $\tmmathbf{T}_t$ remains unchanged, 
see~\eqref{eq:new-contact-stresses}$_2$, while the contact pressure becomes
 \begin{equation}
\begin{array}{lllll}
     P_n & = & \frac{\partial \psi}{\partial u_n} & = & \varepsilon_n  \langle
     u_n - g_0 - \eta \ln (\alpha / \alpha^{\star}) \rangle_+ .
   \end{array} \label{eq:new-pressure_LOG}     
\end{equation}
By an argument similar to the one developed for~\eqref{new-free-energy}, the free-energy density~\eqref{new-free-energy-LOG} can be shown to be consistent with
the first four equations in~\eqref{new-strong-form} and to yield the following normal contact conditions in the limit $\varepsilon_n \rightarrow +
\infty$,
\begin{equation}
  \forall (t, \tmmathbf{x}) \in \mathbb{R}^+ \times \Gamma_c, \hspace{2.0em}
  g_0 - u_n + \eta \ln (\alpha / \alpha^{\star}) \geqslant 0, \qquad \sigma_{n
  \nocomma n} \leqslant 0, \qquad (g_0 - u_n + \eta \ln (\alpha /
  \alpha^{\star})) \sigma_{n \nocomma n} = 0. \label{eq:new-normal-ontact-LOG}
\end{equation}
These are the classical normal contact conditions~\eqref{strong-form}$_4$, up
to the additional penetration~$\eta \ln (\alpha / \alpha^{\star})$, whose
maximum value is expected to scale as $\eta \ln (1 / \alpha^{\star})
\sim 4 \eta$ (with $\alpha^{\star} \approx 0.02$, see set $I$ in
Table~\ref{tab:physical-values}). As expected, the penetration depth is dictated by the asperity height; for this logarithmic variant it is larger by a factor $\text{ln}(1/\alpha^{\star})\approx 4$ than the value $\sim\eta$ predicted by the linear model~\eqref{new-strong-form}$_5$. Eq.~\eqref{eq:new-normal-ontact-LOG}$_3$ shows that active contact
$(- \sigma_{n \nocomma n} = P_n > 0)$ yields $g_0 - u_n + \eta \ln (\alpha /
\alpha^{\star}) = 0$, which proves that the logarithmic model~\eqref{new-free-energy-LOG} is consistent with the exponential scaling law~\eqref{eq:exp-distrib-Pn-alpha}$_1$ and with~\eqref{eq:exp-distrib-Pn-alpha}$_2$ during active plastic flow, as anticipated. We therefore expect it to
predict the penetration depth better than the linear
model~\eqref{new-strong-form}$_5$; Section~\ref{sec:material-point-simulations} provides a more quantitative comparison between the two models.

\paragraph{Dissipation potential} We assume that the evolution of $(\alpha, \tmmathbf{\gamma}_t)$ is governed by the same dissipative mechanisms as in Section~\ref{sec:dissipation}. The presence of the logarithmic term
in~\eqref{new-free-energy-LOG} slightly changes the definition of the
dissipative force conjugate to $\alpha$ compared
with~\eqref{eq:irreversible-forces}$_1$: it is now the scaled \textit{real}
pressure,
\begin{equation}
  - \frac{\partial \psi}{\partial \alpha} = \eta \frac{P_n}{\alpha} .
  \label{eq:irreversible-force-LOG}
\end{equation}
We modify the plastic dissipation potential accordingly, as
\begin{equation}
  \varphi_p (\dot{\alpha}) = \eta \sigma_0  \dot{\alpha} + I_{\mathbb{R}^+}
  (\dot{\alpha}) . \label{eq:plastic-potential-LOG}
\end{equation}
By adding it to the Tresca friction potential in~\eqref{eq:expr-dissipation}$_2$,
we obtain a slightly different expression for the total dissipation potential,
\[ \begin{array}{lll}
     \varphi (\dot{\alpha}, \dot{\tmmathbf{\gamma}}_t ; \alpha) & = &
     \varphi_f (\dot{\tmmathbf{\gamma}}_t ; \alpha) + \varphi_p
     (\dot{\alpha})\\
     & = & \sigma_0 \mu \alpha \| \dot{\tmmathbf{\gamma}}_t \| + \eta
     \sigma_0  \dot{\alpha} + I_{\mathbb{R}^+} (\dot{\alpha}) .
   \end{array} \]
Note that the plastic dissipation is not parametrized by $\alpha$ in this logarithmic variant, in contrast to the linear model~\eqref{eq:expr-dissipation}$_1$. Minimizing $\dot{\psi} + \varphi$ with respect to $(\dot{\alpha},
\dot{\tmmathbf{\gamma}}_t)$ then yields the same evolution laws~\eqref{new-strong-form}$_{6, 7}$,
see~Appendix~\ref{appendix:log-model}; in particular, the definition of the  plastic domain of the asperities remains consistent with~\eqref{new-strong-form}$_6$.

\paragraph{Strong form}This alternative model is thus consistent with all the
equations in~\eqref{new-strong-form}, except for the normal contact condition, which
is replaced by~\eqref{eq:new-normal-ontact-LOG}. For completeness, we give its strong form, valid in the limit $\varepsilon_n, \varepsilon_t
\rightarrow \infty$, for all $t \geqslant 0$,
\begin{equation}
  \left\{\begin{array}{lll}
    \text{div} \tmmathbf{\sigma}+\tmmathbf{f}=\tmmathbf{0}, &\text{in}
    \hspace{0.3em} \Omega & \text{(bulk equilibrium)}\\[2pt]
  
    \tmmathbf{u}=\tmmathbf{u}^{\star}, & \text{on} \hspace{0.3em}
    \Gamma_u   & \text{(Dirichlet bound.)}\\[2pt]
 
    \tmmathbf{\sigma} \cdot \tmmathbf{n}=\tmmathbf{F}, & \text{on}
    \hspace{0.3em} \Gamma_F & \text{(Neumann bound.)}\\
    
    P_n = - \sigma_{n \nocomma n}, \hspace{3em} \tmmathbf{T}_t =
    -\tmmathbf{\sigma}_{n \nocomma t}, &\text{on} \hspace{0.3em}
    \Gamma_c & \text{(contact tractions)}\\[2pt]
   
    g_0 - u_n + \eta \ln
    \left( \frac{\alpha}{\alpha^{\star}} \right)  \geqslant 0, \quad P_n \geqslant 0, \quad (g_0 -
    u_n + \eta \ln
    \left( \frac{\alpha}{\alpha^{\star}} \right) ) P_n = 0, & \text{on} \hspace{0.3em} \Gamma_c
     & \text{(normal contact)}\\[8pt]
  
    \left\{\begin{array}{ll}
      \text{if } \alpha^{\star} \leqslant \alpha < 1, & \quad \eta P_n \leqslant \eta \sigma_0 \alpha,
      \quad \dot{\alpha} \geqslant 0, \quad \eta (P_n -
      \sigma_0 \alpha)  \dot{\alpha} = 0,\\[3pt]
    \alpha = 1 & \quad \text{otherwise} .
    \end{array}\right. &  \text{on} \hspace{0.3em} \Gamma_c &
    \text{(plastic asperities)}\\[15pt]
    
     \| \tmmathbf{T}_t \| \leqslant \mu \sigma_0 \alpha, \qquad
\tmmathbf{T}_t \cdot \dot{\tmmathbf{u}}_t = \mu \sigma_0 \alpha \, \| \dot{\tmmathbf{u}}_t \|
     & \text{on} \hspace{0.3em} \Gamma_c  &
    \text{(Tresca friction)}\\
 
  \end{array}\right. \label{new-strong-form-LOG}
\end{equation}
supplemented by $\tmmathbf{u}(\cdot,0)=\tmmathbf{u}_0$ and $\alpha(\cdot,0)=\alpha^{\star}$.

\paragraph{Global incremental formulation}The plastic potential $\varphi_p (\dot{\alpha})$ in~\eqref{eq:plastic-potential-LOG} is again an exact differential, $\eta \sigma_0 \dot{\alpha} = \mathrm{d} ( \eta \sigma_0 \alpha ) / \mathrm{d} t$ for $\dot{\alpha} \geqslant 0$, so that its integral over a time increment is $\eta \sigma_0 ( \alpha^{(k)} - \alpha^{(k - 1)} )$ exactly, as in~\eqref{eq:exact-integration}. Proceeding as in Section~\ref{sec:incremental-energy}, we obtain the global incremental energy minimization principle for~$k > 0$: 
\textit{for given} $\alpha^{(k - 1)}, \tmmathbf{\gamma}_t^{(k - 1)}$, 
\textit{the current fields} ~$\tmmathbf{u}^{(k)} : \Omega \rightarrow
\mathbb{R}^3$,~$\alpha^{(k)} : \Gamma_c \rightarrow [\alpha^{(k - 1)}, 1],
\tmmathbf{\gamma}_t^{(k)} : \Gamma_c \rightarrow \mathbb{R}^2$ \textit{are solutions
to}
\begin{equation}
  \begin{array}[t]{ll}
    \underset{\alpha^{(k - 1)} \leqslant \alpha \leqslant
    1}{\underset{\tmmathbf{u}=\tmmathbf{u}^{\star} \text{on}
    \Gamma_u}{\underset{\tmmathbf{u}, \alpha,
    \tmmathbf{\gamma}_t}{\text{Min}}}} & 
    \begin{array}[t]{l}  
    \int_{\Omega} \left( \frac{1}{2}
    \tmmathbf{\varepsilon} (\tmmathbf{u}) : \tmmathbf{c}:
    \tmmathbf{\varepsilon} (\tmmathbf{u}) -\tmmathbf{f} \cdot \tmmathbf{u}
    \right) \mathd \tmmathbf{x}- \int_{\Gamma_F} \tmmathbf{F} \cdot
    \tmmathbf{u} \mathd s\\[7pt]
     \quad + \int_{\Gamma_c} \left( \frac{\varepsilon_n}{2}  \left\langle
    u_n - g_0 - \eta \ln \left( \frac{\alpha}{\alpha^{\star}} \right)
    \right\rangle^2_+ + \frac{\varepsilon_t}{2}  \| \tmmathbf{u}_t
    -\tmmathbf{\gamma}_t \|^2 \right) \mathd s\\[7pt]
     \qquad + \int_{\Gamma_c} \left(  \eta \sigma_0  (\alpha - \alpha^{(k - 1)}) +
    \sigma_0 \mu \alpha^{(k - 1)}  \| \tmmathbf{\gamma}_t
    -\tmmathbf{\gamma}_t^{(k - 1)} \| \right) \mathd s.
      \end{array}
  \end{array} \label{eq:global-incremental-LOG}
\end{equation}
The functional in~\eqref{eq:global-incremental-LOG} is convex but not strictly convex; nevertheless, problem \eqref{eq:global-incremental-LOG} has a unique solution, as demonstrated in Appendix~\ref{appendix:incremental-log-model}. The consistency of this
incremental solution with the strong form~\eqref{new-strong-form-LOG} follows by revisiting the arguments below~\eqref{eq:global-incremental}:
 minimization with respect to~$\tmmathbf{u}$ is equivalent to the equilibrium
problem~\eqref{v} and thus yields the first four equations
in~\eqref{new-strong-form-LOG}, while consistency with the normal contact condition~\eqref{new-strong-form-LOG}$_5$ is automatically guaranteed by the use of the free-energy density $\psi$ in~\eqref{new-free-energy-LOG}, in
the limit $\varepsilon_n \rightarrow + \infty$. We prove in Appendix~\ref{appendix:incremental} that minimization with respect to~$\alpha, \tmmathbf{\gamma}_t$ yields the discrete counterparts of the last two equations in~\eqref{new-strong-form-LOG}, the plastic law exactly and the Tresca law with the lagged threshold, in the limit $\varepsilon_t \rightarrow +
\infty$ that imposes $\tmmathbf{\gamma}_t \rightarrow \tmmathbf{u}_t$.

\begin{remark}
  None of the models proposed in this section accounts for geometrical asperity ride-overs. They are not designed to describe microscopic slip, $\|
  \tmmathbf{\gamma}_t \| \sim \eta$, any better than classical Coulomb friction does; for such microscopic slips, the dilatant Coulomb law is likely to perform better.
\end{remark}

\subsection{Lagrange multiplier formulation}\label{LM}
The penalty incremental formulations~\eqref{eq:global-incremental}
and~\eqref{eq:global-incremental-LOG} are equivalent to the contact
problems~\eqref{new-strong-form} and~\eqref{new-strong-form-LOG} only in the limit of infinitely large penalty parameters, $\varepsilon_n, \varepsilon_t
\rightarrow + \infty$. For the finite values used in the numerical examples (Table~\ref{tab:physical-values}), small violations of the constraints
in~\eqref{new-strong-form} and~\eqref{new-strong-form-LOG} are permitted. To
overcome this limitation, in this section we extend the incremental
principles~\eqref{eq:global-incremental} and~\eqref{eq:global-incremental-LOG}
to Lagrange multiplier formulations that enforce the constraints
in~\eqref{new-strong-form},~\eqref{new-strong-form-LOG} exactly.

The contact behavior in~\eqref{new-strong-form} combines
\begin{enumerateroman}
  \item an extended Signorini normal contact condition, with an additional penetration $\eta \alpha$,
  
  \item a Tresca friction whose threshold $\mu \sigma_0 \alpha$ is
  parametrized by~$\alpha$,
  
  \item a perfectly plastic law governing the evolution of~$\alpha$.
\end{enumerateroman}
A constrained minimization (variational inequality) formulation of the classical Signorini problem (no penetration) with Tresca friction is available in the literature, see ~{\cite{duvaut1972,ballard2024}}. Building on these works and on our penalty approach~\eqref{eq:global-incremental}, we propose the alternative incremental formulation: \textit{for given}~$\tmmathbf{u}^{(k - 1)},
\alpha^{(k - 1)}$, \textit{the current fields}~$\tmmathbf{u}^{(k)} : \Omega
\rightarrow \mathbb{R}^3$,~$\alpha^{(k)} : \Gamma_c \rightarrow [\alpha^{(k -
1)}, 1]$ \textit{are solutions to}
\begin{equation}
\begin{array}[t]{ll}
  \underset{\alpha^{(k - 1)} \leqslant \alpha \leqslant 1}{\underset{u_n - g_0
  - \eta \alpha \leqslant 0 \text{ on }
  \Gamma_c}{\underset{\tmmathbf{u}=\tmmathbf{u}^{\star} \text{ on }
  \Gamma_u}{\underset{\tmmathbf{u}, \alpha}{\text{Min}}}}} &
  \begin{array}[t]{l}
  \int_{\Omega} \left( \frac{1}{2} \tmmathbf{\varepsilon} (\tmmathbf{u}) :
  \tmmathbf{c}: \tmmathbf{\varepsilon} (\tmmathbf{u}) -\tmmathbf{f} \cdot
  \tmmathbf{u} \right) \mathd \tmmathbf{x}- \int_{\Gamma_F} \tmmathbf{F} \cdot
  \tmmathbf{u} \mathd s \\[10pt] \qquad + \int_{\Gamma_c} \left( \frac{\eta \sigma_0
  \alpha^2}{2} + \sigma_0 \mu \alpha^{(k - 1)}  \| \tmmathbf{u}_t
  -\tmmathbf{u}_t^{(k - 1)} \| \right) \mathd s.
  \end{array}\label{eq:Lagrange}
  \end{array}
\end{equation}

Problem~\eqref{eq:Lagrange} introduces several modifications of the classical Signorini--Tresca formulation~{\cite{duvaut1972,ballard2024}}, listed below.
\begin{enumerateroman}
  \item A penetration term $\eta \alpha$ is added to the unilateral
  constraint~$u_n - g_0 - \eta \alpha \leqslant 0$.
  
  \item The lagged value~$\alpha^{(k - 1)}$ is used in the friction threshold, for the reasons given in Section~\ref{sec:incremental-energy}.
  
  \item The plastic deformation of the asperities is accounted for exactly as in the penalty approach~\eqref{eq:global-incremental}, that is, by adding the crushing work $\eta \sigma_0 \alpha^2 / 2$ on the contact boundary, together with the irreversibility and cap constraints~$\alpha^{(k - 1)} \leqslant
  \alpha \leqslant 1$.
\end{enumerateroman}
The functional in~\eqref{eq:Lagrange} is strictly convex in $( \tmmathbf{u}, \alpha )$, since the elastic potential is strictly convex in $\tmmathbf{u}$ and $\eta \sigma_0 \alpha^2 / 2$ in $\alpha$, and the admissible set is convex since the constraint $u_n - g_0 - \eta \alpha \leqslant 0$ is affine; \eqref{eq:Lagrange} therefore has a unique solution whenever the prescribed displacement $\tmmathbf{u}^\star$ is compatible with the constraints.

For the logarithmic variant~\eqref{new-strong-form-LOG}-\eqref{eq:global-incremental-LOG} with enhanced microscopic behavior (Section~\ref{sec:log-model}), we adapt~\eqref{eq:Lagrange} as follows: the penetration $\eta \alpha$ is replaced by $\eta \ln (\alpha / \alpha^{\star})$, in order to comply
with~\eqref{new-strong-form-LOG}$_5$, and the crushing work $\eta
\sigma_0 \alpha^2 / 2$ is replaced by the one appearing in the penalty
formulation~\eqref{eq:global-incremental-LOG}. The result reads:
\textit{for given}~$\tmmathbf{u}^{(k - 1)}, \alpha^{(k - 1)}$, 
\textit{the current fields}~$\tmmathbf{u}^{(k)} : \Omega \rightarrow
\mathbb{R}^3$,~$\alpha^{(k)} : \Gamma_c \rightarrow [\alpha^{(k - 1)}, 1]$ \textit{are
solutions to}
\begin{equation}
  \begin{array}[t]{ll}
    \underset{\alpha^{(k - 1)} \leqslant \alpha \leqslant 1}{\underset{u_n -
    g_0 - \eta \ln (\alpha / \alpha^{\star}) \leqslant 0 \text{ on }
    \Gamma_c}{\underset{\tmmathbf{u}=\tmmathbf{u}^{\star} \text{ on }
    \Gamma_u}{\underset{\tmmathbf{u}, \alpha}{\text{Min}}}}} &
    \begin{array}[t]{l}
      \int_{\Omega} \left( \frac{1}{2} \tmmathbf{\varepsilon} (\tmmathbf{u}) :
      \tmmathbf{c}: \tmmathbf{\varepsilon} (\tmmathbf{u}) -\tmmathbf{f} \cdot
      \tmmathbf{u} \right) \mathd \tmmathbf{x}- \int_{\Gamma_F} \tmmathbf{F}
      \cdot \tmmathbf{u}\, \mathd s\\
      \qquad + \int_{\Gamma_c} (\eta \sigma_0  (\alpha - \alpha^{(k - 1)}) +
      \sigma_0 \mu \alpha^{(k - 1)}  \| \tmmathbf{u}_t -\tmmathbf{u}_t^{(k -
      1)} \|) \mathd s.
    \end{array}
  \end{array} \label{eq:Lagrange-LOG}
\end{equation}
In contrast to~\eqref{eq:Lagrange}, the functional in~\eqref{eq:Lagrange-LOG} is convex but not strictly convex in $( \tmmathbf{u}, \alpha )$, as is the penalty functional~\eqref{eq:global-incremental-LOG}; the constraint $u_n - g_0 - \eta \ln ( \alpha / \alpha^\star ) \leqslant 0$ still defines a convex set, since $- \ln$ is convex, and \eqref{eq:Lagrange-LOG} has a unique solution by the pointwise argument in $\alpha$ of Appendix~\ref{LM_proofs}.

\section{Numerical results}\label{sec:numerical-results}
In this section we illustrate numerically the behavior of the two models proposed in~\eqref{eq:global-incremental}
and~\eqref{eq:global-incremental-LOG} in a 2D plane strain setting. Section~\ref{sec:return-mapping} presents an implementation strategy, Section~\ref{sec:material-point-simulations} reports simulations at the material-point level, and Section~\ref{sec:rela-BVP} uses the models to solve two full boundary-value problems and validates their predictions against those of the classical implementation of Coulomb friction~\cite{laursen2003}. The code used in this section is distributed openly as~\cite{supplementary-material-coulomb}.

In this section, the elastic body we consider is homogeneous and isotropic with Young's modulus~$E$ and Poisson's ratio~$\nu$.
Two sets of material parameters are used, given in Table~\ref{tab:physical-values}. The values of $\eta$, $\beta$ and $\delta$ are taken from~\cite{greenwood1966};
the reference area fraction $\alpha^\star$ and the hardness $\sigma_0$ are
estimated from them with the elastic formulas~\eqref{eq:elastic-constants}
(Remark~\ref{rmk:elastic-asperities}), i.e. the hardness is set equal to the elastic
contact hardness of the surface, the scale of the real pressure at which its asperities
yield. The plastic estimate~\eqref{eq:ref-area-for-exp} of $\alpha^\star$, twice
as large, would change only the penetration scale $\eta\ln(1/\alpha^\star)$, from
$4.0\,\eta$ to $3.3\,\eta$.

\begin{table}[H]
\centering
  $\begin{array}{c|c|c|c|c|c|c|c|c|c|c|c}
    & g_0 & E & \nu & \mu & \eta & \beta & \delta & \sigma_0 & \alpha^{\star}
    & \varepsilon_n & \varepsilon_t\\
    \hline
    I & 0 & 10 & 0.375 & 0.8 & 1 \mathe - 6 & 20 \mathe - 6 & 300 \mathe 6 &
    1.47 & 1.88 \mathe - 2 & 1 \mathe 4 & 1 \mathe 4\\
    \hline
    I \nocomma I & 0.1 & \ldots & \ldots & 0.2 & \ldots & \ldots & \ldots &
    \ldots & \ldots & 1 \mathe 6 & 1 \mathe 6
  \end{array}$
  \caption{Values of the parameters used in the simulations of
  Sections~\ref{sec:material-point-simulations} and~\ref{sec:rela-BVP}. The
  values of $\eta,\beta$ and $\delta$ (columns 6 to~8) are taken from~{\cite{greenwood1966}}; $\sigma_0$ and $\alpha^{\star}$ are computed from them with~\eqref{eq:elastic-constants}. The symbol $\ldots$ indicates parameters that
  are identical in sets $I$ and $I \nocomma I$. \label{tab:physical-values}}
\end{table}

\subsection{Implementation strategy}\label{sec:return-mapping}

In classical contact mechanics algorithms, the original contact
problem~\eqref{strong-form} is solved iteratively at each time step
$k$ by a Newton-Raphson procedure~{\cite{laursen2003}}. This allows changes in the geometrical configuration to be tracked and guarantees that the set of
nodes in active contact is correctly identified at each time step. We adopt
a similar strategy for the global incremental penalty
formulations~\eqref{eq:global-incremental}
and~\eqref{eq:global-incremental-LOG}. A superscript $\left( k,
\hspace{0.2em} p \right)$ denotes evaluation at time step $k$ and Newton iteration
$p$. The implementation is presented in a 2D setting, so that the tangential displacement, slip and tangential traction are scalar quantities. 


\paragraph{Local updates}
Like any other internal variables,
$\gamma_t$ and $\alpha$ can be updated locally, node by node,
within a Newton iteration. To do so, we apply two successive return-mapping procedures, see Algorithm~\ref{algo:local_update}, in which the displacement $\tmmathbf{u}^{\left( k, \hspace{0.2em} p \right)}$ is
assumed to be given.

We first solve for $\alpha$, and introduce to this end the function

\begin{equation}
  \pi_n (\alpha) = \varepsilon_n  \left( u^{\left( k, \hspace{0.2em} p
  \right)}_n - g_0 - \eta \times \left\{ \begin{array}{ll}
    0 & \quad \text{classical Signorini~\eqref{strong-form}$_4$}\\
    \alpha & \quad \text{new linear model~~\eqref{eq:new-contact-stresses}}\\
    \ln (\alpha / \alpha^{\star}) & \quad \text{new logarithmic
    model~\eqref{eq:new-pressure_LOG}}
  \end{array} \right\} \right) , \label{eq:expr-of-pi}
\end{equation}
which gives the pressure $P_n$ produced by a real-area fraction~$\alpha$ at a given displacement~$\tmmathbf{u}^{\left( k, \hspace{0.2em} p \right)}$.
Anticipating the final goal of this section, namely to validate the predictions of the new models against those of the classical one~\eqref{strong-form}, we include in the definition the classical
Signorini condition~\eqref{strong-form}$_4$, for which
the pressure is independent of~$\alpha$,
see~\eqref{eq:conventional-contact-stresses}$_1 .$

The return-mapping procedure first assumes the real-area
fraction $\alpha$ to be frozen, that is, it takes $\alpha = \alpha^{(k -
1)}$, where $\alpha^{(k - 1)}$ is the converged solution at the previous time step. This trial solution is then checked against the plastic law~\eqref{new-strong-form}$_6$. If it is not admissible, $\pi_n (\alpha^{(k - 1)}) > \sigma_0
\alpha^{(k - 1)}$, the real-area fraction is allowed to grow, provided the cap
$\alpha = 1$ has not yet been reached, and a corrector~$\alpha$ is sought on
the boundary of the plastic domain,~$\pi_n (\alpha) = \sigma_0 \alpha$. The
current pressure $P_n^{\left( k, \hspace{0.2em} p \right)}$ then follows from
\begin{equation}
  P_n^{\left( k, \hspace{0.2em} p \right)} = \left\langle \pi_n \left(
  \alpha^{\left( k, \hspace{0.2em} p \right)} \right) \right\rangle_+ .
  \label{eq:numerical-contact-pressure}
\end{equation}
This quantity is used to detect the nodes in active contact (active\_set), which satisfy $
P_n^{\left( k, \hspace{0.2em} p \right)} > 0 $.

In accordance with Remark~\ref{rmk:open-gap}, the tangential terms of~\eqref{eq:global-incremental} and~\eqref{eq:global-incremental-LOG} are applied only to the nodes in contact at both $t_{k - 1}$ and $t_k$. Nodes that are not in active contact, and nodes entering contact at the current step, that is, nodes that were not in contact at the previous time step, carry no tangential traction, $T^{\left( k, \hspace{0.2em} p \right)}_t = 0$ (equivalently, $\gamma^{\left( k, \hspace{0.2em} p \right)}_t = u^{\left( k, \hspace{0.2em} p \right)}_t$), and their local update ends here. For the linear model and nodes entering contact for the first time, this is what~\eqref{eq:global-incremental} prescribes, since $\alpha^{(k - 1)} = 0$. For nodes re-entering contact ($\alpha^{(k - 1)} > 0$ by irreversibility) and, in the logarithmic model, for all nodes entering contact ($\alpha^{(k - 1)} \geqslant \alpha^{\star}$), the friction threshold is thus set to zero during the increment in which contact is established: this lag of one time step in the contact status plays the same role as the lag of $\alpha^{(k - 1)}$ in the threshold (Section~\ref{sec:incremental-energy}) and vanishes with the time step. The same rule is applied to the classical Coulomb implementation, for which it is a mere assumption. For the nodes in contact at both time steps, we evaluate the tangential slip $\gamma_t$ by a second return-mapping algorithm. This part of the implementation is identical for both models and coincides with the classical implementation of Tresca friction; Algorithm~\ref{algo:local_update} gives the main steps and {\cite{laursen2003}} a detailed overview. At the end of this step, these nodes are split between stick and slip, and their tangential traction is given by
\begin{equation}
  \begin{array}{lll}
    T_t^{\text{stick}} & = &T_t^{(k - 1)} +
    \varepsilon_t  \left( u_t^{\left( k, \hspace{0.2em} p \right)}
    -u_t^{(k - 1)} \right),\\
    T^{\text{slip}}_t & = & \frac{T_t^{\text{stick}}}{|
    T_t^{\text{stick}} |} \times \left\{ \begin{array}{ll}
     \mu P_n^{\left( k, \hspace{0.2em} p
    \right)} & \quad \text{classical Coulomb~\eqref{eq:coulomb-friction-for-penalty}}\\
    \mu \sigma_0 \alpha^{(k-1)} & \quad \text{new model (linear and logarithmic)~\eqref{eq:Tresca-lagged}} \end{array} \right\},
  \end{array}  \label{eq:numerical-tangential-stresses}
\end{equation}
where $u_t^{(k - 1)}$ and $T_t^{(k - 1)}$ are the
converged solutions at the previous time step $t_{k - 1}$. The expression~\eqref{eq:numerical-tangential-stresses} for the tangential stress is consistent with the
definition $T_t =
\varepsilon_t  (u_t -\gamma_t)$ of~\eqref{eq:new-contact-stresses}$_2$ used so far. In stick, $\gamma^{\left( k, \hspace{0.2em} p \right)}_t
=\gamma_t^{(k - 1)}$, so that
\begin{equation*}
T^{\left( k, \hspace{0.2em} p \right)}_t -T_t^{(k - 1)}
= \varepsilon_t  \left( u^{\left( k, \hspace{0.2em} p \right)}_t - \gamma^{(k,p)}_t
\right) - \varepsilon_t  \left( u^{(k - 1)}_t -
\gamma^{(k,p)}_t \right) = \varepsilon_t  \left(
u^{\left( k, \hspace{0.2em} p \right)}_t -u^{(k - 1)}_t
\right) =T_t^{\text{stick}} -T_t^{(k - 1)}.
\end{equation*}
In slip, $\gamma^{\left( k, \hspace{0.2em} p \right)}_t
=\gamma_t^{(k - 1)} + \varepsilon_t^{-1} ( T_t^{\text{stick}}  - T_t^{\text{slip}} )$, see Algorithm~\ref{algo:local_update}, and 

\begin{equation*}
T^{\left( k, \hspace{0.2em} p \right)}_t = \varepsilon_t (u^{\left( k, \hspace{0.2em} p \right)}_t - \gamma^{\left( k, \hspace{0.2em} p \right)}_t) = \varepsilon_t (u^{\left( k, \hspace{0.2em} p \right)}_t - \gamma_t^{(k - 1)}) - ( T_t^{\text{stick}}  - T_t^{\text{slip}} )~=~T_t^{\text{slip}}.
\end{equation*}

\paragraph{Global Newton loop} 
In contrast to $\gamma_t$ and
$\alpha$, the observable~$\tmmathbf{u}$ is solved for globally, by an iterative
Newton-Raphson procedure, see Algorithm~\ref{algo:full-Newton-loop}. Minimization with
respect to~$\tmmathbf{u}$ in~\eqref{eq:global-incremental}
and~\eqref{eq:global-incremental-LOG} yields the same weak
form~\eqref{eq:usual-weak-form} as that used in classical contact algorithms~{\cite{laursen2003}}, so we follow the same strategy: at each time step $k$ we solve a linearized weak form, parametrized by an index $p \geqslant 0$ on which we iterate. In iteration $p$ the unknown is the displacement increment $\Delta
\tmmathbf{u}^{\left( k, \hspace{0.2em} p \right)}$, \[ \Delta \tmmathbf{u}^{\left( k, \hspace{0.2em} p \right)}
   =\tmmathbf{u}^{\left( k, \hspace{0.2em} p + 1 \right)}
   -\tmmathbf{u}^{\left( k, \hspace{0.2em} p \right)}, \] which satisfies the linearized weak form
\begin{equation}
  \forall \delta \tmmathbf{u} \in \mathcal{C}_0, \hspace{3em} \phantom{\left[
  \frac{\partial}{} \right]} \delta \Phi \left[ \tmmathbf{u}^{\left( k,
  \hspace{0.2em} p \right)} ; \delta \tmmathbf{u} \right] + \Delta \delta \Phi
  \left[ \tmmathbf{u}^{\left( k, \hspace{0.2em} p \right)} ; \delta
  \tmmathbf{u} \right] \cdot \Delta \tmmathbf{u}^{\left( k, \hspace{0.2em} p
  \right)} =0. \label{newtom-generic}
\end{equation}
The residual $\delta \Phi \left[ \tmmathbf{u}^{\left( k, \hspace{0.2em} p
\right)} ; \delta \tmmathbf{u} \right]$ and the Jacobian $\Delta \delta \Phi
\left[ \tmmathbf{u}^{\left( k, \hspace{0.2em} p \right)} ; \delta \tmmathbf{u}
\right]$ are given by
\begin{equation}
  \begin{array}{cll}
    \delta \Phi \left[ \tmmathbf{u}^{\left( k, \hspace{0.2em} p \right)} ;
    \delta \tmmathbf{u} \right] & = & \int_{\Omega} \left(
    \tmmathbf{\varepsilon} \left( \tmmathbf{u}^{\left( k, \hspace{0.2em} p
    \right)} \right) : \tmmathbf{c}: \tmmathbf{\varepsilon} (\delta
    \tmmathbf{u}) -\tmmathbf{f} \cdot \delta \tmmathbf{u} \right) \mathd
    \tmmathbf{x}- \int_{\Gamma_F} \tmmathbf{F} \cdot \delta \tmmathbf{u}
    \,\mathd s\\
    &  & \qquad + \int_{\Gamma_c} \left( P_n^{\left( k, \hspace{0.2em} p
    \right)} \delta u_n +T_t^{\left( k, \hspace{0.2em} p \right)}
 \delta u_t \right) \mathd s,\\
    \Delta \delta \Phi \left[ \tmmathbf{u}^{\left( k, \hspace{0.2em} p
    \right)} ; \delta \tmmathbf{u} \right] \cdot \Delta \tmmathbf{u}^{\left(
    k, \hspace{0.2em} p \right)} & = & \int_{\Omega} \tmmathbf{\varepsilon}
    (\Delta \tmmathbf{u}^{\left( k, \hspace{0.2em} p \right)}) : \tmmathbf{c}: \tmmathbf{\varepsilon} (\delta
    \tmmathbf{u}) \mathd \tmmathbf{x}+ \int_{\Gamma_c} (\Delta P^{\left( k, \hspace{0.2em} p \right)}_n \delta u_n
    + \Delta T^{\left( k, \hspace{0.2em} p \right)}_t \delta u_t)\, \mathd s,
  \end{array} \label{contact-residual}
\end{equation}
where the pressure $P_n^{\left( k, \hspace{0.2em} p \right)}$ and the tangential traction
$T_t^{\left( k, \hspace{0.2em} p \right)}$ are given by the local update~\eqref{eq:expr-of-pi}-\eqref{eq:numerical-tangential-stresses}. Their linearizations $\Delta P^{\left( k, \hspace{0.2em} p \right)}_n$, $\Delta T^{\left( k, \hspace{0.2em} p \right)}_t$ yield an additional
contribution $\int_{\Gamma_c} (\Delta P^{\left( k, \hspace{0.2em} p \right)}_n \delta u_n + \Delta T^{\left( k, \hspace{0.2em} p \right)}_t \delta u_t)\, \mathd s$ to the stiffness matrix, a surface integral that we approximate by a sum of nodal contributions,
\begin{equation}
  \int_{\Gamma_c} \left( \Delta P^{\left( k, \hspace{0.2em} p \right)}_n
  \delta u_n + \Delta T^{\left( k, \hspace{0.2em} p \right)}_t \delta u_t
  \right) \mathd s = \underset{N \in \Gamma_c}{\sum}
  \left(\begin{array}{c}
    \left( \delta u_n \right)_N\\
    \left( \delta u_t \right)_N
  \end{array}\right)^T \tmmathbf{K}_N 
  \left(\begin{array}{c}
    (\Delta u^{\left( k, \hspace{0.2em} p \right)}_n)_N\\
    (\Delta u^{\left( k, \hspace{0.2em} p \right)}_t)_N
  \end{array}\right) \Delta s_N, \label{eq:contact-stiffness-contribution}
\end{equation}
where $\left( \Delta
\tmmathbf{u}^{\left( k, \hspace{0.2em} p \right)} \right)_N$  and $(\delta
\tmmathbf{u})_N$ are the displacement increment and the virtual displacement of
node $N$ at time step $k$, iteration $p$, and $\Delta s_N$ is the average size of the boundary facets to which $N$ is connected. The stiffness matrix
$\tmmathbf{K}^{\left( k, \hspace{0.2em} p \right)}_N$ 
depends on the contact model; for
the linear variant of our model~\eqref{eq:global-incremental} it reads
\begin{equation}
  \begin{array}{lll}
    \tmmathbf{K}^{\left( k, \hspace{0.2em} p \right)}_N & = &
    \left(\begin{array}{cc}
      1 & 0\\
      0 & 0
    \end{array}\right) \times \left\{ \begin{array}{ll}
      \sigma_0 \varepsilon_n / (\sigma_0 + \varepsilon_n \eta) & \text{ if } N
      \in \text{active\_set and } N \in \text{grow\_set}\\
      \varepsilon_n & \text{ if } N \in \text{active\_set and } N \notin
      \text{grow\_set}\\
      0 & \text{ otherwise}
    \end{array} \right\} \\[2em]
    &  & \qquad + \left(\begin{array}{cc}
      0 & 0\\
      0 & 1
    \end{array}\right) \times \left\{ \begin{array}{ll}
      \varepsilon_t & \text{ if } N \in \text{stick\_set}\\
      0 & \text{ otherwise}
    \end{array} \right\},
  \end{array} \label{contact-stiffness-matrix}
\end{equation}
where the dependence on the time step and iteration indices $k, p$ is hidden
in the definition of the node sets (active\_set, grow\_set, {\textdots}), see Algorithm~\ref{algo:local_update}. The expressions of $\tmmathbf{K}^{\left( k,
\hspace{0.2em} p \right)}_N$ for the logarithmic variant~\eqref{eq:global-incremental-LOG} and for the classical implementation of Coulomb friction~\eqref{strong-form} are given in Appendix~\ref{appendix:contact-stiffness-matrix}. Both variants of our model produce symmetric contact stiffnesses, which is a
direct consequence of the variational structure underlying the model; in
particular, the tangential contact stiffness vanishes in the slip regime, as for classical Tresca friction. In the classical implementation, by contrast, in the slip regime the contact stiffness contains an additional off-diagonal entry $\pm \mu \varepsilon_n$, which breaks its symmetry.

The above procedure is repeated for increasing values of $p$ until the displacement increment~$\Delta \tmmathbf{u}^{\left(
    k, \hspace{0.2em} p \right)}$ and the residual~$\delta \Phi \left[ \tmmathbf{u}^{\left( k,
  \hspace{0.2em} p \right)} ; \delta \tmmathbf{u} \right]$ have converged.

\subsection{Contact material-point response}\label{sec:material-point-simulations}
To evaluate the local contact constitutive response, we first perform material-point simulations under a prescribed displacement, which amounts to running only the local update routine of Algorithm~\ref{algo:local_update}. For both proposed models~\eqref{eq:global-incremental} and~\eqref{eq:global-incremental-LOG}, we track the internal state variables $(\alpha, \gamma_t)$ and the resulting contact tractions $(P_n, T_t)$ for the three loading cases of Table~\ref{tab:loading-cases}. Where a target normal pressure is specified, the normal displacement $u_n$ is controlled directly so as to achieve it. The response of the classical Coulomb model is also evaluated for comparison. All plots in this subsection use the material parameters of set $II$ in Table~\ref{tab:physical-values}. 

\begin{table}[htbp]
\resizebox{\textwidth}{!}{%
\centering
\begin{tabular}{c|c|c|c|c}
steps & 1 & 2 & 3 & 4\\\hline
normal load
 & $\begin{array}{l}u_n-g_0:\ 0\to8\text{e}-6\\ u_t=0\end{array}$ & & &\\\hline
$\begin{array}{c}\text{shear loop under}\\ \text{increasing pressure}\\
P_3>P_2>P_1\end{array}$
 & $\begin{array}{l}P_n:\ 0\to P_1\\ u_t=0\end{array}$
 & $\begin{array}{l}P_n=P_1\\ u_t:\ 0\to0.05\\ \phantom{u_t:\ }\to-0.05\to0
    \end{array}$
 & $\begin{array}{c}\text{redo 1-2 with}\\ P_n:\ P_1\to P_2\end{array}$
 & $\begin{array}{c}\text{redo 1-2 with}\\ P_n:\ P_2\to P_3\end{array}$\\\hline
$\begin{array}{c}\text{shear loop with}\\ \text{pressure decrease}\\
P_1>P_2\end{array}$
 & $\begin{array}{l}P_n:\ 0\to P_1\\ u_t=0\end{array}$
 & $\begin{array}{l}P_n=P_1\\ u_t:\ 0\to0.05\\ \phantom{u_t:\ }\to-0.05\to0
    \end{array}$
 & $\begin{array}{c}\text{redo 1-2 with}\\ P_n:\ P_1\to P_2\end{array}$
 &
\end{tabular}%
}
  \caption{Loading cases for material-point simulations.}\label{tab:loading-cases}
\end{table}

\begin{algorithm}[H]
\label{algo:local_update}
 \caption{\textbf{local update, Newton iteration $p$}\quad (time step
  $k > 0$)}
  
  \underline{\textbf{trial frozen step:}}   $  \alpha^{\left( k, \hspace{0.2em} p \right)} = \alpha^{(k - 1)}$

  \If{$ \pi_n (\alpha^{(k - 1)}) - \sigma_0 \alpha^{(k - 1)} > 0$}{
  
  \underline{\textbf{corrector grow step:}}\newline \uIf{$ \pi_n (1) > \sigma_0$}{ cap  
  $\alpha^{\left( k, \hspace{0.2em} p \right)} = 1$ }
  
\Else{solve $\pi_n \left(\alpha^{\left( k, \hspace{0.2em} p \right)}
\right) = \sigma_0 \alpha^{\left( k, \hspace{0.2em} p
\right)}$ for $\alpha^{\left( k, \hspace{0.2em} p \right)}$;
  
  append $N$ to grow\_set;}}
  
  evaluate the pressure $P^{\left( k, \hspace{0.2em} p
  \right)}_n=\langle \pi_n\left(\alpha^{\left( k, \hspace{0.2em} p
  \right)} \right)\rangle_+$, see~\eqref{eq:numerical-contact-pressure};
  
  \If{$P^{\left( k, \hspace{0.2em} p
  \right)}_n > 0$}{append $N$ to active\_set; 
  
  \uIf{$N$ already in contact at $t_{k - 1}$}{
  
  \underline{\textbf{trial stick step:}} 
  
  frozen slip $\gamma^{\left( k, \hspace{0.2em} p
  \right)}_t = \gamma^{(k-1)}_t$; 
  
  trial tangential traction
  $T^{\left( k, \hspace{0.2em} p
  \right)}_t=T^{\text{stick}}_t$, see~\eqref{eq:numerical-tangential-stresses}$_1$;
  
  \uIf{ $ 
  \| T_t^{\text{stick}} \| - 
  \left\{
  \begin{array}{ll}
     \mu P_n^{\left( k, \hspace{0.2em} p
    \right)} & \quad \text{classical Coulomb~\eqref{eq:coulomb-friction-for-penalty}}\\
    \mu \sigma_0 \alpha^{(k-1)} & \quad \text{new model (linear and logarithmic)~\eqref{eq:Tresca-lagged}} \end{array} 
    \right\} > 0      
  $
  }{
  \underline{\textbf{corrector slip step:}}
  
  append $N$ to slip\_set;
  
  correct the slip $\gamma^{\left( k, \hspace{0.2em} p \right)}_t
=\gamma_t^{(k - 1)} + \varepsilon_t^{-1} ( T_t^{\text{stick}}  - T_t^{\text{slip}} )$;

  correct the stress $T^{\left( k, \hspace{0.2em} p
  \right)}_t=T^{\text{slip}}_t$, see~\eqref{eq:numerical-tangential-stresses}$_2$;}
  
  \Else{append $N$ to stick\_set;}}
  }
 
\end{algorithm} 

\begin{algorithm}[H]
 
 $\tmmathbf{u}^{\left( k, \hspace{0.2em} 0 \right)} =\tmmathbf{u}^{(k - 1)}$
  
  $p = 0$
  
  \While{$p <$max\_iterations}{
  
  \For{each node $N$ in $\Gamma_c$}{
  
  active\_set, grow\_set, stick\_set, slip\_set $=$
  local\_update ($N$)
  
  using~\eqref{eq:expr-of-pi}-\eqref{eq:numerical-tangential-stresses} and~\eqref{eq:contact-stiffness-contribution}-\eqref{contact-stiffness-matrix}, update the residual and the stiffness entries corresponding to
  node $N$}
  
  solve~\eqref{newtom-generic}-\eqref{contact-residual} for $\Delta \tmmathbf{u}^{\left( k, \hspace{0.2em} p \right)}$;
  
  \If{$\norm{ \Delta \tmmathbf{u}^{\left( k, p \right)}} < \text{tol}$ and $\norm{\text{residual}} < \text{tol}$}{\textbf{break}}

  $\tmmathbf{u}^{\left( k, \hspace{0.2em} p + 1 \right)}
  =\tmmathbf{u}^{\left( k, \hspace{0.2em} p \right)} + \Delta
  \tmmathbf{u}^{\left( k, \hspace{0.2em} p \right)}$

  $p = p + 1$
  
  }
 \caption{\textbf{Newton-Raphson loop at time step $k>0$}} \label{algo:full-Newton-loop}
\end{algorithm}

\begin{figure}[H]\includegraphics[width=\textwidth]{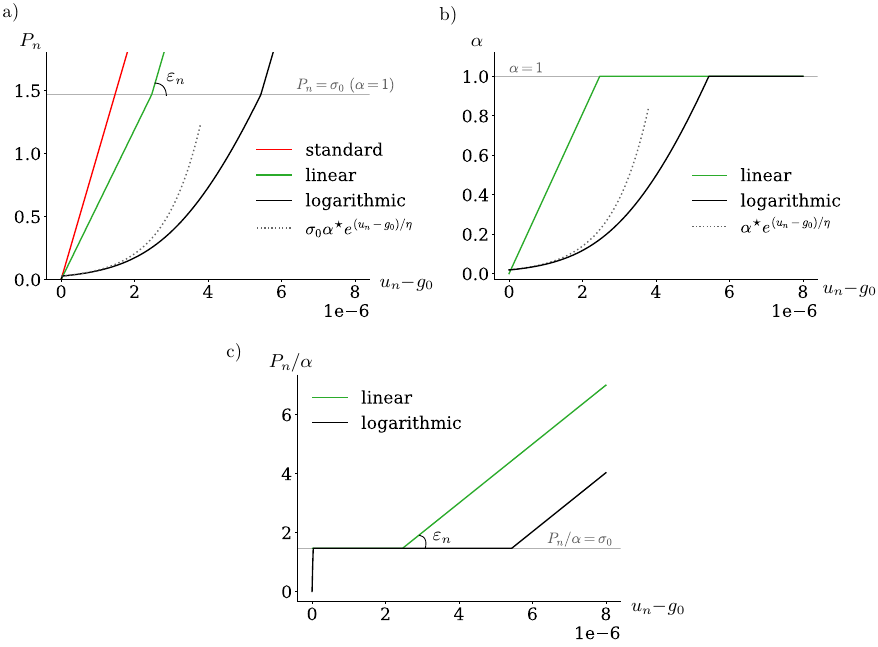}
  \caption{Material-point response of the two models~\eqref{eq:global-incremental} and~\eqref{eq:global-incremental-LOG} to an imposed normal penetration, see the second row of Table~\ref{tab:loading-cases}. a) Contact pressure against normal penetration. For comparison, the response of the classical contact law~\eqref{eq:conventional-contact-stresses}$_1$ is also plotted. b) Real-area fraction against normal penetration. c) Real pressure against normal penetration.} \label{fig:normal-load}
\end{figure}

\paragraph{Normal load}
We first consider the response to pure normal indentation ($u_t = 0$, second line of Table~\ref{tab:loading-cases}). Because the normal and tangential fields are decoupled in formulations~\eqref{eq:global-incremental} and~\eqref{eq:global-incremental-LOG}, this loading produces no shear response. Figure~\ref{fig:normal-load} shows the resulting contact pressure~$P_n$~(a), real-area fraction~$\alpha$~(b), and real pressure~$P_n/\alpha$~(c) against the normal penetration $u_n - g_0$. Whereas the classical contact model maintains a strictly linear pressure evolution $P_n = \varepsilon_n(u_n - g_0)$, in agreement with~\eqref{eq:expr-of-pi}$_1$-\eqref{eq:numerical-contact-pressure}, both variants of the proposed model exhibit an additional normal penetration caused by micro-asperity compaction. As anticipated in the previous section, this additional penetration reaches a maximum of approximately~$\eta$ for the linear model and~$4 \eta$ for the logarithmic one, attained at the yield cap $P_n = \sigma_0$ ($\alpha = 1$). Beyond the cap, Figure~\ref{fig:normal-load}b confirms that $\alpha$ saturates, so that both variants resume the linear elastic slope~$\varepsilon_n$ characteristic of the classical contact model. The logarithmic variant reproduces the exponential scaling~\eqref{eq:exp-distrib-Pn-alpha} between the normal penetration and the contact parameters ($\alpha, P_n$) up to the penalty penetration $P_n/\varepsilon_n$, which shifts its curves to the right of the dotted exponential laws in Figures~\ref{fig:normal-load}a and~\ref{fig:normal-load}b (from the classical rough-surface contact models \cite{greenwood1966, greenwood1967}) by up to $\sigma_0 / \varepsilon_n \approx 1.5\, \eta$ for set~$I\nocomma I$ and disappears in the limit $\varepsilon_n \rightarrow + \infty$. Figure~\ref{fig:normal-load} shows that the real pressure $P_n/\alpha$ reaches the plastic threshold $\sigma_0$ almost immediately, that is, for very small normal displacements; once all asperities are crushed ($\alpha = 1$), the real pressure coincides with the nominal one and follows the elastic stiffening slope $\varepsilon_n$.

\paragraph{Tangential cycle under increasing pressure}
 To evaluate the shear response under monotonically increasing loads, we apply the two-step loading protocol of Table~\ref{tab:loading-cases} (row 3) at three target normal pressure levels $P_1 < P_2 < P_3$. Each level involves two sequential steps: the normal pressure is first raised to $P_i$ at zero tangential displacement ($u_t = 0$); then the pressure is kept constant and a complete tangential displacement cycle is applied. Because normal compaction and shear response are constitutively decoupled, the first step generates no tangential stress and the second leaves the normal contact state unaltered.

\begin{figure}[H]{\includegraphics[width=\textwidth]{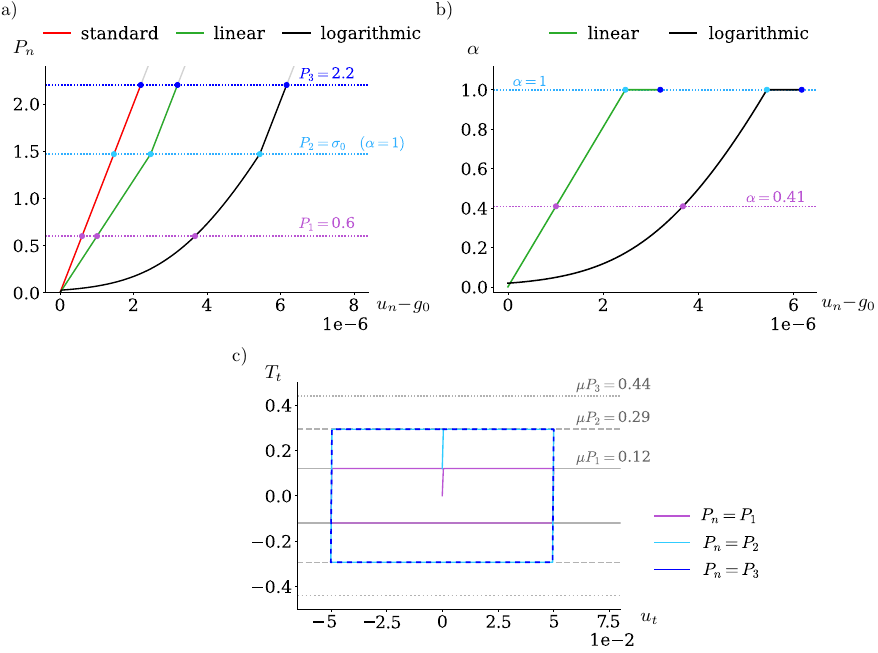}}
  \caption{Material-point response of the two models~\eqref{eq:global-incremental} and~\eqref{eq:global-incremental-LOG} to tangential cycles at monotonically increasing pressures, see the third row of Table~\ref{tab:loading-cases}. a) Contact pressure against normal penetration. The horizontal dashed lines mark the pressures at which a tangential cycle is applied; they are reached in increasing order, first $P_1$, then $P_2$, finally $P_3$. b) Real-area fraction against normal penetration. The horizontal dashed lines mark the values of the real-area fraction at which a tangential cycle is applied. c) Tangential traction response to an imposed tangential cycle, for each pressure marked in plot a). The plot is valid for both formulations~\eqref{eq:global-incremental} and~\eqref{eq:global-incremental-LOG}. The horizontal gray lines indicate the extremal values $\mu P_i$ ($i\in\{1,2,3\}$) expected for Coulomb friction.}
  \label{fig:tangential-cycle-increasing-pressure}
\end{figure}

Figures~\ref{fig:tangential-cycle-increasing-pressure}a and~\ref{fig:tangential-cycle-increasing-pressure}b plot the contact pressure $P_n$ and the real-area fraction $\alpha$ as functions of the normal penetration $u_n - g_0$. These trajectories follow the pure normal loading paths of Figure~\ref{fig:normal-load}, the horizontal dashed lines marking the constant pressure and real-area fraction at which tangential cycling is conducted. The targets are chosen such that $P_2 = \sigma_0$ lies precisely on the yield cap. Figure~\ref{fig:tangential-cycle-increasing-pressure}c shows the  tangential traction~$T_t$ during the tangential cycling phase at each pressure level $P_i$, with the horizontal gray lines marking the slip thresholds of classical Coulomb friction. Since both model variants~\eqref{eq:global-incremental} and~\eqref{eq:global-incremental-LOG} predict identical shear behavior under monotonically increasing normal pressures, their responses overlap in a single set of curves.

 In the elastic stick regime, the large penalty stiffness ($\varepsilon_t = 10^6$; set $II$ in Table~\ref{tab:physical-values}) produces near-vertical tangential stress-displacement slopes. In the slip regime, the behavior depends on the pressure level: for pressures up to the plastic threshold ($P_1, P_2 \le \sigma_0$), the tangential traction coincides with classical Coulomb friction~($\mu P_n$); once the pressure exceeds the yield cap ($P_3 > \sigma_0$), the response saturates at the Tresca limit $\mu \sigma_0$, which caps the maximum shear capacity. This transition reflects the underlying surface topography: fully crushed asperities ($\alpha = 1$) have already reached the maximum contact area and therefore cannot sustain a higher frictional resistance.

\begin{figure}[H]{\includegraphics[width=\textwidth]{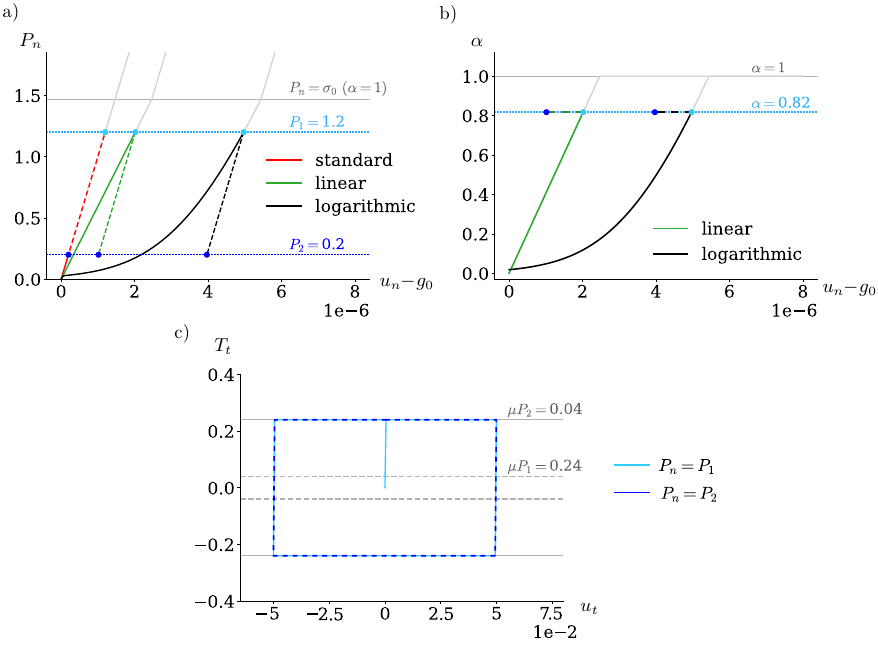}}
  \caption{Material-point response of the two models~\eqref{eq:global-incremental} and~\eqref{eq:global-incremental-LOG} to tangential cycles at
decreasing pressures, see the fourth row of Table~\ref{tab:loading-cases}. a) Contact pressure against normal penetration. The horizontal dashed lines indicate the values of pressure at which a tangential cycle is applied, and the inclined dashed lines represent unloading from $P_1$ to the lower pressure $P_2<P_1$. b) Real-area fraction against normal penetration. During unloading from $P_1$ to $P_2<P_1$ (thick dashed lines), $\alpha$ remains constant at $0.82$ (cyan horizontal line). c) Tangential traction response to an imposed tangential cycle, for each pressure marked in plot a). The plot is valid for both formulations~\eqref{eq:global-incremental} and~\eqref{eq:global-incremental-LOG}. The horizontal gray lines indicate the extremal values $\mu P_i$ ($i\in\{1,2\}$) expected for Coulomb friction.}
  \label{fig:tangential-cycle-pressure-decrease}
\end{figure}

 \paragraph{Tangential cycle with pressure decrease}
To evaluate the history dependence during normal unloading, we apply the same two-step protocol as in the previous paragraph, but with decreasing pressure levels ($\sigma_0 > P_1 > P_2$), see the fourth line of Table~\ref{tab:loading-cases}. Figures~\ref{fig:tangential-cycle-pressure-decrease}a and~\ref{fig:tangential-cycle-pressure-decrease}b trace the evolution of $P_n$ and $\alpha$ against~$u_n - g_0$, the thin horizontal dashed lines indicating the $(P_n, \alpha)$ states at which tangential cycling is conducted, while Figure~\ref{fig:tangential-cycle-pressure-decrease}c shows the tangential traction against the tangential displacement, with the horizontal gray lines denoting the classical Coulomb thresholds. In all panels, the unloading path ($P_1 \to P_2$) and its tangential cycle are drawn with thick dashed lines.

The loading phase up to $P_1$ is identical to that of the previous paragraph. During the unloading phase~$P_1 \to P_2$, by contrast, the plastic deformation of the micro-asperities is irreversible, so that the real-area fraction remains frozen at its peak historical value ($\alpha = 0.82$; Figure~\ref{fig:tangential-cycle-pressure-decrease}b). The normal pressure therefore drops along an elastic linear unloading path of slope $\varepsilon_n$ for both model variants, matching classical Coulomb behavior. This microscopic irreversibility produces a distinct memory effect in the tangential response (Figure~\ref{fig:tangential-cycle-pressure-decrease}c): while the friction threshold at $P_1$ matches classical Coulomb ($\mu P_1$), subsequent tangential cycling at the lower pressure $P_2 < P_1$ retains the higher threshold $\mu P_1$ rather than $\mu P_2$. Unlike classical Coulomb friction, whose threshold follows instantaneously the normal force, both proposed formulations exhibit a path-dependent friction threshold. Memory effects of this kind are known in other frictional systems. In rock friction, a step change of the normal stress produces an immediate change of the resistance to sliding followed by a transient evolution as sliding proceeds, which rate-and-state laws describe through the evolution of the state variable, a measure of the effective contact time, over a characteristic slip distance~\cite{linker1992}. In granular materials, the memory of the maximum past pressure, known as overconsolidation, is lost as shear-induced dilatancy or compaction drives the material towards the critical state~\cite{schofield1968,agajanian2026}. In both cases, however, the memory fades with slip, whereas here it is permanent, since slip leaves $\alpha$ unchanged.
 
 In the next subsection, we extend the analysis to full boundary-value problems solved by finite elements, and show that this memory effect does not disrupt the agreement with classical Coulomb predictions in two classical benchmark tests.

\subsection{2D boundary-value problems}
\label{sec:rela-BVP}

We implement the full procedure
of Section~\ref{sec:return-mapping} in a finite element framework based on the DOLFINx
library~{\cite{baratta2023,scroggs2022a, scroggs2022b,alnaes2014}}. 
In the following, we use this implementation to validate the predictions of our model on two boundary-value problems.

\paragraph{Generalized Hertz benchmark test}

\begin{figure}[H]
\centering
 {\includegraphics[width=\textwidth]{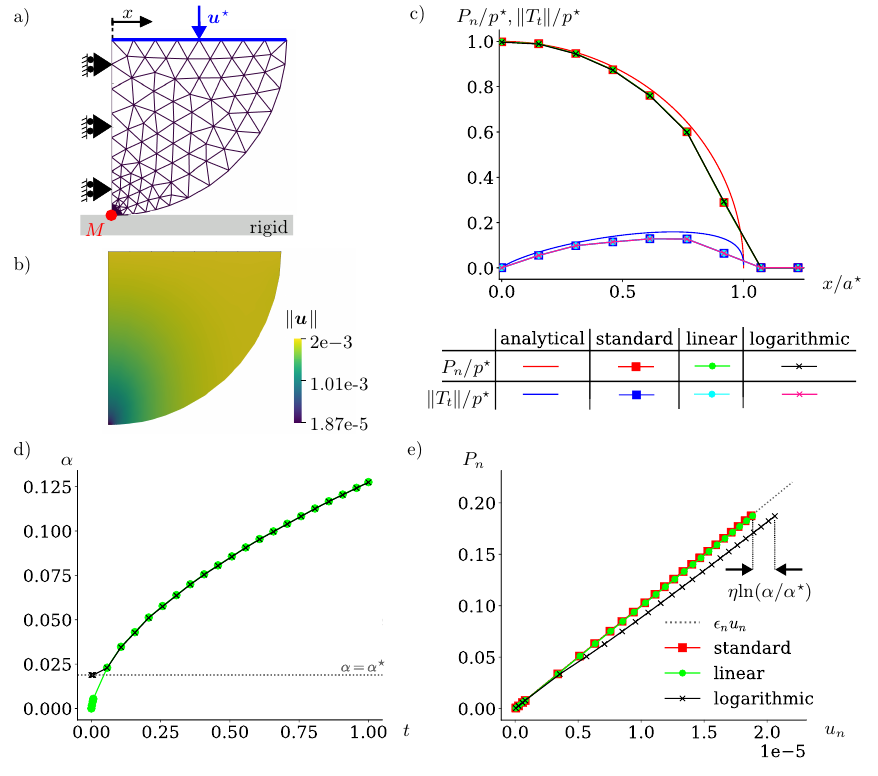}}
  \caption{Generalized Hertz benchmark test. a)~Loading case and mesh. The displacement $\tmmathbf{u}^{\star}$ prescribed on the top boundary is given in~\eqref{eq:imposed-indentation-displacement}. We highlight a node~$M$ for which we track the real-area fraction, pressure and normal displacement in subplots d and e. b)~Deformed shape at the last time step, computed with the classical implementation of Coulomb friction. c)~Scaled contact tractions $(P_n/p^{\star},|T_t|/p^{\star})$ against the scaled abscissa $x
/a^{\star}$, with $p^{\star}$ and $a^{\star}$ given in~\eqref{eq:p0}-\eqref{eq:a0}, for all three models (linear, logarithmic, classical Coulomb). The predictions are also compared with the approximate theoretical solution in~\eqref{eq:Hertz}--\eqref{eq:q-theo}. d)~Evolution in time of the real-area fraction $\alpha$ at node~$M$. The horizontal gray line indicates the initialization of the logarithmic variant; the linear model is initialized at $\alpha=0$. e)~Contact pressure $P_n$ against normal displacement $u_n$ at node $M$. }
  \label{fig:hertz-indentation-test}
\end{figure}

We consider a half-disk of unit radius ($r = 1$) brought into frictional contact with a horizontal rigid plane under plane-strain conditions. Exploiting the symmetry of the boundary-value problem about the plane $x = 0$, we simulate only a quarter-disk, with symmetry boundary conditions on its left edge (Figure~\ref{fig:hertz-indentation-test}a). The material parameters are those of set~$I$ in Table~\ref{tab:physical-values}. The initial gap between the half-disk and the substrate is zero ($g_0 = 0$), and contact is enforced by prescribing the displacement $\tmmathbf{u} = \tmmathbf{u}^{\star}$ on the top boundary of the quarter-disk. At each time step $t_k$, the prescribed displacement is
 \begin{equation}
   \tmmathbf{u}^{\star} = (0, - 1\mathe-6 - 0.002 t_k), \label{eq:imposed-indentation-displacement}
\end{equation}
where the discrete time grid $t_k$ ($k \ge 0$) is given by
\[ t_k \in \{ 0, 0.002, 0.005, 0.007, \ldots, 0.007 + j \Delta t, \ldots, 1 \}  , \]
with $1\leqslant j \leqslant 19$ and $\Delta t = 0.05$. The first time step is frictionless, which is a consequence of the lag for the linear model, an approximation for the logarithmic one ($\mu \sigma_0 \alpha^{\star}\neq 0$) and an assumption for the classical model. The quarter-disk  is discretized with first-order triangular elements whose typical size ranges from $0.005$ in the highly refined contact region to $1$ in the bulk (Figure~\ref{fig:hertz-indentation-test}a).

Numerical solutions are computed for both variants of the proposed formulation and, for comparison, for the classical Coulomb friction model. The classical and the proposed linear model achieve quadratic convergence of the Newton scheme, whereas the proposed logarithmic model shows a slightly slower convergence rate (Appendix~\ref{appendix:convergence-plots}, Figure~\ref{fig:convergence-hertz}), which we attribute to the stronger nonlinearity of the logarithm. The deformed shape at the last time step, computed with the classical Coulomb implementation, is shown in Figure~\ref{fig:hertz-indentation-test}b.

An approximate analytical solution for this problem is available in the literature~{\cite{nowell1988}}: the contact pressure follows a classical Hertzian profile,
\begin{equation}P^{\text{theo}} = p^{\star} \sqrt{1 - \left( \frac{x}{a^{\star}} \right)^2}, 
\label{eq:Hertz}
\end{equation}
where the maximum pressure is \begin{equation}
    p^{\star} = (E a^{\star}) / [2 r (1 - \nu^2)],
    \label{eq:p0}
\end{equation} and the total contact width is $2 a^{\star}$, the semi-width~$a^{\star}$ being given explicitly in terms of the total vertical force $F$ applied to the top boundary of the half-disk,
\begin{equation}
    a^{\star} = \sqrt{\frac{4 F}{\pi} \frac{(1 - \nu^2) r}{E}}.
    \label{eq:a0}
\end{equation}

In our numerical implementation the displacement is prescribed on the top boundary and the resulting total reaction force $F$ (i.e. twice the reaction force on the simulated quarter-disk) is extracted via numerical boundary integration within DOLFINx. 

The tangential traction produced under full stick by the elastic mismatch between the cylinder and the rigid plane, obtained by neglecting the influence of the shear traction on the pressure, reads~\cite{spence1968,johnson1985}
\begin{equation}
  T^{\text{theo}}_t = \frac{\kappa p^{\star}}{\pi} \left[ \sqrt{1 - \xi^2} \, \ln \frac{1 + \xi}{1 - \xi} + 2 \xi \ln \frac{1 + \sqrt{1 - \xi^2}}{| \xi |} \right], \qquad \xi = \frac{x}{a^{\star}}, \qquad \kappa = \frac{1 - 2 \nu}{2 (1 - \nu)} .
  \label{eq:q-theo}
\end{equation}
For a finite friction coefficient, slip occurs in two zones at the edges of the contact region, whose width decreases exponentially with $\mu / \kappa$~\cite{spence1975,nowell1988}; for $\mu = 0.8$ and $\nu = 0.375$ ($\kappa = 0.2$) it is of the order of $10^{-4} a^{\star}$, far below the mesh size, which is why the full-stick expression~\eqref{eq:q-theo} serves as reference.
As in~{\cite{delorenzis2011}}, we take $\mu = 0.8$ (set~$I$, Table~\ref{tab:physical-values}), for which the theoretical solution serves as a baseline in Figure~\ref{fig:hertz-indentation-test}c. In this figure, the normalized contact tractions $P_n / p^{\star}$ and $|T_t| / p^{\star}$ predicted by the theoretical solution and by all three numerical models (linear, logarithmic and classical Coulomb) are plotted against the rescaled horizontal coordinate $x / a^{\star}$. The three numerical predictions match closely, and all three are in good agreement with the theoretical prediction.  The small difference between the numerical and the theoretical curves can be attributed to the approximate nature of the theoretical solution and to artifacts of the penalty regularization at the values $\varepsilon_n = \varepsilon_t = 1\mathe4$ (set~$I$, Table~\ref{tab:physical-values}).

Figures~\ref{fig:hertz-indentation-test}d and~\ref{fig:hertz-indentation-test}e take a more local view, tracking the evolution in time of the real-area fraction~$\alpha$ and the pressure-penetration path~$P_n(u_n)$ at a representative node $M$ inside the contact region (Figure~\ref{fig:hertz-indentation-test}a). Apart from the difference in initial state ($\alpha = 0$ for the linear variant against $\alpha = \alpha^{\star}$ for the logarithmic one), the real-area fraction follows identical trajectories for the two variants. Since the macroscopic contact tractions match across all three models (Figure~\ref{fig:hertz-indentation-test}c), the progressive growth of $\alpha$ shows up in Figure~\ref{fig:hertz-indentation-test}e as an additional micro-scale asperity penetration, consistent with the material-point observations of Section~\ref{sec:material-point-simulations}.

\begin{remark}
Quasi-quadratic convergence with the logarithmic formulation~\eqref{eq:global-incremental-LOG} was obtained only for penalty values up to $\varepsilon_n = \varepsilon_t = 1\mathe4$; for the higher values of set~$I\nocomma I$ the convergence is approximately linear (Figure~\ref{fig:convergence-sliding}b). This motivated the choice of set $I$ for the present test, in which a theoretical solution serves as reference. Penalty values of this magnitude may produce numerical approximations coarser than the physical normal gaps predicted by the logarithmic law, so that the resolution of the logarithmic model may not be fully exploited in this particular penalty test. Such limitations are inherent to penalty regularization and can be circumvented with the Lagrange multiplier framework presented in Section~\ref{LM}.
\end{remark}

\paragraph{Half-disk sliding on a rigid plane}

\begin{figure}[H]
\centering
 {\includegraphics[width=\textwidth]{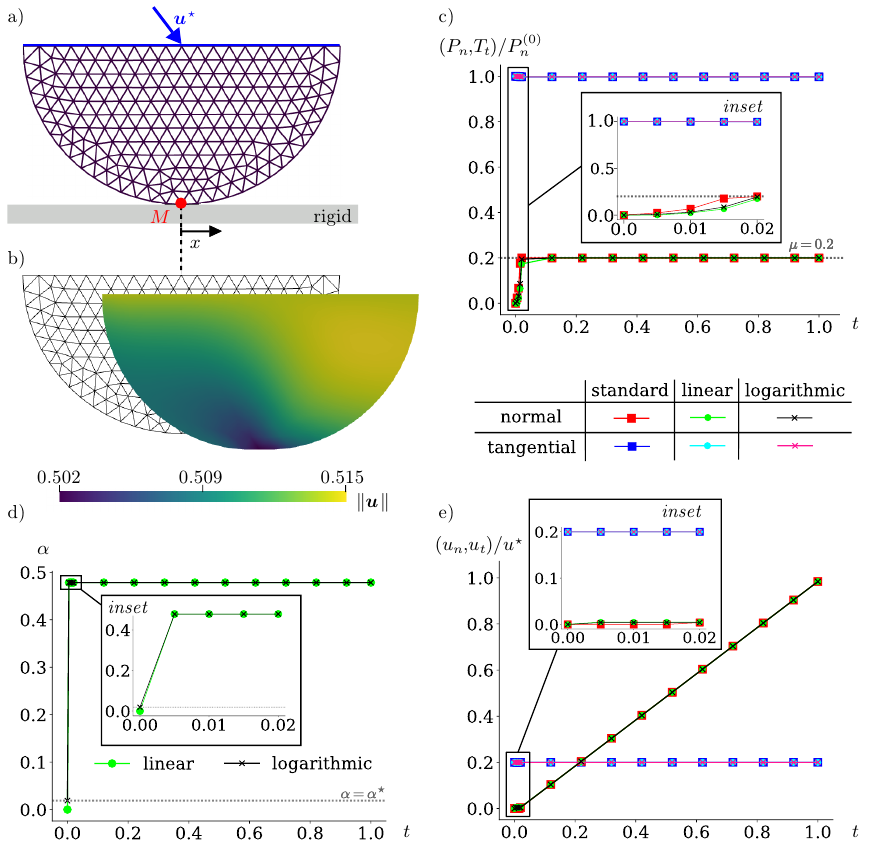}}
  \caption{Half-disk sliding on a rigid plane. a) Loading case and mesh. The displacement $\tmmathbf{u}^{\star}$ prescribed on the top boundary is given in~\eqref{eq:imposed-displacement}. We highlight a node~$M$ for which we track the real-area fraction, the displacement and the contact tractions in subplots c,~d and~e. b) Deformed shape at the last time step, computed with the classical implementation of Coulomb friction. c)~Evolution in time of the rescaled contact traction amplitudes $P_n, T_t$ at node $M$. The stresses are rescaled by the initial pressure $P_n^{(0)}$ given by the classical implementation. The horizontal gray line highlights the value of the friction coefficient~$\mu=0.2$, which in slip conditions coincides with the rescaled tangential traction amplitude, since the rescaled pressure is constant, $P_n/P_n^{(0)}=1$. d)~Evolution in time of the real-area fraction $\alpha$ at node~$M$. The horizontal gray line indicates the initialization of the logarithmic variant; the linear model is initialized at $\alpha=0$. e)~Evolution in time of the rescaled displacements $(u_n/u^{\star}, u_t/u^{\star})$ at node~$M$, with $u^{\star}$ defined in~\eqref{eq:imposed-displacement}.}
  \label{fig:half-disk-sliding}
\end{figure}

In this second example, a half-disk of unit radius ($r = 1$) is brought into contact with a horizontal rigid plane and then dragged along the horizontal direction (Figure~\ref{fig:half-disk-sliding}a). We again assume plane-strain conditions. The material parameters are those of set $I \nocomma I$ in Table~\ref{tab:physical-values}, and the domain is discretized with first-order triangular elements of representative size approximately $0.1$ (Figure~\ref{fig:half-disk-sliding}a). 
Contact is driven by prescribing on the top boundary the displacement
\begin{equation}
  \tmmathbf{u}^{\star} = (t_k u^{\star}, - 0.12) \hspace{1em} \text{with} \hspace{1em}  u^{\star} = 0.5.
  \label{eq:imposed-displacement}
\end{equation}
 The discrete time sequence $t_k$ is specified by
\[t_k \in \{ 0, 0.005, 0.01, 0.015, 0.02, 0.02 + j \Delta t, \ldots, 1 \},\]
with $1 \leqslant j \leqslant 9$ and $\Delta t = 0.1$. At the initial step $(k=0)$, the internal variables keep their initial values, $\alpha^{(0)} = 0$ or $\alpha^{\star}$ and $\gamma_t^{(0)} = 0$, so that the three models coincide there: the body is pushed vertically downward and contact is established without friction. In the subsequent steps the vertical displacement of the top boundary is held fixed and the half-disk is dragged horizontally across the plane. Figure~\ref{fig:half-disk-sliding}b shows the deformed configuration at the final time step, computed with the classical Coulomb implementation.

The refined initial time increments (of 0.005) are needed because the
friction threshold of the proposed models at $t_k$ is that of the state at $t_{k - 1}$ (Section~\ref{sec:incremental-energy}). At $t_1$ it is the threshold of the initial state, $\mu \sigma_0 \alpha^{(0)}$, which vanishes for the linear model and equals $\mu \sigma_0 \alpha
^{\star} \approx 0.008\, P_n^{(0)}$ for the logarithmic one: the nodes in contact slide almost freely during the first increment, by at most $0.005 u^{\star}$. From $t_2$ on, the
threshold $\mu \sigma_0 \alpha^{(1)} = \mu P_n^{(1)}$ coincides with the Coulomb one, since the pressure is constant in this test (Figure~\ref{fig:half-disk-sliding}c). The deviation
from Coulomb friction is thus confined to the first increment and proportional to its size.

Numerical solutions are computed for both model variants and for classical Coulomb friction. For the classical and the proposed linear model the Newton scheme again converges quadratically, whereas convergence is approximately linear for the logarithmic formulation, probably because of the stronger nonlinearity combined with the very high penalty parameters $\varepsilon_n = \varepsilon_t = 1 \mathe 6$ (set $I \nocomma I$, Table~\ref{tab:physical-values}). Convergence plots are provided in Appendix~\ref{appendix:convergence-plots} (Figure~\ref{fig:convergence-sliding}).

Figure~\ref{fig:half-disk-sliding} shows the evolution in time of the rescaled contact traction amplitudes $P_n / P_n^{(0)}$ and $T_t / P_n^{(0)}$ (subplot c), of the real-area fraction $\alpha$ (subplot d), and of the rescaled displacements $u_n/u^{\star},u_t/u^{\star}$ (subplot e) at a node in contact (node~$M$, see Figure~\ref{fig:half-disk-sliding}a). Displacements are rescaled by the maximum imposed value $u^{\star} = 0.5$ of~\eqref{eq:imposed-displacement}, and contact tractions by the initial normal pressure $P_n^{(0)}$, evaluated at $t_0 = 0$ with the classical Coulomb model. 

The classical model predicts a nearly constant normal displacement and contact pressure over time, a behavior accurately reproduced by both proposed formulations. In both variants, the real-area fraction $\alpha$ jumps at the first time step $t_1$ and then settles on a stable asymptotic plateau (Figure~\ref{fig:half-disk-sliding}d). This evolution of $\alpha$ induces small micro-asperity penetrations, but its effect on the normal displacement and pressure is imperceptible at the macroscopic scale.

The jump of $\alpha$ at $t_1$ also explains the minor mismatch in tangential traction between the classical and the proposed approaches during the first time steps (inset of Figure~\ref{fig:half-disk-sliding}c): the classical model, whose threshold is the current $\mu P_n$, resists
sliding from $t_1$, whereas the proposed models start resisting at $t_2$, so that their
tangential traction is the classical one delayed by one increment. This transient effect remains minor and localized: once the steady sliding regime is established, around $t_5 = 0.12$, all three models give an identical response.

\section{Conclusions}
We have proposed an associated model of Coulomb friction, i.e.~a friction model compatible with the framework of GSM: the model derives from a free-energy density and a dissipation potential, and the evolution of the internal variables obeys normality with respect to a convex domain in the space of their dissipative forces. Our key idea is to introduce, alongside the slip, a second internal variable~$\alpha$, the real-area fraction, which describes the state of contact at the microscopic level. The frictional behavior takes the form of a simple Tresca friction whose threshold is parametrized by~$\alpha$, whereas the evolution of~$\alpha$ is driven by the plastic deformation of the asperities, in accordance with the classical results on the real area of contact of rough surfaces~\cite{bowden1954,greenwood1966,greenwood1967}. The two dissipative processes are associated and only weakly coupled through~$\alpha$; normality holds in the space of the dissipative forces $(\eta P_n, \tmmathbf{T}_t)$ rather than in that of the contact tractions, so that slip remains purely tangential. The model thus inherits all the convenient properties of the GSM framework: it is thermodynamically consistent, variational in both the rate and the incremental sense, and upon spatial discretization it yields a symmetric stiffness matrix.

The new variable produces an additional asperity-scale penetration in the classical Signorini condition. We have explored two alternative expressions for this penetration: a linear one, particularly suited to numerical implementation, and a logarithmic one, which follows the exponential law of the classical rough-surface contact models {\cite{greenwood1966,greenwood1967}}. For each variant we have given a
continuous and an incremental penalty formulation, which describe the contact of a deformable body with a flat rigid
obstacle in 3D, as well as an incremental formulation based on Lagrange multipliers. For validation purposes, we have implemented the penalty formulations in a 2D setting and performed simulations both at the material-point level and at the structural level. In all cases, the model predictions agree closely with those of the classical Coulomb law as long as the pressure does not decrease and remains below the saturation threshold~$\sigma_0$; in the incremental problems the friction threshold lags the pressure by one time step, a discrepancy of order $\Delta t$ that vanishes with the time step. For significant pressure decreases, the model exhibits a memory effect, which originates from the irreversibility of the asperity deformation; above the saturation threshold, the frictional traction saturates at the Tresca limit $\mu \sigma_0$.

In future work, we plan to further study the memory effect and possibly mitigate it by taking into account the elastic deformation of the asperities alongside plasticity, which would introduce additional $\alpha$-dependent terms
in the contact free-energy density. These various microscopic energy contributions could also be derived by boundary-layer analysis. The evolution of $\alpha$ could further be enriched to account for the aging and renewal of contacts that underlie rate-and-state friction~\cite{dieterich1994,baumberger2006}. 
Adapting the implementation to the Lagrange multiplier formulation, and to more complex 3D problems with non-flat and/or deformable obstacles, is also left for future work. We further hope that the key idea of this paper, namely the recovery of associativity, and hence GSM-compatibility, by uncovering hidden internal variables, can be adapted to other non-associated laws, including non-associated plasticity.

\section{Declaration on the use of generative AI in the manuscript preparation process}
During the preparation of this work, the authors used Anthropic Claude for brainstorming and for improving the English language. After using this tool, the authors reviewed and edited the paper as needed and
take full responsibility for its contents.

\appendix\section{Dissipation potentials and evolution
laws}\label{appendix:dissipation-potentials}

Section~\ref{appendix:dilatant-coulomb} justifies the expression of the dissipation potential of dilatant Coulomb friction announced in Section \ref{sec:lack-of-normality}. Sections~\ref{appendix:linear-model}-\ref{appendix:log-model} then show that the
evolution laws~\eqref{eq:Tresca-reformulation} and~\eqref{new-strong-form}$_6$ proposed above are equivalent to the continuous minimization principle~\eqref{eq:new-max-dissipation-principle}.

\subsection{Dilatant Coulomb friction}\label{appendix:dilatant-coulomb}

We look for a convex, lower semicontinuous
and positively one-homogeneous dissipation potential $\varphi (\dot{\gamma}_n,
\dot{\tmmathbf{\gamma}}_t)$ consistent with the dissipative forces~\eqref{eq:dissipative-forces-dilatant} and the normality rule~\eqref{eq:associated-flow-rule}, with the Coulomb friction domain $\|
\tmmathbf{T}_t \| \leqslant \mu
P_n$~of \eqref{eq:coulomb-friction-for-penalty}$_1$, and with $P_n = \varepsilon_n 
\langle u_n - g_0 - \gamma_n \rangle_+ \geqslant 0$. Such a potential is the support function of its elastic domain, i.e. the Legendre–Fenchel transform of the indicator function of that domain
\[ \varphi (\dot{\gamma}_n, \dot{\tmmathbf{\gamma}}_t) = \underset{P_n,
   \tmmathbf{T}_t}{\sup} \enspace \dot{\gamma}_n P_n +
   \dot{\tmmathbf{\gamma}}_t \cdot \tmmathbf{T}_t - I_{\{ \| \tmmathbf{T}_t \|
   \leqslant \mu P_n \}} - I_{\{ 0 \leqslant P_n \}}, \]
where $I_{\{ a \leqslant b \}}$ denotes the indicator function of the convex set $\left\{ (a, b), \enspace a \leqslant b \right\}$, 
\[ I_{\{ a \leqslant b \}} = \left\{\begin{array}{ll}
     + \infty & \text{if } a > b,\\
     0 & \text{if } a \leqslant b.
   \end{array}\right. \]
This supremum can be computed as
\[ \begin{array}{lll}
     \varphi (\dot{\gamma}_n, \dot{\tmmathbf{\gamma}}_t) & = & \underset{0
     \leqslant P_n}{\underset{\| \tmmathbf{T}_t \| \leqslant \mu
     P_n}{\underset{P_n, \tmmathbf{T}_t}{\sup}}} \enspace \dot{\gamma}_n P_n +
     \dot{\tmmathbf{\gamma}}_t \cdot \tmmathbf{T}_t\\
     & = & \underset{0 \leqslant P_n}{\underset{P_n, \tmmathbf{T}_t}{\sup}}
     \enspace (\dot{\gamma}_n + \| \dot{\tmmathbf{\gamma}}_t \| \mu) P_n .
   \end{array} \]
It can be expressed as the indicator function
\begin{equation}
  I_{\left\{ \hspace{0.em} \dot{\gamma}_n + \| \dot{\tmmathbf{\gamma}}_t \|
  \mu \leqslant 0 \right\}} = \left\{\begin{array}{ll}
    + \infty & \text{if } \dot{\gamma}_n + \| \dot{\tmmathbf{\gamma}}_t \| \mu
    > 0,\\
    0 & \text{if } \dot{\gamma}_n + \| \dot{\tmmathbf{\gamma}}_t \| \mu
    \leqslant 0,
  \end{array}\right. \label{eq:dilatant-potential-dissip}
\end{equation}
hence the expression announced in Section \ref{sec:lack-of-normality} for the dissipation potential $\varphi (\dot{\gamma}_n,
\dot{\tmmathbf{\gamma}}_t)$,
\begin{equation}
  \varphi (\dot{\gamma}_n, \dot{\tmmathbf{\gamma}}_t) = I_{\left\{
  \hspace{0.em} \dot{\gamma}_n + \| \dot{\tmmathbf{\gamma}}_t \| \mu \leqslant
  0 \right\}} . \label{eq:dilatant-pot}
\end{equation}
\subsection{New model (linear version)}\label{appendix:linear-model}

Our aim is to prove that the continuous minimization principle~\eqref{eq:new-max-dissipation-principle},
\[ \underset{\dot{\alpha},
   \dot{\tmmathbf{\gamma}}_t}{\text{Min}} \enspace \dot{\psi} (u_n,
   \tmmathbf{u}_t, \alpha, \tmmathbf{\gamma}_t, \dot{u}_n,
   \dot{\tmmathbf{u}}_t, \dot{\alpha}, \dot{\tmmathbf{\gamma}}_t) + \varphi
   (\dot{\alpha}, \dot{\tmmathbf{\gamma}_t} ; \alpha), \]
together with the free-energy density $\psi$ of~\eqref{new-free-energy}
and the dissipation potential~\eqref{eq:new-dissipation-potential}-\eqref{eq:expr-dissipation}, is consistent with the evolution
laws~\eqref{eq:Tresca-reformulation} and~\eqref{new-strong-form}$_6$. The expression of the derivative
\[ \dot{\psi} (u_n, \tmmathbf{u}_t, \alpha, \tmmathbf{\gamma}_t, \dot{u}_n,
   \dot{\tmmathbf{u}}_t, \dot{\alpha}, \dot{\tmmathbf{\gamma}}_t) =
   \frac{\partial \psi (u_n, \tmmathbf{u}_t, \alpha,
   \tmmathbf{\gamma}_t)}{\partial u_n}  \dot{u}_n + \frac{\partial \psi
   (\ldots)}{\partial \tmmathbf{u}_t} \cdot \dot{\tmmathbf{u}}_t + \frac{\partial
   \psi (\ldots)}{\partial \alpha}  \dot{\alpha} + \frac{\partial \psi
   (\ldots)}{\partial \tmmathbf{\gamma}_t} \cdot \dot{\tmmathbf{\gamma}}_t, \]
and the decomposition of the dissipation potential $\varphi (\dot{\alpha},
\dot{\tmmathbf{\gamma}_t} ; \alpha) = \varphi_p (\dot{\alpha} ; \alpha) +
\varphi_f (\dot{\tmmathbf{\gamma}_t} ; \alpha)$
of~\eqref{eq:new-dissipation-potential} show that the minimization over
$\dot{\alpha}$ and $\dot{\tmmathbf{\gamma}}_t$ can be
performed independently,
\[ \begin{array}{l}
     \underset{\dot{\alpha}}{\text{Min}} \enspace \frac{\partial
     \psi (u_n, \tmmathbf{u}_t, \alpha, \tmmathbf{\gamma}_t)}{\partial \alpha}
     \dot{\alpha} + \varphi_p (\dot{\alpha} ; \alpha),\\
      \underset{\dot{\tmmathbf{\gamma}}_t}{\text{Min}} \enspace
     \frac{\partial \psi (u_n, \tmmathbf{u}_t, \alpha,
     \tmmathbf{\gamma}_t)}{\partial \tmmathbf{\gamma}_t}  \cdot \dot{\tmmathbf{\gamma}}_t +
     \varphi_f (\dot{\tmmathbf{\gamma}_t} ; \alpha) .
   \end{array} \]
The first terms are linear in $\dot{\alpha}$ and $\dot{\tmmathbf{\gamma}}_t$, respectively, while $\varphi_p
(\dot{\alpha} ; \alpha)$ and $\varphi_f (\dot{\tmmathbf{\gamma}_t} ; \alpha)$ are convex, proper and lower semicontinuous functions of $\dot{\alpha}$ and $\dot{\tmmathbf{\gamma}}_t$, respectively, 
see~\eqref{eq:expr-dissipation}$_1$-\eqref{eq:indicator-function}; being positively one-homogeneous, they are not strictly convex. Each of the above problems is therefore a convex minimization problem, and by the generalized Fermat rule of convex analysis $\dot{\alpha}$, respectively $\dot{\tmmathbf{\gamma}}_t$, is a minimizer 
\textit{if and only if} the following first-order stationarity condition holds: for a \textit{given} $\alpha$,
\begin{equation}
    - \frac{\partial \psi (u_n, \tmmathbf{u}_t, \alpha,
    \tmmathbf{\gamma}_t)}{\partial \alpha} \in \partial \varphi_p
    (\dot{\alpha} ; \alpha), \hspace{4em} 
    - \frac{\partial \psi (u_n, \tmmathbf{u}_t, \alpha,
    \tmmathbf{\gamma}_t)}{\partial \tmmathbf{\gamma}_t} \in \partial \varphi_f
    (\dot{\tmmathbf{\gamma}}_t ; \alpha),
  \label{eq:first-stationrity}
\end{equation}
where $\partial \varphi_p$, $\partial \varphi_f$ are subdifferentials.

For the plastic potential $\varphi_p (\dot{\alpha} ; \alpha) = \eta \sigma_0
\alpha \dot{\alpha} + I_{\mathbb{R}^+} (\dot{\alpha})$
in~\eqref{eq:expr-dissipation}$_1$, the subdifferential is the set
of $g \in \mathbb{R}$ such that
\[ \eta \sigma_0 \alpha \dot{\alpha}^{\dag} + I_{\mathbb{R}^+}
   (\dot{\alpha}^{\dag}) - \eta \sigma_0 \alpha \dot{\alpha} -
   I_{\mathbb{R}^+} (\dot{\alpha}) \geqslant g (\dot{\alpha}^{\dag} -
   \dot{\alpha}) \quad \forall \dot{\alpha}^{\dag} \in \mathbb{R}. \]
The definition of the indicator function $I_{\mathbb{R}^+}$ in~\eqref{eq:indicator-function} imposes that there is no solution for $\dot{\alpha} < 0$, in which case the subdifferential is empty. Besides, when~$\dot{\alpha} \geqslant 0$, the inequality holds trivially for $\dot{\alpha}^{\dag}
< 0$, where its left-hand side is $+ \infty$, so that it needs to be imposed only for $\dot{\alpha}^{\dag} \geqslant 0$ and simplifies to: for a given $\dot{\alpha}
\geqslant 0$,
\[\forall \dot{\alpha}^{\dag} \geqslant 0, \hspace{4em} (g - \eta \sigma_0 \alpha)  (\dot{\alpha}^{\dag} -
   \dot{\alpha}) \leqslant 0. \]
When $\dot{\alpha} > 0$, we can always find two scalars
$\dot{\alpha}^{\dag}_1$, $\dot{\alpha}^{\dag}_2$ such that $\dot{\alpha} >
\dot{\alpha}^{\dag}_1 \geqslant 0$ and $\dot{\alpha}^{\dag}_2 > \dot{\alpha} >
0$, which forces $g = \eta \sigma_0 \alpha$. When $\dot{\alpha} = 0$, we have
$\dot{\alpha}^{\dag} - \dot{\alpha} \geqslant 0$ and the inequality is
satisfied for any $g \leqslant \eta \sigma_0 \alpha$. Collecting the different
cases, we obtain
\[ \partial \varphi_p (\dot{\alpha} ; \alpha) = \left\{\begin{array}{ll}
     \emptyset & \text{if } \dot{\alpha} < 0,\\
     \eta \sigma_0 \alpha & \text{if } \dot{\alpha} > 0,\\
     (- \infty, \eta \sigma_0 \alpha] & \text{if } \dot{\alpha} = 0.
   \end{array}\right. \]
Combining this with the stationarity condition~\eqref{eq:first-stationrity}$_1$ and with the expression of the dissipative force conjugate to $\alpha$, $- \partial \psi / \partial \alpha = \eta P_n$ in~\eqref{eq:irreversible-forces}$_1$, we finally obtain the plastic evolution law anticipated in the first line of~\eqref{new-strong-form}$_{6}$,
\begin{equation}
    \forall (t, \tmmathbf{x}) \in \mathbb{R}^+ \times \Gamma_c, \hspace{3.5em}
   \eta P_n \leqslant \eta \sigma_0 \alpha, \hspace{3em} \dot{\alpha}
   \geqslant 0, \hspace{3em} \eta (P_n - \sigma_0 \alpha)  \dot{\alpha} = 0.
   \label{eq:evolution-law-linear-model}
\end{equation} 
Following similar arguments, the subdifferential of the friction potential $
\varphi_f (\dot{\tmmathbf{\gamma}_t} ; \alpha) = \sigma_0 \mu \alpha \|
\dot{\tmmathbf{\gamma}}_t \|$ of~\eqref{eq:expr-dissipation}$_2$ can be
expressed as
\[ \partial \varphi_f (\dot{\tmmathbf{\gamma}}_t ; \alpha) =
   \left\{\begin{array}{ll}
     \sigma_0 \mu \alpha \frac{\dot{\tmmathbf{\gamma}}_t}{\|
     \dot{\tmmathbf{\gamma}}_t \|} & \text{if } \| \dot{\tmmathbf{\gamma}}_t \|
     > 0,\\
     \left\{ \tmmathbf{\tau}, \enspace \| \tmmathbf{\tau} \| \leqslant
     \sigma_0 \mu \alpha \right\} & \text{if } \| \dot{\tmmathbf{\gamma}}_t \|
     = 0.
   \end{array}\right. \]
Combining this with the stationarity condition \eqref{eq:first-stationrity}$_2$ and with the expression
of the dissipative force conjugate to $\tmmathbf{\gamma}_t$,
$- \partial \psi / \partial \tmmathbf{\gamma}_t = \tmmathbf{T}_t$ in \eqref{eq:irreversible-forces}$_2$, we
obtain $\tmmathbf{T}_t \in \partial \varphi_f ( \dot{\tmmathbf{\gamma}}_t ; \alpha )$,
that is, either $\dot{\tmmathbf{\gamma}}_t \neq \tmmathbf{0}$ and
$\tmmathbf{T}_t = \sigma_0 \mu \alpha \, \dot{\tmmathbf{\gamma}}_t / \| \dot{\tmmathbf{\gamma}}_t \|$,
or $\dot{\tmmathbf{\gamma}}_t = \tmmathbf{0}$ and
$\| \tmmathbf{T}_t \| \leqslant \sigma_0 \mu \alpha$. In both cases
$\| \tmmathbf{T}_t \| \leqslant \mu \sigma_0 \alpha$ and
$\tmmathbf{T}_t \cdot \dot{\tmmathbf{\gamma}}_t = \mu \sigma_0 \alpha \, \| \dot{\tmmathbf{\gamma}}_t \|$;
conversely, these two conditions imply
$\tmmathbf{T}_t \in \partial \varphi_f ( \dot{\tmmathbf{\gamma}}_t ; \alpha )$ by
the Cauchy--Schwarz inequality (this is the Fenchel characterization of the
subdifferential of a support function, see \cite[Theorem~23.5]{rockafellar1970}).
We thus recover the Tresca friction law \eqref{eq:Tresca-reformulation} exactly.
   
\subsection{New model (logarithmic version)}\label{appendix:log-model}
In this section we aim to recover the evolution
laws~\eqref{eq:Tresca-reformulation} and~\eqref{new-strong-form-LOG}$_{6}$ (hence~\eqref{new-strong-form-LOG}$_{7}$ in the limit $\varepsilon_t \rightarrow +\infty$), starting from the continuous minimization principle~\eqref{eq:new-max-dissipation-principle},
\[ \underset{\dot{\alpha},
   \dot{\tmmathbf{\gamma}}_t}{\text{Min}} \enspace \dot{\psi} (u_n,
   \tmmathbf{u}_t, \alpha, \tmmathbf{\gamma}_t, \dot{u}_n,
   \dot{\tmmathbf{u}}_t, \dot{\alpha}, \dot{\tmmathbf{\gamma}}_t) + \varphi
   (\dot{\alpha}, \dot{\tmmathbf{\gamma}_t} ; \alpha), \]
together with the free-energy density $\psi$ of~\eqref{new-free-energy-LOG} and the dissipation
potential~\eqref{eq:new-dissipation-potential},~\eqref{eq:expr-dissipation}$_2$,~\eqref{eq:plastic-potential-LOG}.
The proof is very similar to the one given in Section~\ref{appendix:linear-model}
for the linear model~\eqref{new-strong-form}-\eqref{eq:expr-dissipation}.

For the frictional behavior the proof is unchanged, since the tangential
elastic term of~\eqref{new-free-energy-LOG} and the friction dissipation
potential $\varphi_f$ of \eqref{eq:expr-dissipation} are the same in the two
models; the Tresca friction~\eqref{eq:Tresca-reformulation} follows immediately.

For the normal plastic behavior, the stationarity condition
~\eqref{eq:first-stationrity}$_1$ remains valid. By comparison with the
linear model~\eqref{new-strong-form}-\eqref{eq:expr-dissipation}, the two
terms entering this condition~\eqref{eq:first-stationrity}$_1$ are modified as
follows: the dissipative force conjugate to~$\alpha$ is divided by a factor
$\alpha$ ($- \partial \psi / \partial \alpha = \eta P_n /
\alpha$, see~\eqref{eq:irreversible-force-LOG}), while the first term of the
dissipation potential is divided by the same factor $\alpha$ ($\varphi_p
(\dot{\alpha}) = \eta \sigma_0  \dot{\alpha} + I_{\mathbb{R}^+}
(\dot{\alpha})$, see~\eqref{eq:plastic-potential-LOG}), and so is its
subdifferential. These factors are strictly positive, $\alpha\geqslant\alpha^{\star}>0$ (see~\eqref{eq:alpha-domain-LOG}) and  compensate each other. Consequently, the evolution law given by the logarithmic model coincides with that of the linear model~\eqref{eq:evolution-law-linear-model}, which is precisely the plastic law in the first line of~\eqref{new-strong-form-LOG}$_{6}$.

\section{Incremental
formulations and consistency with evolution laws}\label{appendix:incremental}

Section~\ref{appendix:incremental-linear-model} shows that the functional of the
global incremental problem~\eqref{eq:global-incremental} is strictly convex, so that \eqref{eq:global-incremental} has a unique solution, and
that the local incremental problem~\eqref{eq:local-incremental} is the
discrete analogue of the continuous minimization
principle~\eqref{eq:new-max-dissipation-principle}: its stationarity conditions are the plastic law~\eqref{new-strong-form}$_6$, satisfied exactly by the current real-area fraction, and the Tresca law~\eqref{eq:Tresca-reformulation} with the lagged threshold $\mu \sigma_0 \alpha^{(k - 1)}$, the discrete counterparts of the evolution laws derived from~\eqref{eq:new-max-dissipation-principle} in Appendix~\ref{appendix:dissipation-potentials}. Section~\ref{appendix:incremental-log-model} revisits this analysis for the logarithmic
model~\eqref{new-free-energy-LOG}-\eqref{eq:global-incremental-LOG}. Section~\ref{LM_proofs} shows that the Lagrange multiplier formulations~\eqref{eq:Lagrange} and~\eqref{eq:Lagrange-LOG} are consistent with the contact problems~\eqref{new-strong-form} and~\eqref{new-strong-form-LOG} and have a unique solution.

Throughout, $\Gamma_u$ has positive measure, so that the elastic potential is strictly convex and coercive on the set of admissible displacements (Section~\ref{sec:objective}). The incremental functionals  are all convex (as shown below), continuous and coercive, and the existence of a minimizer on the closed convex admissible set follows by standard arguments~\cite{duvaut1972}; we therefore focus on uniqueness and on the stationarity conditions.

\subsection{Linear model}\label{appendix:incremental-linear-model}

\paragraph{Convexity and uniqueness}We write the functional of the global incremental problem~\eqref{eq:global-incremental} as
\begin{equation}
  \Phi (u_n, \tmmathbf{u}_t, \alpha, \tmmathbf{\gamma}_t) = \int_{\Omega}
  \left( \frac{1}{2} \tmmathbf{\varepsilon} (\tmmathbf{u}) : \tmmathbf{c}:
  \tmmathbf{\varepsilon} (\tmmathbf{u}) -\tmmathbf{f} \cdot \tmmathbf{u}
  \right) \mathd \tmmathbf{x}- \int_{\Gamma_F} \tmmathbf{F} \cdot \tmmathbf{u}
  \,\mathd s + \int_{\Gamma_c} \phi^c \mathd s, \label{eq:total-nrj}
\end{equation}
with
\begin{equation}
  \phi^c (u_n, \tmmathbf{u}_t, \alpha, \tmmathbf{\gamma}_t) =
  \frac{\varepsilon_n}{2}  \langle u_n - g_0 - \eta \alpha \rangle^2_+ +
  \frac{\varepsilon_t}{2}  \| \tmmathbf{u}_t -\tmmathbf{\gamma}_t \|^2 +
  \frac{\eta \sigma_0 \alpha^2}{2} + \sigma_0 \mu \alpha^{(k - 1)}  \|
  \tmmathbf{\gamma}_t -\tmmathbf{\gamma}_t^{(k - 1)} \| .
  \label{eq:contact-nrj-density}
\end{equation}
Every term of \eqref{eq:total-nrj}--\eqref{eq:contact-nrj-density} is
convex: the elastic potential by \eqref{eq:bulk-NRJ_CVX}, the contact terms $\frac{\varepsilon_n}{2} \langle u_n - g_0 - \eta \alpha \rangle_+^2$ and $\frac{\varepsilon_t}{2} \| \tmmathbf{u}_t - \tmmathbf{\gamma}_t \|^2$ as convex functions of affine arguments, $\eta \sigma_0 \alpha^2 / 2$
trivially, and the friction term as the norm of an affine function of
$\tmmathbf{\gamma}_t$. Hence $\Phi$ is convex on the convex admissible set
\[
  \mathcal{A} = \bigl\{ ( \tmmathbf{u}, \alpha, \tmmathbf{\gamma}_t ) \in
  H^1 ( \Omega )^3 \times L^2 ( \Gamma_c ) \times L^2 ( \Gamma_c )^2 :
  \ \tmmathbf{u} = \tmmathbf{u}^\star \text{ on } \Gamma_u, \
  \alpha^{(k - 1)} \leqslant \alpha \leqslant 1 \text{ a.e. on } \Gamma_c
  \bigr\} .
\]
The functional is moreover strictly convex. Two distinct points of $\mathcal{A}$ differ in $\tmmathbf{u}$, in $\tmmathbf{\gamma}_t$ or in $\alpha$, and along the segment joining them one term of $\Phi$ is strictly convex: the elastic potential if they differ in $\tmmathbf{u}$ (Section~\ref{sec:objective}); otherwise the term $\frac{\varepsilon_t}{2} \| \tmmathbf{u}_t - \tmmathbf{\gamma}_t \|^2$, at fixed $\tmmathbf{u}_t$, if they differ in $\tmmathbf{\gamma}_t$; otherwise the term $\eta \sigma_0 \alpha^2 / 2$. Since the other terms are convex, $\Phi$ is strictly convex along the segment. Problem~\eqref{eq:global-incremental} therefore has a unique solution.


\paragraph{Consistency with local evolution laws}The same argument shows that the contact term~$\phi^c$ of the functional~\eqref{eq:total-nrj} is jointly strictly convex in $(\alpha, \tmmathbf{\gamma}_t)$. By construction, this term coincides up to the constant $-\eta \sigma_0\left(\alpha^{(k-1)}\right)^2/2$ with the functional entering the local incremental formulation~\eqref{eq:local-incremental},
see~\eqref{new-free-energy},~\eqref{eq:expr-dissipation}.

Convex analysis thus implies that the local incremental minimization under constraint~\eqref{eq:local-incremental} is fully equivalent to the local first-order stationarity conditions
\begin{equation}
  \forall \tmmathbf{x} \in \Gamma_c, \qquad \frac{\partial \phi^c (u^{(k)}_n,
  \tmmathbf{u}^{(k)}_t, \alpha, \tmmathbf{\gamma}_t)}{\partial \alpha} +
  \tilde{\lambda}_2 - \tilde{\lambda}_1 = 0, \hspace{4em} \tmmathbf{0}\in\partial_{\tmmathbf{\gamma}_t}
  \phi^c (u^{(k)}_n, \tmmathbf{u}^{(k)}_t, \alpha,
  \tmmathbf{\gamma}_t),
  \label{1st-order-stationarity}
\end{equation}
together with the Karush-Kuhn-Tucker conditions
\begin{equation}
  \begin{array}{lll}
    \tilde{\lambda}_1 \geqslant 0, & \qquad \alpha^{(k - 1)} \leqslant \alpha,
    & \qquad \tilde{\lambda}_1  (\alpha^{(k - 1)} - \alpha) = 0,\\
    \tilde{\lambda}_2 \geqslant 0, & \qquad \alpha \leqslant 1, & \qquad
    \tilde{\lambda}_2  (\alpha - 1) = 0.
  \end{array} \label{eq:KKT1}
\end{equation}
The above equations make use of Lagrange
multipliers~$\tilde{\lambda}_1$,~$\tilde{\lambda}_2$ to account for the
constraints $\alpha^{(k - 1)} \leqslant \alpha \leqslant 1$.

Using the expression of $\phi^c$ in~\eqref{eq:contact-nrj-density},
that of the contact tractions in~\eqref{eq:new-contact-stresses} and the
definition of the subdifferential, these conditions read explicitly
\begin{equation}
  \forall \tmmathbf{x} \in \Gamma_c, \qquad \eta P_n = \eta \sigma_0 \alpha +
  \tilde{\lambda}_2 - \tilde{\lambda}_1, \hspace{2em} \tmmathbf{T}_t \in
  \left\{\begin{array}{ll}
    \sigma_0 \mu \alpha^{(k - 1)}  \frac{\tmmathbf{\gamma}_t
    -\tmmathbf{\gamma}_t^{(k - 1)}}{\| \tmmathbf{\gamma}_t
    -\tmmathbf{\gamma}_t^{(k - 1)} \|} & \text{if } \| \tmmathbf{\gamma}_t
    -\tmmathbf{\gamma}_t^{(k - 1)} \| > 0,\\
    \left\{ \tmmathbf{\tau}, \enspace \| \tmmathbf{\tau} \| \leqslant \sigma_0
    \mu \alpha^{(k - 1)} \right\} & \text{if } \| \tmmathbf{\gamma}_t
    -\tmmathbf{\gamma}_t^{(k - 1)} \| = 0.
  \end{array}\right. \label{eq:discrete-tangential-evolution}
\end{equation}
By the argument at the end of Appendix~\ref{appendix:linear-model}, the second condition in \eqref{eq:discrete-tangential-evolution} is
equivalent to: $\forall \tmmathbf{x} \in \Gamma_c$,
\begin{equation}
\| \tmmathbf{T}_t \| \leqslant \mu \sigma_0 \alpha^{(k - 1)}, \qquad
\tmmathbf{T}_t \cdot \bigl( \tmmathbf{\gamma}_t - \tmmathbf{\gamma}_t^{(k - 1)} \bigr)
= \mu \sigma_0 \alpha^{(k - 1)} \, \bigl\| \tmmathbf{\gamma}_t - \tmmathbf{\gamma}_t^{(k - 1)} \bigr\| ,   \label{eq:Tresca-lagged}  
\end{equation}
which is the discrete analogue of the Tresca friction law~\eqref{eq:Tresca-reformulation} at time $t_k$.

The evolution of $\alpha$ is dictated by~\eqref{eq:discrete-tangential-evolution}$_1$ and~\eqref{eq:KKT1}:
\begin{itemizedot}
  \item when $P_n < \sigma_0 \alpha$, we must have $\tilde{\lambda}_1 >
  \tilde{\lambda}_2 \geqslant 0$ and $\alpha = \alpha^{(k - 1)}$;
  
  \item when $P_n = \sigma_0 \alpha$, we have $\tilde{\lambda}_1 = \tilde{\lambda}_2$, which yields $\left\{\begin{array}{l}
    \alpha^{(k - 1)} \leqslant \alpha \leqslant 1 \text{ \quad if } \tilde{\lambda}_1 = \tilde{\lambda}_2 = 0,\\
    \alpha^{(k - 1)} = \alpha = 1 \text{ \quad if } \tilde{\lambda}_1 = \tilde{\lambda}_2 > 0;
  \end{array}\right. $
  
  \item when $P_n > \sigma_0 \alpha$, we must have $\tilde{\lambda}_2 >
  \tilde{\lambda}_1 \geqslant 0$ and $\alpha = 1$.
\end{itemizedot}
This behavior is a discrete equivalent of the plastic law~\eqref{new-strong-form}$_6$.

As announced above Eq. \eqref{eq:global-incremental}, the plastic threshold is formulated in terms of the current
area fraction~$\alpha$, while the friction threshold
in~\eqref{eq:discrete-tangential-evolution} makes use of the lagged
value~$\alpha^{(k - 1)}$. Because~$\alpha^{(k - 1)}$ is \textit{known},
the tangential dissipation $\sigma_0 \mu \alpha^{(k - 1)}  \|
\tmmathbf{\gamma}_t -\tmmathbf{\gamma}_t^{(k - 1)} \|$ does not appear in the
expression of $\partial \phi^c / \partial \alpha$
in~\eqref{1st-order-stationarity} and the normal and tangential behaviors are
\textit{fully decoupled}. 
With the current $\alpha$ in the friction threshold instead, the
stationarity condition \eqref{eq:discrete-tangential-evolution}$_1$ would become
$\eta P_n = \eta \sigma_0 \alpha
+ \sigma_0 \mu \| \tmmathbf{\gamma}_t - \tmmathbf{\gamma}_t^{(k - 1)} \|
+ \tilde{\lambda}_2 - \tilde{\lambda}_1$: the slip within the step would
raise the crushing threshold by a term of order $\Delta t$ with no
counterpart in \eqref{new-strong-form}$_6$, and the incremental functional would lose its
convexity in $( \alpha, \tmmathbf{\gamma}_t )$, since
$( \alpha, \tmmathbf{\gamma}_t ) \mapsto
\alpha \| \tmmathbf{\gamma}_t - \tmmathbf{\gamma}_t^{(k - 1)} \|$ is not
jointly convex, 
see Section~\ref{sec:incremental-energy}.

\subsection{Logarithmic model}\label{appendix:incremental-log-model}

\paragraph{Convexity and uniqueness}
As in Appendix~\ref{appendix:incremental-linear-model}, every term of \eqref{eq:global-incremental-LOG} is convex, the normal contact term as the composition of the convex non-decreasing function $\langle \cdot \rangle_+^2$ with the convex function $( u_n, \alpha ) \mapsto u_n - g_0 - \eta \ln ( \alpha / \alpha^\star )$, and the functional is convex on $\mathcal{A}$. Unlike its counterpart \eqref{eq:global-incremental}, however, it is \emph{not} strictly convex: where the gap is open, $u_n - g_0 - \eta \ln ( \alpha / \alpha^\star ) < 0$, its dependence on $\alpha$ reduces to the affine term $\eta \sigma_0 ( \alpha - \alpha^{(k - 1)} )$. The uniqueness of the minimizer is nevertheless obtained in two steps.

The elastic potential is strictly convex in $\tmmathbf{u}$ (Section~\ref{sec:objective}) and the term $\frac{\varepsilon_t}{2} \| \tmmathbf{u}_t - \tmmathbf{\gamma}_t \|^2$ is strictly convex in $\tmmathbf{\gamma}_t$ at fixed $\tmmathbf{u}_t$; hence all minimizers of \eqref{eq:global-incremental-LOG} share the same displacement $\tmmathbf{u}^{(k)}$ and the same slip $\tmmathbf{\gamma}_t^{(k)}$ (if there were, e.g., two minimizers with different displacements, their midpoint would be admissible and, by convexity of all terms and strict convexity of the elastic potential, the functional would take a strictly lower value there, which is a contradiction). For this displacement, $\alpha$ minimizes pointwise on $\Gamma_c$ the function
\[
  f ( \alpha ) = \frac{\varepsilon_n}{2} \Bigl\langle u_n^{(k)} - g_0 - \eta \ln \frac{\alpha}{\alpha^\star} \Bigr\rangle_+^2 + \eta \sigma_0 \alpha, \qquad \alpha^{(k - 1)} \leqslant \alpha \leqslant 1,
\]
whose derivative is $f' ( \alpha ) = \eta ( \sigma_0 \alpha - P_n ) / \alpha$, with $P_n = \varepsilon_n \langle u_n^{(k)} - g_0 - \eta \ln ( \alpha / \alpha^\star ) \rangle_+$ the contact pressure~\eqref{eq:new-pressure_LOG}. In the contact range $\alpha < \alpha_c := \alpha^\star \mathe^{( u_n^{(k)} - g_0 ) / \eta}$, $f$ is strictly convex, since there $f'' ( \alpha ) = \eta ( \varepsilon_n \eta + P_n ) / \alpha^2 > 0$, while for $\alpha \geqslant \alpha_c$ it reduces to $\eta \sigma_0 \alpha$, which is strictly increasing. A convex function on an interval has a unique minimizer unless it is constant on a subinterval, which $f$ is not; hence $\alpha^{(k)}$ is unique. Explicitly, $\alpha^{(k)} = \alpha^{(k - 1)}$ if $P_n \leqslant \sigma_0 \alpha$ at $\alpha = \alpha^{(k - 1)}$, $\alpha^{(k)} = 1$ if $P_n > \sigma_0$ at $\alpha = 1$, and otherwise $\alpha^{(k)}$ is the unique root of $P_n = \sigma_0 \alpha$ in $( \alpha^{(k - 1)}, 1 )$: the three cases of the return mapping in Algorithm~\ref{algo:local_update}.

Problem \eqref{eq:global-incremental-LOG} therefore admits a unique solution although its functional is only convex. The same situation, and the same pointwise argument in $\alpha$, occur for the Lagrange multiplier formulation \eqref{eq:Lagrange-LOG}, see Appendix~\ref{LM_proofs}.

\paragraph{Consistency with local evolution laws}
For fixed $\tmmathbf{u}^{(k)}$, the minimization of \eqref{eq:global-incremental-LOG} with respect to $( \alpha, \tmmathbf{\gamma}_t )$ is carried out pointwise on $\Gamma_c$, and its first-order stationarity conditions take the form \eqref{1st-order-stationarity}--\eqref{eq:KKT1}, with $\phi^c$ replaced by the integrand of the contact integral in \eqref{eq:global-incremental-LOG} and with multipliers $\hat{\lambda}_1$, $\hat{\lambda}_2$ for the constraints $\alpha^{(k - 1)} \leqslant \alpha \leqslant 1$. The terms depending on $\tmmathbf{\gamma}_t$ are the same as in the linear model, so the second condition yields the evolution law~\eqref{eq:discrete-tangential-evolution}$_2$, the discrete analogue of~\eqref{eq:Tresca-reformulation}. The first condition reads
\begin{equation*}
    \forall \tmmathbf{x} \in \Gamma_c, \hspace{3em} \eta \dfrac{P_n}{\alpha} = \eta \sigma_0  +
  \hat{\lambda}_2 - \hat{\lambda}_1.
\end{equation*}
Multiplied by $\alpha$, which is strictly positive by~\eqref{eq:alpha-domain-LOG}, and with $\tilde{\lambda}_j = \alpha \hat{\lambda}_j$ $( j = 1, 2 )$, it becomes the condition~\eqref{eq:discrete-tangential-evolution}$_1$ of the linear model, and $\tilde{\lambda}_1$, $\tilde{\lambda}_2$ satisfy the same Karush-Kuhn-Tucker conditions~\eqref{eq:KKT1} as $\hat{\lambda}_1$, $\hat{\lambda}_2$. The analysis of the linear model (bullet points below~\eqref{eq:Tresca-lagged}) therefore applies, and the incremental formulation~\eqref{eq:global-incremental-LOG} yields a discrete plastic law consistent with the continuous one in~\eqref{new-strong-form-LOG}$_6$.

\subsection{Lagrange multiplier formulation} \label{LM_proofs}
It remains to be shown that the solutions of~\eqref{eq:Lagrange}
and~\eqref{eq:Lagrange-LOG} are consistent with the contact problems~\eqref{new-strong-form} and~\eqref{new-strong-form-LOG} respectively, in discrete
form. We give the proof below for the logarithmic variant~\eqref{eq:Lagrange-LOG}, establishing uniqueness along the way; the proof extends naturally to the linear
model~\eqref{new-strong-form},~\eqref{eq:Lagrange}, and that extension is left to the reader.

The Lagrange multipliers introduced below are understood in the discrete setting, where the constraints of~\eqref{eq:Lagrange-LOG} are finitely many inequalities on the nodal values and multipliers exist~\cite{rockafellar1970}.

To prove the consistency between~\eqref{eq:Lagrange-LOG} and~\eqref{new-strong-form-LOG}, we introduce Lagrange multipliers $p_n$, $\tmmathbf{t}_t$, $\lambda_1$, $\lambda_2$ and consider the Lagrangian
\begin{equation}
  \begin{array}{lll}
    \mathcal{L} (\tmmathbf{u}, \alpha, p_n, \tmmathbf{t}_{t,} \lambda_1,
    \lambda_2) & = & \int_{\Omega} \left( \frac{1}{2} \tmmathbf{\varepsilon}
    (\tmmathbf{u}) : \tmmathbf{c}: \tmmathbf{\varepsilon} (\tmmathbf{u})
    -\tmmathbf{f} \cdot \tmmathbf{u} \right) \mathd \tmmathbf{x}-
    \int_{\Gamma_F} \tmmathbf{F} \cdot \tmmathbf{u} \mathd s + \int_{\Gamma_c}
    \eta \sigma_0  (\alpha - \alpha^{(k - 1)}) \mathd s\\
    & &\quad  + \int_{\Gamma_c} (\tmmathbf{t}_t \cdot (\tmmathbf{u}_t
    -\tmmathbf{u}_t^{(k - 1)}) + p_n  (u_n - g_0 - \eta \ln (\alpha /
    \alpha^{\star})) ) \mathd s \\
    &  & \qquad  + \int_{\Gamma_c} ( \lambda_1  (\alpha^{(k - 1)} - \alpha) + \lambda_2 
    (\alpha - 1)) \mathd s.
  \end{array} \label{eq:expr-lagrangian}
\end{equation}
The constrained minimization~\eqref{eq:Lagrange-LOG} is then equivalent to minimizing this Lagrangian with respect to~$(\tmmathbf{u}, \alpha)$,
\begin{equation}
  \underset{\tmmathbf{u}=\tmmathbf{u}^{\star} \text{ on }
  \Gamma_u}{\underset{\tmmathbf{u}, \alpha}{\text{Min}}}
  \mathcal{L} (\tmmathbf{u}, \alpha, p_n, \tmmathbf{t}_{t,} \lambda_1,
  \lambda_2), \label{eq:min-augmented-energy}
\end{equation}
while the Lagrange multipliers satisfy the optimality conditions
\begin{equation}
  \begin{array}{lll}
    p_n \geqslant 0, & \qquad u_n - g_0 - \eta \ln (\alpha / \alpha^{\star})
    \leqslant 0, & \qquad p_n  (u_n - g_0 - \eta \ln (\alpha /
    \alpha^{\star})) = 0,\\
    \lambda_1 \geqslant 0, & \qquad \alpha^{(k - 1)} \leqslant \alpha, &
    \qquad \lambda_1  (\alpha^{(k - 1)} - \alpha) = 0,\\
    \lambda_2 \geqslant 0, & \qquad \alpha \leqslant 1, & \qquad \lambda_2 
    (\alpha - 1) = 0,
  \end{array} \label{eq:KKT}
\end{equation}
and $\tmmathbf{t}_t \in \sigma_0 \mu \alpha^{(k-1)}\partial\left\Vert \cdot \right\Vert(\tmmathbf{u}_t-\tmmathbf{u}_t^{(k-1)})$ (see Appendix~\ref{appendix:linear-model}), that is,
\begin{equation}
  \tmmathbf{t}_t \ \left\{\begin{array}{ll}
    =\sigma_0 \mu \alpha^{(k - 1)}  \frac{\tmmathbf{u}_t -\tmmathbf{u}_t^{(k -
    1)}}{\| \tmmathbf{u}_t -\tmmathbf{u}_t^{(k - 1)} \|} & \text{if } \|
    \tmmathbf{u}_t -\tmmathbf{u}_t^{(k - 1)} \| > 0,\\
    \in \left\{ \tmmathbf{\tau}, \enspace \| \tmmathbf{\tau} \| \leqslant \sigma_0
    \mu \alpha^{(k - 1)} \right\} & \text{if } \| \tmmathbf{u}_t
    -\tmmathbf{u}_t^{(k - 1)} \| = 0.
  \end{array}\right. \label{tresca-reformulated}
\end{equation}
The convex minimization problem~\eqref{eq:min-augmented-energy} is equivalent to the first-order stationarity condition, obtained by taking the first variation of $\mathcal{L}$ with respect to $( \tmmathbf{u}, \alpha )$,
\begin{equation}
  \begin{array}{llll}
    \forall \delta \tmmathbf{u} \in \mathcal{C}_0, \enspace \forall \delta
    \alpha, \qquad & 0 & = & \int_{\Omega} \left(\tmmathbf{\varepsilon} (\tmmathbf{u}) : \tmmathbf{c}:
    \tmmathbf{\varepsilon} (\delta \tmmathbf{u}) -\tmmathbf{f} \cdot\delta
    \tmmathbf{u} \right) \mathd \tmmathbf{x}- \int_{\Gamma_F} \tmmathbf{F}
    \cdot \delta \tmmathbf{u} \mathd s\\
    &  &  & \quad + \int_{\Gamma_c} ((\eta \sigma_0 - \eta p_n / \alpha -
    \lambda_1 + \lambda_2) \delta \alpha +\tmmathbf{t}_t \cdot \delta
    \tmmathbf{u}_t + p_n \delta u_n) \mathd s
  \end{array} \label{eq:lagrangian-stationarity}
\end{equation}
where $\mathcal{C}_0$ is the set of kinematically admissible perturbations defined in~\eqref{eq:kinematic-admissible-set}.

As is standard in elasticity, we integrate the first term
of~\eqref{eq:lagrangian-stationarity} by parts and then apply the fundamental lemma of the calculus of variations. This yields the equilibrium equation~\eqref{new-strong-form-LOG}$_1$, the Neumann and Dirichlet boundary conditions~\eqref{new-strong-form-LOG}$_{2, 3}$, and the following set of conditions on the contact boundary,
\begin{equation}
  \begin{array}{ll}
    \forall \tmmathbf{x} \in \Gamma_c, \hspace{3em} & p_n = - \sigma_{n
    \nocomma n},\\
    & \tmmathbf{t}_t = -\tmmathbf{\sigma}_{n \nocomma t},\\
    & \eta (\sigma_0 - p_n / \alpha) - \lambda_1 + \lambda_2 = 0.
  \end{array} \label{stationarity-augmented-energy}
\end{equation}
The first two lines show that the Lagrange multipliers $p_n$ and
$\tmmathbf{t}_t$ play the same role as the contact tractions $P_n$,
$\tmmathbf{T}_t$ in the penalty formulation,
see~\eqref{new-strong-form-LOG}$_4$, so that Eq.~\eqref{eq:KKT}$_1$ corresponds to the normal contact condition~\eqref{new-strong-form-LOG}$_5$. The tangential stress $\tmmathbf{t}_t$ satisfies the threshold law \eqref{tresca-reformulated}, which is the discrete counterpart of the Tresca friction \eqref{new-strong-form-LOG}$_7$. Finally, Eqs.~\eqref{stationarity-augmented-energy}$_3$ and~\eqref{eq:KKT}$_{2, 3}$, upon the rescaling $\tilde\lambda_j=\alpha\lambda_j$ of Appendix~\ref{appendix:incremental-log-model}, are identical to~\eqref{eq:discrete-tangential-evolution}$_1$ and~\eqref{eq:KKT1}
in Appendix~\ref{appendix:incremental-linear-model}, and the bullet points at the end of Section~\ref{appendix:incremental-linear-model} show that these conditions are the discrete 
counterpart of the plastic law \eqref{new-strong-form-LOG}$_6$. This proves that the solution of the incremental problem~\eqref{eq:Lagrange-LOG} satisfies
all the constraints of~\eqref{new-strong-form-LOG}.

We further show that this solution is unique. The functional in~\eqref{eq:Lagrange-LOG} is strictly convex in $\tmmathbf{u}$ on the convex admissible set (Section~\ref{sec:objective}), so that all solutions share the same displacement $\tmmathbf{u}^{(k)}$. For this displacement, the penetration constraint reads $\alpha \geqslant \alpha^{\star} \mathe^{(u_n^{(k)} - g_0) / \eta}$ pointwise on $\Gamma_c$, while the only $\alpha$-dependent term of~\eqref{eq:Lagrange-LOG}, $\eta \sigma_0 (\alpha - \alpha^{(k - 1)})$, is strictly increasing; hence
\begin{equation*}
\alpha^{(k)} = \max \bigl( \alpha^{(k - 1)}, \, \alpha^{\star} \mathe^{(u_n^{(k)} - g_0) / \eta}
\bigr) \qquad \text{on } \Gamma_c,
\end{equation*}
which is unique. In the Lagrange multiplier formulation the real-area fraction thus follows the exponential law~\eqref{eq:exp-distrib-Pn-alpha}$_1$ exactly wherever it grows, the plastic law~\eqref{stationarity-augmented-energy}$_3$ being carried by the multiplier $p_n$. The same argument gives $\alpha^{(k)} = \max ( \alpha^{(k - 1)}, (u_n^{(k)} - g_0) / \eta )$ for the linear variant~\eqref{eq:Lagrange}, since $\eta \sigma_0 \alpha^2 / 2$ is increasing for $\alpha \geqslant 0$.

\section{Contact contribution to the stiffness matrix}\label{appendix:contact-stiffness-matrix}

In this appendix we derive the expression of the contact stiffness
$\tmmathbf{K}^{\left( k, \hspace{0.2em} p \right)}_N$ for the logarithmic variant
of the new model~\eqref{eq:global-incremental-LOG} and for the classical
implementation of Coulomb friction~\eqref{strong-form} (as presented
in~{\cite{laursen2003}}). This operator allows us to evaluate the stiffness
contribution of a node $N$ located on the boundary of the domain,
see~\eqref{eq:contact-stiffness-contribution}. It collects the work of the
pressure increment $\Delta P^{\left( k, \hspace{0.2em} p \right)}_n$, with
$P_n^{\left( k, \hspace{0.2em} p \right)}$ given
in~\eqref{eq:expr-of-pi}--\eqref{eq:numerical-contact-pressure}, and that of
the tangential traction $\Delta T^{\left( k, \hspace{0.2em} p \right)}_t$, \
with $T^{\left( k, \hspace{0.2em} p \right)}_t$ given
in~\eqref{eq:numerical-tangential-stresses} for stick and slip nodes
($T^{\left( k, \hspace{0.2em} p \right)}_t = 0$~otherwise).

\subsection{New model, logarithmic variant}

For nodes in active contact, Eq.~\eqref{eq:numerical-contact-pressure} gives
$P_n^{\left( k, \hspace{0.2em} p \right)} = \pi_n \left(
\alpha^{\left( k, \hspace{0.2em} p \right)} \right) > 0$ and
by~\eqref{eq:expr-of-pi}$_3$ we have
\begin{equation}
  \Delta P^{\left( k, \hspace{0.2em} p \right)}_n = \varepsilon_n  \left(
  \Delta u^{\left( k, \hspace{0.2em} p \right)}_n - \eta \frac{\Delta
  \alpha^{\left( k, \hspace{0.2em} p \right)}}{\alpha^{\left( k,
  \hspace{0.2em} p \right)}} \right) . \label{linearization-pressure}
\end{equation}
The real-area fraction $\alpha^{\left( k, \hspace{0.2em} p \right)}$ is
determined by the return-mapping procedure described in Algorithm~\ref{algo:local_update}. It is
constant for the trial state ($\alpha^{\left( k, \hspace{0.2em} p \right)} =
\alpha^{(k - 1)}$), for the cap regime ($\alpha^{\left( k, \hspace{0.2em} p
\right)} = 1$), and for nodes that are not in contact ($\alpha^{\left(
k, \hspace{0.2em} p \right)} = \alpha^{(k - 1)}$); in these cases we thus have
$\Delta \alpha^{\left( k, \hspace{0.2em} p \right)} = 0$. In the remaining regime
of active plastic flow, we have $P_n^{\left( k, \hspace{0.2em} p \right)} =
\sigma_0 \alpha^{\left( k, \hspace{0.2em} p \right)}$, whose linearization
reads $\Delta P_n^{\left( k, \hspace{0.2em} p \right)} = \sigma_0 \Delta
\alpha^{\left( k, \hspace{0.2em} p \right)}$. Combining this
with~\eqref{linearization-pressure}, we obtain $\Delta \alpha^{\left( k,
\hspace{0.2em} p \right)} = \varepsilon_n \alpha^{\left( k, \hspace{0.2em} p \right)} \Delta u^{\left( k, \hspace{0.2em} p
\right)}_n / \left( \sigma_0 \alpha^{\left( k, \hspace{0.2em} p \right)} +
\varepsilon_n \eta \right)$ and
\[ \Delta P^{\left( k, \hspace{0.2em} p \right)}_n = \frac{\varepsilon_n
   \sigma_0 \alpha^{\left( k, \hspace{0.2em} p \right)}}{\sigma_0
   \alpha^{\left( k, \hspace{0.2em} p \right)} + \varepsilon_n \eta} \Delta
   u^{\left( k, \hspace{0.2em} p \right)}_n, \]
for nodes in active plastic flow.

Only nodes that are in contact at both the current and the previous time steps may
have a non-zero tangential contribution; for the others $T_t^{\left( k,
\hspace{0.2em} p \right)} = 0$, which implies $\Delta T_t^{\left( k,
\hspace{0.2em} p \right)} = 0$, see Algorithm~\ref{algo:local_update}. For these nodes the contribution depends on their state: stick or slip. In stick, the tangential
stress is given by $T^{\text{stick}}$ in~\eqref{eq:numerical-tangential-stresses}$_1$ and $\Delta T_t^{\left( k, \hspace{0.2em} p \right)} = \Delta T^{\text{stick}} = \varepsilon_t \Delta u_t$. In slip, the tangential traction is constant, equal to~$T^{\text{slip}}$ in~\eqref{eq:numerical-tangential-stresses}$_2$, with $T^{\text{stick}} / |
T^{\text{stick}} | = \pm 1$; this yields $\Delta T_t^{\left( k,
\hspace{0.2em} p \right)} = \Delta T^{\text{slip}} = 0$.

All these results can be collected into a single contact matrix
$\tmmathbf{K}_N^{\left( k, \hspace{0.2em} p \right)}$ of the form
\begin{equation}
\setlength{\arraycolsep}{3.5pt}
  \begin{array}{lll}
    \tmmathbf{K}^{\left( k, \hspace{0.2em} p \right)}_N & = &
    \left(\begin{array}{cc}
      1 & 0\\
      0 & 0
    \end{array}\right) \times \left\{ \begin{array}{ll}
      \sigma_0 \varepsilon_n \alpha^{\left( k, \hspace{0.2em} p \right)} /
      \left( \sigma_0 \alpha^{\left( k, \hspace{0.2em} p \right)} +
      \varepsilon_n \eta \right) & \text{ if } N \in \text{active\_set and } N
      \in \text{grow\_set}\\
      \varepsilon_n & \text{ if } N \in \text{active\_set and } N \notin
      \text{grow\_set}\\
      0 & \text{ otherwise}
    \end{array} \right\}\\[2em]
    &  & \qquad + \left(\begin{array}{cc}
      0 & 0\\
      0 & 1
    \end{array}\right) \times \left\{ \begin{array}{ll}
      \varepsilon_t & \text{ if } N \in \text{stick\_set}\\
      0 & \text{ otherwise}
    \end{array} \right\},
  \end{array} \label{contact-stiffness-matrix-LOG}
\end{equation}
where the definition of the different node sets can be found in Algorithm~\ref{algo:local_update}.

Very similar arguments can be used to justify the expression of the contact
stiffness given in~\eqref{contact-stiffness-matrix} for the linear variant of
the model.

\subsection{Classical Coulomb}\label{C2}

As indicated by~\eqref{eq:expr-of-pi}$_1$ and~\eqref{eq:numerical-contact-pressure}, the
classical model is recovered by taking $\alpha^{\left( k, \hspace{0.2em} p
\right)} = 0$. The pressure increment for nodes in active contact is thus
obtained by setting $\Delta \alpha^{\left( k, \hspace{0.2em} p \right)} = 0$
in~\eqref{linearization-pressure},
\[ \Delta P^{\left( k, \hspace{0.2em} p \right)}_n = \varepsilon_n \Delta
   u^{\left( k, \hspace{0.2em} p \right)}_n . \]
In addition, the tangential traction $T_t^{\left( k, \hspace{0.2em} p \right)}$
and its linearization differ from those of the new model in the slip regime
only, see~\eqref{eq:numerical-tangential-stresses}. The tangential traction for
slip nodes is then given by \ $T_t^{\left( k, \hspace{0.2em} p \right)} =
T_t^{\text{slip}} = \mu P_n^{\left( k, \hspace{0.2em} p \right)}
T_t^{\text{stick}} / | T_t^{\text{stick}} |$, and combining this with
the equation above we obtain
\[ \Delta T_t^{\left( k, \hspace{0.2em} p \right)} = \mu \varepsilon_n 
   \frac{T_t^{\text{stick}}}{| T_t^{\text{stick}} |} \Delta u^{\left( k,
   \hspace{0.2em} p \right)}_n . \]
This yields a contact matrix of the form
\begin{equation}
  \setlength{\arraycolsep}{3.5pt}
  \begin{array}{lll}
    \tmmathbf{K}^{\left( k, \hspace{0.2em} p \right)}_N & = &
    \left(\begin{array}{cc}
      1 & 0\\ 
      0 & 0
    \end{array}\right) \times \left\{ \begin{array}{ll}
      \varepsilon_n & \text{ if } N \in \text{active\_set}\\
      0 & \text{ otherwise}
    \end{array} \right\}\\[1.0em]
    & \quad &  + \left(\begin{array}{cc}
      0 & 0\\ 
      0 & 1
    \end{array}\right) \times \left\{ \begin{array}{ll}
      \varepsilon_t & \text{ if } N \in \text{stick\_set}\\
      0 & \text{ otherwise}
    \end{array} \right\} + \left(\begin{array}{cc}
      0 & 0\\
      1 & 0
    \end{array}\right) \times \left\{ \begin{array}{ll}
      \mu \varepsilon_n  \frac{T_t^{\text{stick}}}{| T_t^{\text{stick}}|} &
      \text{ if } N \in \text{slip\_set}\\
      0 & \text{ otherwise}
    \end{array} \right\}.
  \end{array}  \label{contact-stiffness-matrix-standard}
\end{equation}
Because of the last term, this stiffness matrix is non-symmetric, in contrast to
that of the new model~\eqref{contact-stiffness-matrix}.
\section{Convergence plots for boundary-value problems} \label{appendix:convergence-plots}
This appendix documents the convergence of the numerical solutions for the two boundary-value problems presented in Section~\ref{sec:rela-BVP}. To this end, we plot the norm of the numerical residual defined in~\eqref{contact-residual}$_1$, against the Newton iteration index $p$, for each time step $k\geqslant0$. Convergence curves are given for both variants of the proposed model and for the classical Coulomb friction implementation, see Section~\ref{sec:return-mapping}.
\subsection{Generalized Hertz benchmark test}
\begin{figure}[H]
    \centering
    \includegraphics[width=0.99\linewidth]{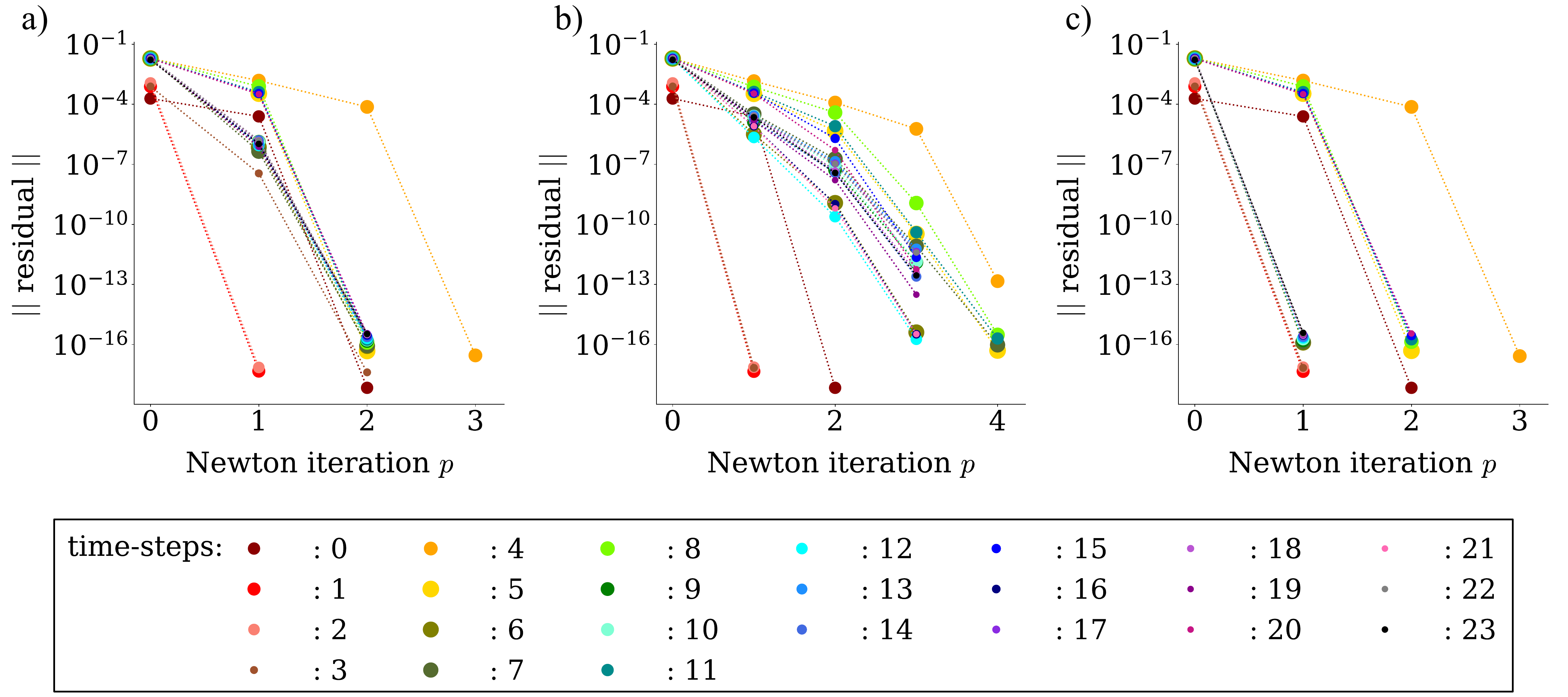}
    \caption{Residual convergence plot for the generalized Hertz benchmark test: a) our linear model~\eqref{eq:global-incremental}, b)~our logarithmic model~\eqref{eq:global-incremental-LOG} and~c) the classical implementation of Coulomb friction.}
    \label{fig:convergence-hertz}
\end{figure}
\subsection{Half-disk sliding on a rigid plane}
\begin{figure}[H]
    \centering
    \includegraphics[width=0.99\linewidth]{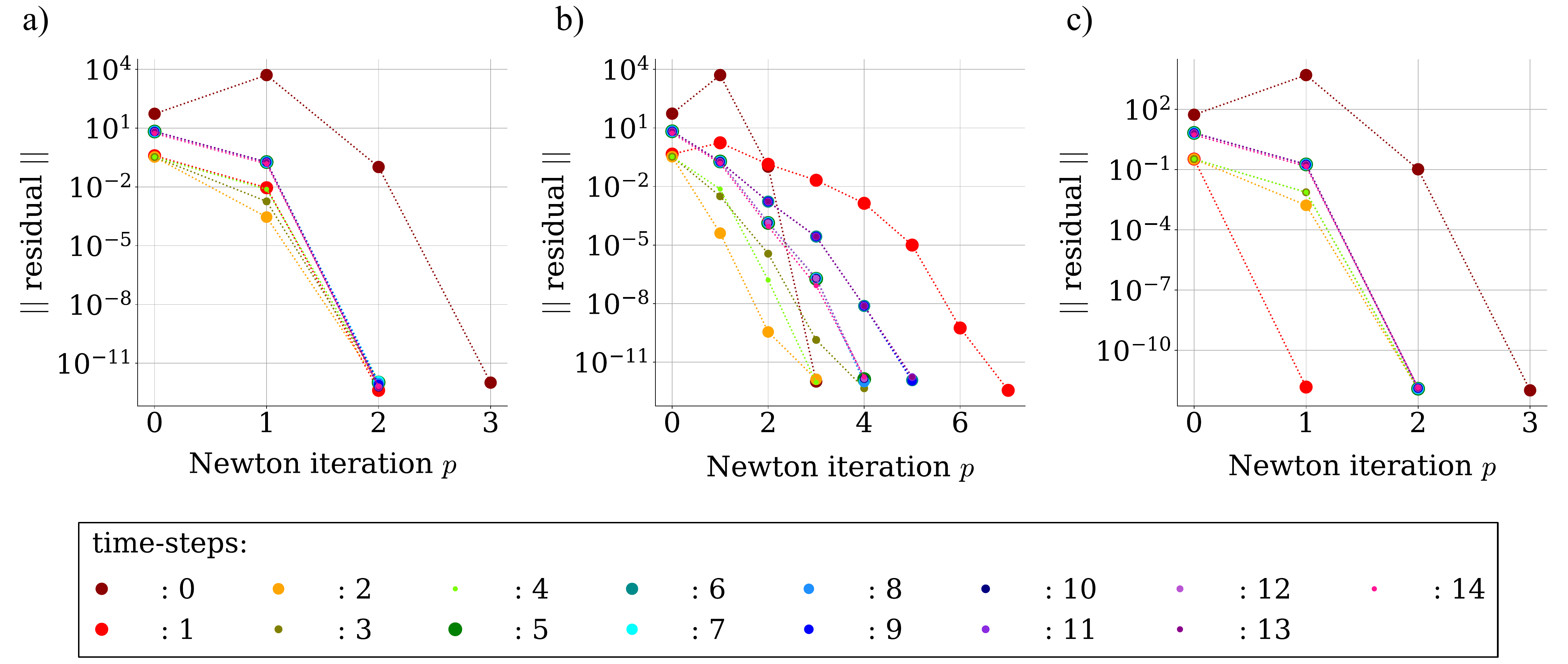}
    \caption{Residual convergence plot for the half-disk sliding on a rigid plane: a) our linear model~\eqref{eq:global-incremental}, b)~our logarithmic model~\eqref{eq:global-incremental-LOG} and~c) the classical implementation of Coulomb friction.}
    \label{fig:convergence-sliding}
\end{figure}

\sloppy
\bibliographystyle{unsrt}
\bibliography{working-biblio}

\end{document}